\documentclass[12pt,pre,pas,showpacs,reprint]{revtex4-1}
\usepackage{graphicx}
\usepackage{geometry}
\usepackage{mathrsfs}
\usepackage{subfigure}
\usepackage{amssymb}
\usepackage{indentfirst}
\usepackage{color}
\usepackage{cancel}
\usepackage{amsmath}
\usepackage{hyperref}
\usepackage[]{appendix}
\usepackage{float}
\begin{document}
\title{\LARGE\textbf{Precise algorithms to compute surface correlation functions of two-phase heterogeneous media and their applications}}
\author{Zheng Ma}
\affiliation{Department of Physics, Princeton University, Princeton, New Jersey 08544, USA}

\author{Salvatore Torquato}
\email{torquato@electron.princeton.edu} 
\affiliation{Department of Chemistry, 
Department of Physics, Princeton Institute for the Science and Technology of Materials,\\
and Program in Applied and Computational Mathematics, Princeton University,
Princeton, New Jersey 08544, USA}
\begin{abstract}
The quantitative characterization of the microstructure of random heterogeneous media in $d$-dimensional Euclidean space $\mathbb{R}^d$ via a variety of $n$-point correlation 
functions is of great importance, since the respective infinite set determines the effective physical properties of the media. In particular, surface-surface $F_{ss}$ and surface-void $F_{sv}$ correlation functions (obtainable from radiation scattering experiments) contain crucial interfacial information that enables one to estimate transport properties of the media (e.g., the mean survival time and fluid permeability) and complements the information content of the conventional two-point correlation function. However, the current technical difficulty involved in sampling surface correlation functions has been a stumbling block in their widespread use. We first present a concise derivation of the small-$r$ behaviors of these functions, which are linked to the \textit{mean curvature} of the system. Then we demonstrate that one can reduce the computational complexity of the problem, without sacrificing accuracy, by extracting the necessary interfacial information from a cut of the $d$-dimensional statistically homogeneous and isotropic system with an infinitely long line. Accordingly, we devise algorithms based on this idea and test them for two-phase media in continuous and discrete spaces. Specifically for the exact benchmark model of overlapping spheres, we find excellent agreement between numerical and exact results. We compute surface correlation functions and corresponding local surface-area variances for a variety of other model microstructures, including hard spheres in equilibrium, decorated ``stealthy" patterns, as well as snapshots of evolving pattern formation processes (e.g., spinodal decomposition). It is demonstrated that the precise determination of surface correlation functions provides a powerful means to characterize a wide class of complex multiphase microstructures.     
    
\end{abstract}

\maketitle
\newpage

\section{INTRODUCTION}

\indent Random heterogeneous media are ubiquitous and arise in many applications
in physics, materials science, biology, and geophysics. Examples of such media include composites \cite{torquato2013random}, porous materials \cite{zohdi2008introduction, sahimi2003heterogeneous, vasseur2017sphere}, biological tissues \cite{gibson1999cellular, gevertz2008novel}, and even cosmological structures \cite{peebles1993principles}. The quantitative characterization of the structure via higher-order correlation functions of these complex media is of importance in many
fields \cite{torquato2013random, chiu2013stochastic}. In general, an infinite set of correlation functions are required to exactly determine the effective physical properties of the media \cite{torquato2013random, zohdi2008introduction, sahimi2003heterogeneous}. However, such complete structural information about the medium is generally not available and hence one must settle for reduced information in the form of lower-order correlation functions. The study of these descriptors has proved fruitful and new applications involving these descriptors are constantly coming up, including in reconstructions using state-of-the-art techniques such as neural networks \cite{mosser2017reconstruction, lubbers2017inferring}.\\
\indent There are a variety of two-point structural descriptors, including the two-point correlation function $S_2$ \cite{jiao2007modeling, jiao2008modeling}, two-point cluster function $C_2$ \cite{jiao2009superior}, surface-surface correlation function $F_{ss}$ \cite{torquato1986interfacial}, and surface-void correlation function $F_{sv}$ \cite{torquato1986interfacial}, as well as the pore-size density function $P(\delta)$ \cite{torquato2013random}, that are practically accessible via computer simulations or imaging techniques. Among them, the most well-known descriptor is the standard two-point correlation function $S_2$, which can be obtained from scattering experiments \cite{debye1949scattering, debye1957scattering}. This quantity has been employed to characterize the microstructure and physical properties of heterogeneous materials \cite{torquato2013random}, reconstruct the microstructure of heterogeneous materials \cite{jiao2007modeling, jiao2008modeling}, and recently, to quantify the hyperuniformity of two-phase systems \cite{PhysRevE.94.022122, torquato2016disordered, doi:10.1063/1.4989492}. Although knowledge of the two-point correlation function $S_2$ has proved to be extremely useful, the corresponding correlation functions that characterize the interface of two-phase media such as the specific surface $s$, surface-surface correlation function $F_{ss}(r)$, and surface-void correlation function $F_{sv}(r)$, which contain crucial structural information, have received considerably less attention, especially $F_{ss}$ and $F_{sv}$. This is due partly to the fact that these surface correlation functions are not as easy to sample as $S_2$, which we remedy in this paper, as described in Sec. IV.\\
\begin{figure*}[]
\centering
\subfigure[]{
\includegraphics[width=7.5cm, height=6.5cm]{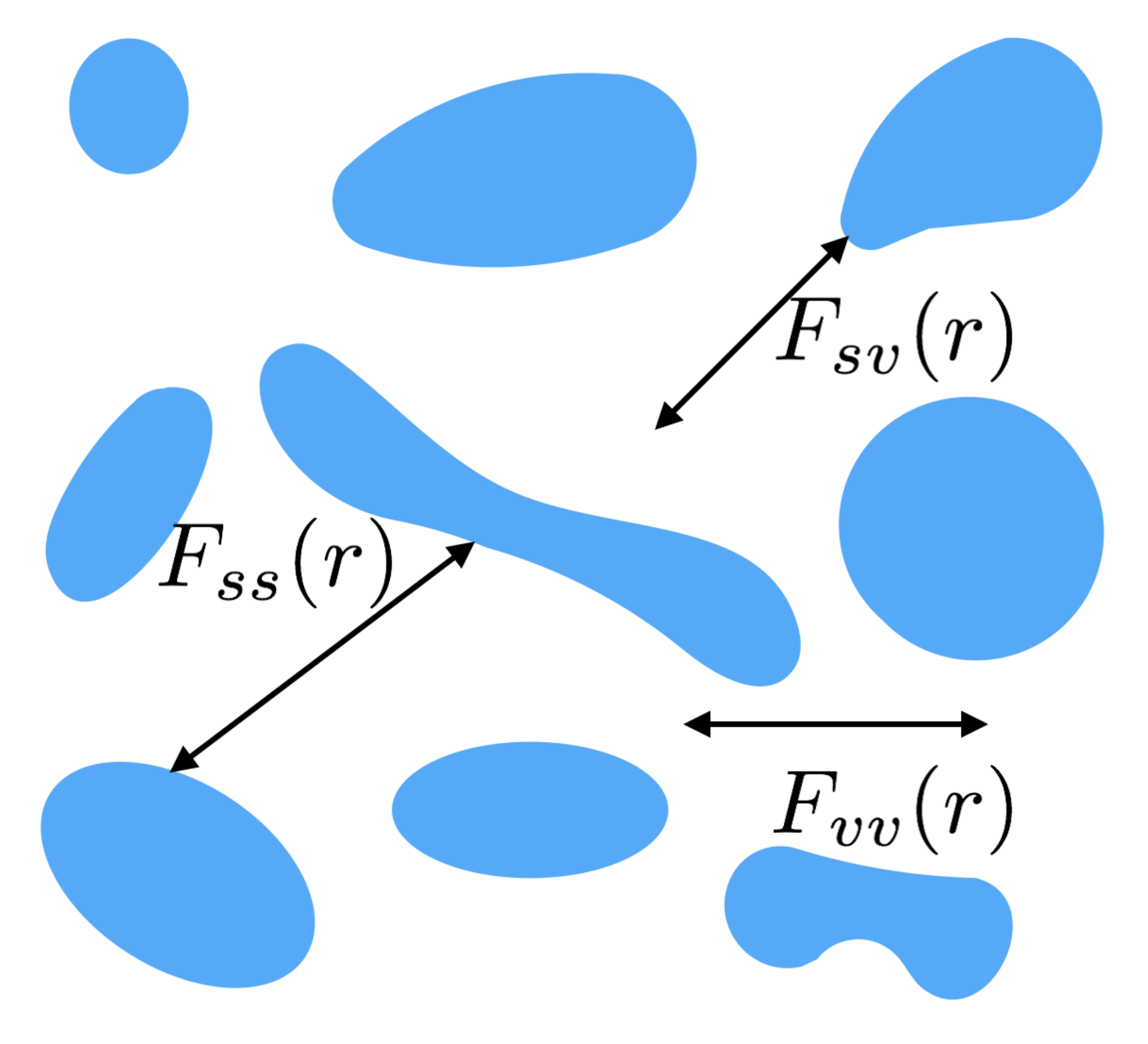}
}
\subfigure[]{
\includegraphics[width=9cm, height=6.75cm]{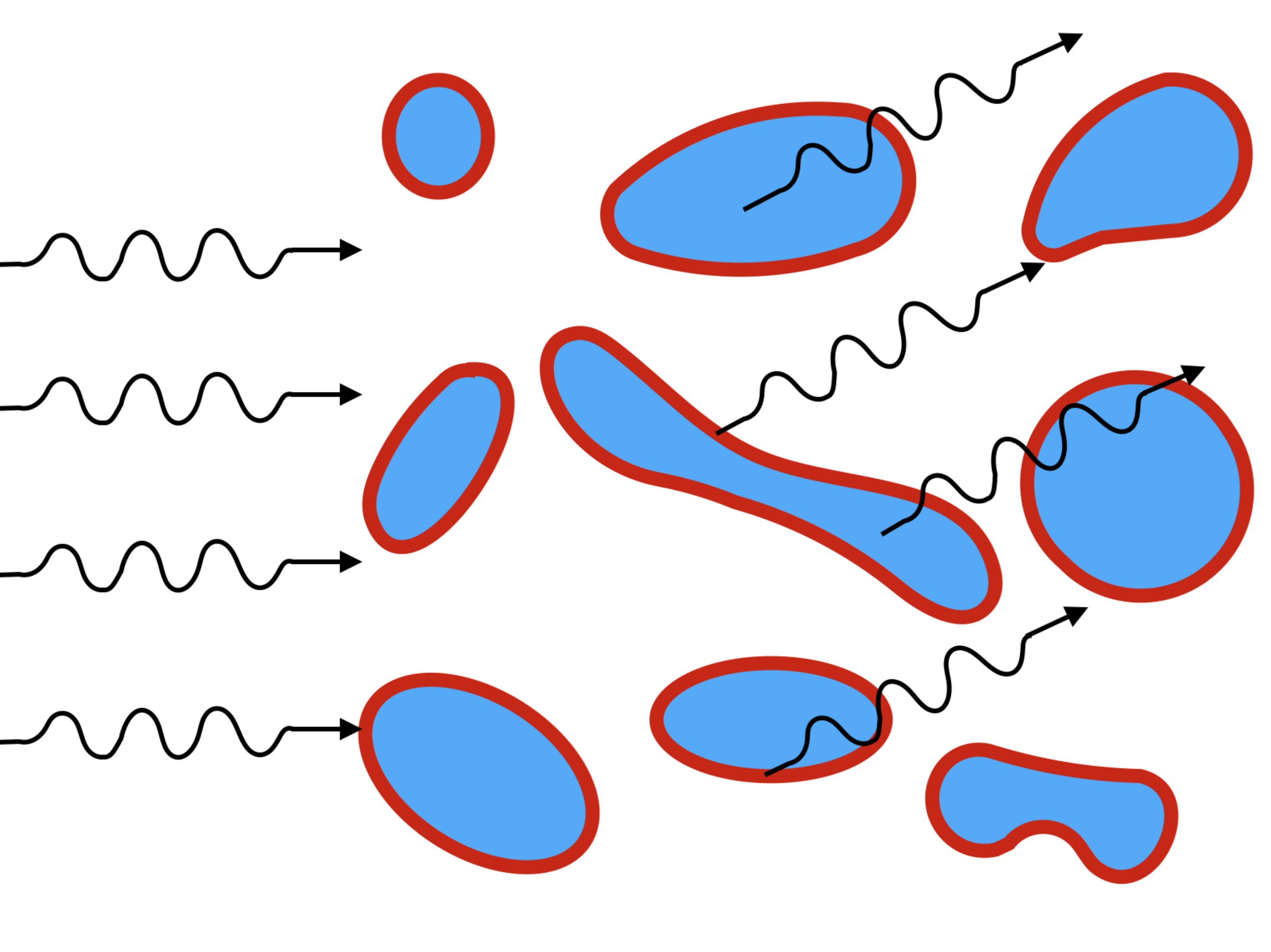}
}
\caption{(a) A schematic two-phase medium showing the surface-surface correlation function $F_{ss}(r)$, surface-void correlation function $F_{sv}(r)$, and void-void correlation function $F_{vv}(r)$ [or $S_2(r)$], where the blue phase is the ``solid" phase. (b) Corresponding schematic showing the scattering of radiation by a two-phase medium, where the blue phase indicates the bulk and the red region indicates the interface.}
\label{fig:cartoon}
\end{figure*}
\indent While the two-point correlation function $S_2(r)$ contains important structural
information, it is usually insufficient to determine both the structure and physical properties of heterogeneous media \cite{yeong1998reconstructing, jiao2009superior, torquato2013random}. It has been shown that supplementing $S_2$ with surface correlation functions can lead to improved reconstructions
of two-phase media \cite{torquato2013random, zachary2011improved, guo2014accurate}.\\ 
\indent These surface correlation functions determine rigorous upper bounds on the fluid permeability $k$ of porous media and mean survival time $\tau$ associated with diffusion-controlled reactions among traps. These two-point ``interfacial-surface" bounds have been shown to be much sharper than the so-called two-point ``void" bound involving $S_2$ alone, reflecting the importance of surface correlation functions. For isotropic media in three-dimensional Euclidean space, these involve the following two key integrals:
\begin{equation}
I_1=\int_{0}^{\infty} [\frac{\phi_1^2}{s^2}F_{ss}(r)-\frac{2\phi_1}{s}F_{sv}(r)+F_{vv}(r)]r dr,
\end{equation} 
\begin{equation}
I_2=\int_{0}^{\infty} [F_{vv}(r)-\phi_1^2]r dr.
\end{equation} 
where $F_{vv}$ is just another way to write $S_2$ for the void phase if we focus on porous media, i.e., $F_{vv} \equiv S_2$, and $\phi_1$ is the volume fraction of the void phase. A schematic plot of the correlation functions is shown in Fig. \ref{fig:cartoon}(a). For statistically isotropic media, the two-point ``interfacial-surface"  upper bound for the fluid permeability is given by \cite{torquato2013random}
\begin{equation}
k \leq \frac{2}{3}I_1,
\end{equation} 
while the two-point ``void" bound is
\begin{equation}
k \leq \frac{2}{3\phi_2^2} I_2,
\end{equation}
where $\phi_2$ is the volume fraction of the solid phase. Similarly, the analogous bounds on the mean survival time are given by \cite{torquato2013random}
\begin{equation}
\tau \leq \frac{I_1}{\phi_1 \mathcal D},
\end{equation} 
and 
\begin{equation}
\tau \leq \frac{I_2}{\phi_1 \phi_2^2 \mathcal D},
\end{equation}
where $\mathcal D$ is the diffusion coefficient of the reactant. The fact that the key integrals in these bounds on the fluid permeability are the same as those for the mean survival time is more than a 
coincidence. Indeed, $k$ is rigorously bounded from above in terms of $\tau$ for general media \cite{torquato1990relationship}. \\
\indent Moreover, two-point correlation functions determine local volume-fraction and local surface-area fluctuations as measured by the relevant variances. These variances enable one to generalize the concept of hyperuniformity \cite{PhysRevE.94.022122}, which was originally conceived in the context of point configurations, namely, it refers to the anomalous suppression of density fluctuations on large length scales \cite{torquato2003local, zachary2009hyperuniformity}. Notably, it is proven that sphere packings will inherit the hyperuniformity of the underlying point pattern \cite{PhysRevE.94.022122}. The local volume-fraction variance $\sigma_{_V}^2(R)$ within a spherical observation window of radius $R$ in $d$-dimensional Euclidean space $\mathbb{R}^d$ is given by \cite{lu1990local} 
\begin{equation} \label{auto01}
\sigma_{_V}^2(R)=\frac {1} {v_1(R)} \int_{\mathbb R^d} \chi_{_V}(\mathbf r)\alpha(r;R)d\mathbf r,
\end{equation}
where 
\begin{equation} \label{auto02}
\chi_{_V}(\mathbf r)=F_{vv}(\mathbf r)-\phi_1^2
\end{equation}
is the autocovariance function associated with $S_2(\mathbf r)$, and $v_1(R)$ is the volume of a $d$-dimensional sphere of radius $R$, and $\alpha(r;R)$ is the scaled intersection volume, the ratio of the intersection volume of two spherical windows of radius $R$ whose centers are separated by a distance $r$ to the volume of a spherical window. A two-phase system is hyperuniform with respect to volume-fraction variances if $\sigma_{_V}^2(R)$ decreases more rapidly than $R^{-d}$ for large $R$ \cite{PhysRevE.94.022122}, or equivalently
\begin{equation} \label{auto4}
\lim_{|\mathbf k|\rightarrow 0} \tilde\chi_{_V}(\mathbf k)=0,
\end{equation}
where $\tilde\chi_{_V}(\mathbf k)$ is the Fourier transform of $\chi_{_V}(\mathbf r)$. Similarly, the local surface-area variance $\sigma_{_S}^2(R)$ has been defined by \cite{lu1990local}
\begin{equation} \label{auto}
\sigma_{_S}^2(R)=\frac {1} {s^2v_1(R)} \int_{\mathbb R^d} \chi_{_S}(\mathbf r)\alpha(r;R)d\mathbf r,
\end{equation}
where
\begin{equation} \label{auto2}
\chi_{_S}(\mathbf r)=F_{ss}(\mathbf r)-s^2
\end{equation}
is the autocovariance function associated with $F_{ss}(\mathbf r)$. A two-phase system is hyperuniform with respect to surface-area variances if $\sigma_{_S}^2(R)$ decreases more rapidly than $R^{-d}$ for large $R$ \cite{PhysRevE.94.022122}, or equivalently
\begin{equation} \label{auto3}
\lim_{|\mathbf k|\rightarrow 0} \tilde\chi_{_S}(\mathbf k)=0,
\end{equation}
where $\tilde\chi_{_S}(\mathbf k)$ is the Fourier transform of $\chi_{_S}(\mathbf r)$. It has been suggested that surface-area fluctuations are more sensitive microstructural measures for heterogeneous media than corresponding volume-fraction fluctuations in some cases \cite{torquato2016disordered, PhysRevE.94.022122}. Results obtained in this paper further support this conclusion.\\
\indent Similar to the two-point correlation function $S_2$, surface correlation functions can be related to and obtained from the scattering intensity as well \cite{strey1994microemulsion, dietrich1995scattering}. In the most general case that involves scattering from both the bulk and the surface [see Fig. \ref{fig:cartoon}(b) for a schematic plot], the scattering intensity can be written as 
\begin{equation} \label{scatter}
I(k) = c_1 \mathcal { \tilde F}(S_2) + c_2 \mathcal  { \tilde F}(F_{ss}) + c_3 \mathcal { \tilde F}(F_{sv}),
\end{equation}
where $c_1, c_2, c_3$ are certain coefficients. When the scattering from the surface is comparable to the bulk, one must consider all these three terms to determine the scattering intensity, while if only bulk or surface scattering is dominant, then one should only care about the corresponding correlation function, this interpretation can potentially provide a general way to understand hyperuniformity in two-phase media.\\ 
\indent The rest of the paper is organized as follows: in Sec. II, we provide necessary definitions and background. In Sec. III, we present 
a concise and simple derivation of the small-$r$ behavior of the two-point surface correlation function, which involves the \textit{mean curvature} of the entire system. In Sec. IV, we introduce and describe a general algorithm that enables the efficient computation of $F_{ss}$ and $F_{sv}$. We verify the accuracy of our algorithm by applying it to overlapping spheres for which we have exact results \cite{torquato2013random}. In Sec. V, we show how to apply the algorithm to treat digitized two-phase media, which is of practical importance. Using Gaussian random fields as an example, we will demonstrate that the image resolution and some drop-out in sampling are crucial in order to obtain reliable results. In Sec. VI we explicitly show results of overlapping spheres, hard spheres in equilibrium and decorated stealthy point patterns. In Sec. VII we explicitly show results of patterns from spinodal decomposition and patterns from the Swift-Hohenberg equation. Using these examples, we demonstrate how surface correlation functions will be very useful for microstructural characterization and can be superior to $S_2$ in certain cases. Finally, in Sec. VIII, we make concluding remarks and discuss the implications of our findings.

\section{BACKGROUND AND DEFINITIONS}

\indent A two-phase random medium is a domain of space $\mathcal V \subseteq \mathbb{R}^d$ that is partitioned into two disjoint regions that make up $\mathcal V$: a phase 1 region $\mathcal V_1$ of volume fraction $\phi_1$ and a phase 2 region $\mathcal V_2$ of volume fraction $\phi_2$ \cite{torquato2013random}. The phase indicator function $\mathcal I^{(i)}(\mathbf x)$ for a given realization is defined as
\begin{equation} \label{indicator}
\mathcal I^{(i)}(\mathbf x)=\left \{ \begin{aligned} & 1, & \mathbf x \in \mathcal V_i,\\ & 0, & \mathbf x \notin \mathcal V_i. \end{aligned}\right.
\end{equation}
For statistically homogeneous media, the volume fraction for phase $i$
\begin{equation}
\phi_{i}=\left\langle \mathcal I^{(i)}(\mathbf x)\right\rangle
\end{equation}
is a constant. The two-point correlation function is defined as
\begin{equation} \label{s2}
S_2^{(i)}(\mathbf x_1,\mathbf x_2)=\left\langle \mathcal I^{(i)}(\mathbf x_1)\mathcal I^{(i)}(\mathbf x_2)\right\rangle.
\end{equation} 
For homogeneous media, this quantity only depends on the relative displacement vector $\mathbf r\equiv \mathbf x_2-\mathbf x_1$. The two-point correlation function simplifies as $S_2(\mathbf x_1,\mathbf x_2)=S_2(\mathbf r)$. If the system is also statistically isotropic, then $S_2(r)$ depends only on the radial distance $r =|\bf{r}|$.\\
\indent The interface indicator function is defined as \cite{torquato2013random}
\begin{equation} \label{indicator1}
\mathcal M(\mathbf x)=|\nabla \mathcal I^{(1)}(\mathbf x)|=|\nabla \mathcal I^{(2)}(\mathbf x)|.
\end{equation} 
The specific surface is the expected area of the interface per unit volume, and for homogeneous media is simply the ensemble average of the interface indicator function, i.e.,
\begin{equation} \label{sss}
s=\left\langle \mathcal M(\mathbf x) \right\rangle.
\end{equation} 
The surface-surface correlation function measures the correlation between two points on the interface, and for homogeneous media is defined as
\begin{equation} \label{deffss}
F_{ss}(\mathbf r)=\left\langle \mathcal M(\mathbf x)\mathcal M(\mathbf x+\mathbf r) \right\rangle.
\end{equation}
The surface-void correlation function measures the correlation between one point on the interface and the other in the void phase, and for homogeneous media is defined as 
\begin{equation} \label{deffsv}
F_{sv}(\mathbf r)=\left\langle \mathcal M(\mathbf x)\mathcal I^{(void)}(\mathbf x+\mathbf r) \right\rangle.
\end{equation}
Higher-order surface correlation functions are similarly defined \cite{torquato2013random}, but the focus in this paper will be the two-point varieties.\\
\indent Closed-form expressions for the two-point surface correlation functions are very limited. The most notable one is for the model of overlapping spheres \cite{torquato2013random, chiu2013stochastic}, which is generated by circumscribing spheres of radius $a$ around each point in a Poisson point process with density $\rho$. The space interior to the spheres is the solid phase and the space exterior is the void phase \cite{torquato2013random}. For statistically homogeneous overlapping spheres in three dimensions, we have
\begin{align}
\label{overfss}
&F_{ss}(r)=S_2(r)\\\nonumber 
&\left\{\frac {9\eta^2}{a^2}[1-(\frac{1}{2}-\frac{r}{4a})\Theta(2a-r)]^2+\frac{3\eta}{2ra}\Theta(2a-r)\right\},
\end{align}
and
\begin{equation} \label{overfsv}
F_{sv}(r)=\frac {3\eta}{a}[1-(\frac{1}{2}-\frac{r}{4a})\Theta(2a-r)]S_2(r),
\end{equation}
where $r=|{\bf r}|$ is a radial distance, $\eta=\rho v_1(a)$ is a reduced density and $\Theta(x)$ is the Heaviside step function. Here,
\begin{equation} 
S_2(r)=\exp(\frac{-\eta v_2(r;a)}{v_1(a)})
\end{equation} is the two-point correlation function for the ``void'' phase. These relations were first given by Doi \cite{doi1976new}.\\
\indent Using the canonical function $H_n$ \cite{torquato2013random}, Torquato derived the following expressions for surface correlation functions for hard spheres:
\begin{equation} \label{hardfss}
F_{ss}(r)=\frac{s}{2r}\Theta(2a-r)+s^2+\rho^2  \delta \otimes \delta \otimes h,
\end{equation}
and
\begin{equation} \label{hardfsv}
F_{sv}(r)=s-\frac{s}{2}(1-\frac{r}{2a}) \Theta(2a-r)-s\eta-\rho^2 m \otimes \delta \otimes h,
\end{equation}
where $s=d\eta/a$ is the specific surface, $\delta$ is the radial Dirac delta function, and $m$ is the sphere indicator function. The quantity $h(\mathbf r)$ is the total correlation function defined as $h(\mathbf r)=g_2(\mathbf r)-1$, where $g_2(\mathbf r)$ is the pair correlation function, and $\otimes$ denotes the convolution of two functions.\\
\indent The impenetrability constraint alone is not sufficient to specify the hard-sphere model; a hard-sphere
system can be in equilibrium or be derived from an infinite number of nonequilibrium ensembles \cite{torquato2013random}. The pair correlation function is generally not known for nontrivial hard-sphere models for all densities, an exception being the ``ghost" random sequential addition packing model \cite{torquato2006exactly}. For $d=3$, Torquato used the Percus-Yevick approximation and the Verlet-Weis correction to evaluate these functions for statistically isotropic systems of hard spheres in equilibrium \cite{torquato1986interfacial}. These are useful benchmark results that will be used in Sec VI. To date, numerical evaluations of the surface correlation functions have been limited to hard spheres in equilibrium \cite{torquato1986interfacial, seaton1986spatial} and maximally random jammed sphere packings \cite{klatt2016characterization}.

\section{SOME THEORETICAL REMARKS ON SURFACE CORRELATION FUNCTIONS}
\subsection{The small-$r$ behavior of $F_{ss}$ and $F_{sv}$ in general}
\begin{figure*}[]
\centering
\subfigure[\ $F_{ss}(r)$]{
\includegraphics[width=8cm, height=5.5cm]{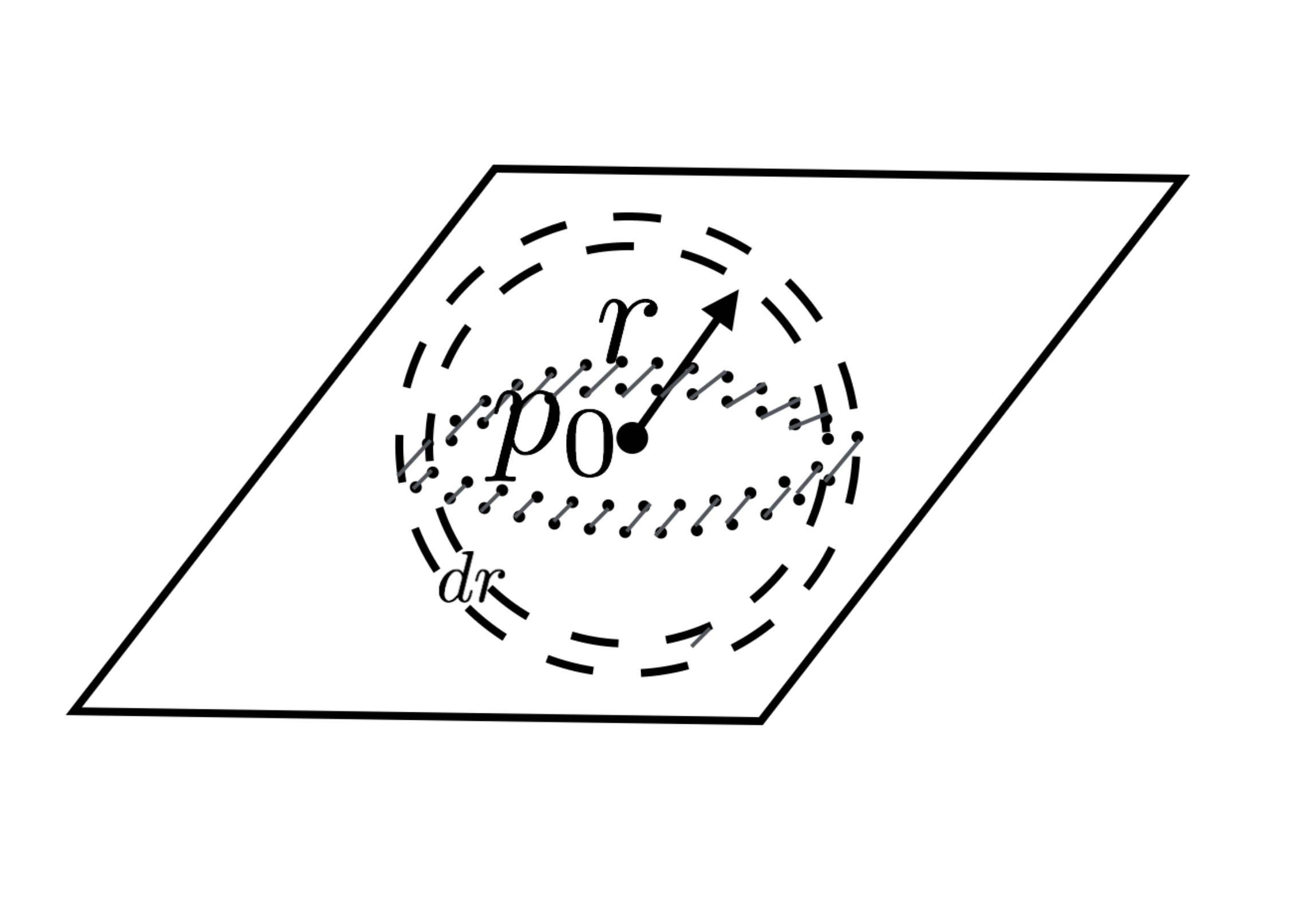}
}
\subfigure[\ $F_{sv}(r)$]{
\includegraphics[width=6.8cm, height=5.9cm]{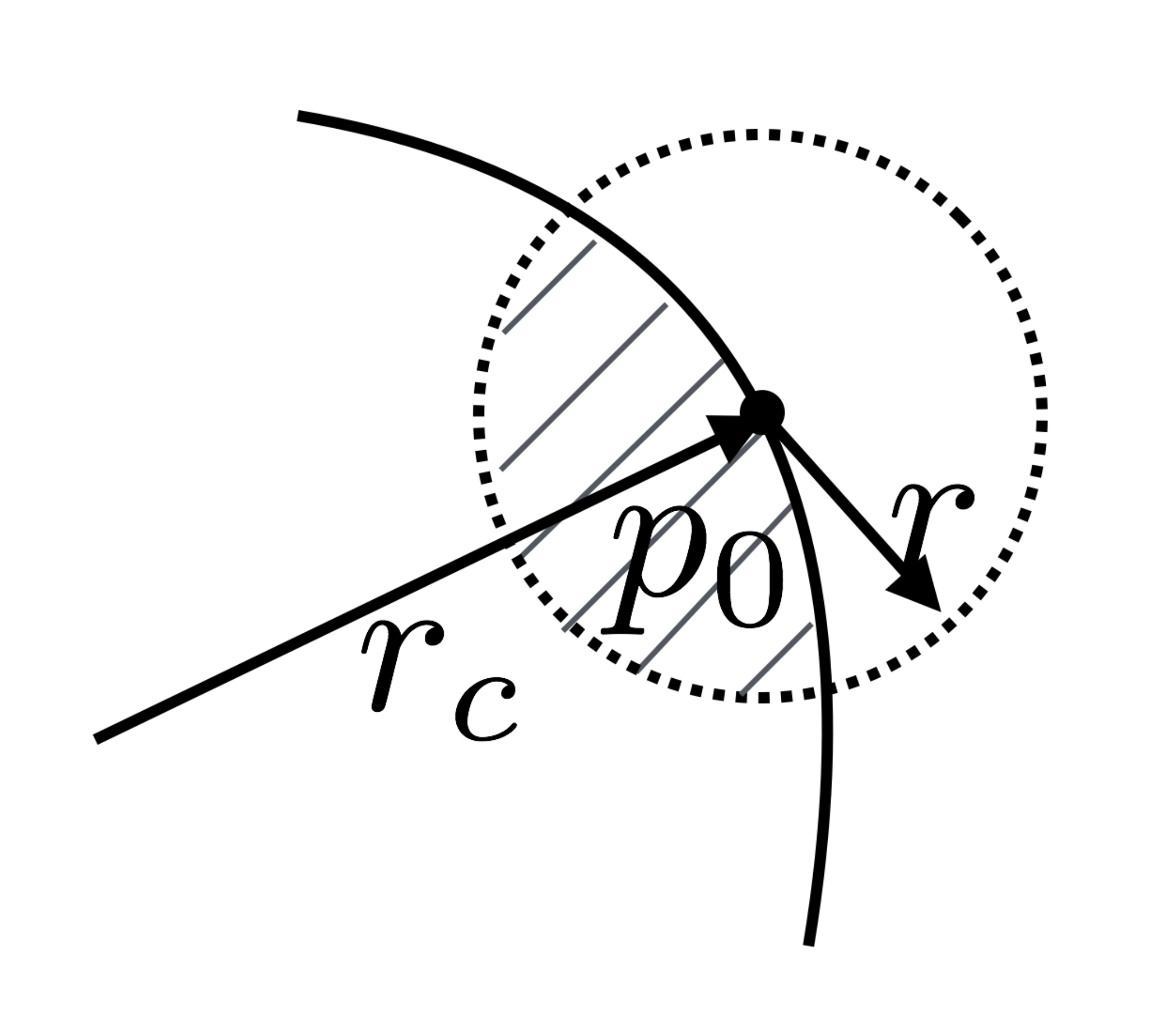}
}
\caption{(a) Schematic that illustrates the small-$r$ asymptotic behavior of the surface-surface correlation function in three dimensions in which the vicinity of the $p_0$ is approximated by a plane, where the area of interface contained in the shell is shaded. (b) Schematic that illustrates the small-$r$ asymptotic behavior of the surface-void correlation function in two dimensions, where $r_c$ is the local radius of curvature of the interface.} 
\label{fig:derive}
\end{figure*}

\indent Debye and co-workers \cite{debye1949scattering, debye1957scattering} showed that the slope of $S_2(r)$ at the origin ($r=0$) is directly proportional to the specific surface $s$, which enables people to obtain the surface area of the whole system by measuring the tail of a scattering profile. The small-$r$ behavior of the two-point surface correlation functions have been derived previously by taking higher-order derivatives of $S_2(r)$ of a dilated interface and then letting the thickness go to zero \cite{teubner1990scattering, ciccariello1981correlation}. Here we present a much simpler derivation based on a probabilistic interpretation of the surface correlation functions.\\    
\indent We restrict ourselves to the discussion of systems with interfaces that are differentiable everywhere. This assumption enables us to approximate the vicinity of a point on the interface with planes or spheres in the following discussion.\\
\indent The small-$r$ behavior of the surface-surface correlation function is straightforward to obtain. First, randomly pick a reference point $p_0$ on the interface (with specific surface $s$ for the entire system). Second, consider a concentric shell with radius $r$ to $r+dr$ around the reference point, then a local specific surface of the shell can be defined as $dr \rightarrow 0$. The quantity $F_{ss}(r)$ is then the product of the specific surface $s$ of the system and the average local specific surface over the interfaces (the local specific surface is defined at every point on interfaces, thus can be integrated to compute the average). We present a schematic plot that elucidates the derivation in Fig. \ref{fig:derive}(a) in three dimensions. When $r$ is very small, the vicinity of $p_0$ is basically flat for the zeroth-order approximation [see the quadrangle in Fig. \ref{fig:derive}(a)] and there is no other interface intersecting with the shell. As shown in Fig. \ref{fig:derive}(a), the area of interface contained in the shell is $2\pi r dr$, and the volume of the shell is $4\pi r^2 dr$, so 
\begin{equation}
F_{ss}(r) \sim s\times \left\langle \frac{2\pi r dr} {4 \pi r^2 dr} \right\rangle=\frac{s}{2r}, \quad  r \rightarrow 0.
\end{equation}
For $d=2$, following the same method we have 
\begin{equation}
F_{ss}(r) \sim s\times \left\langle \frac{2 dr}{2\pi r dr} \right\rangle=\frac{s}{\pi r}, \quad  r \rightarrow 0.
\end{equation}
Since the zeroth-order approximation of $F_{ss}(r)$ is divergent as $r \rightarrow 0$, we will not discuss higher-order finite correction terms here \cite{teubner1990scattering}.

\indent The determination of the small-$r$ behavior of the surface-void correlation function is more involved. Again, we randomly pick a reference point $p_0$ on the interface. Next consider a ``test" sphere of radius $r$ centered at the reference point. We denote by $P_{sv}(p_0)$ the following conditional probability: given a point $p_0$ on the interface, the probability that a uniformly and randomly placed vector $\bf r$ emanating from $p_0$ lands in the void phase. The quantity $F_{sv}(r)$ is then the product of the specific surface of the system $s$ and the average of $P_{sv}(p_0)$ over the interfaces, i.e., $F_{sv}=s\left\langle P_{sv}(p_0) \right\rangle $. Since it is more convenient to illustrate the basic idea behind the computation in two dimensions and the result can be easily generalized to three dimensions, a schematic plot that elucidates our approach is illustrated in Fig. \ref{fig:derive}(b) in two dimensions, where the shaded area is the part of the small ``test" sphere contained in the solid phase. In $d=2$, under aforementioned assumptions, we can approximate the interface with a circular arc of radius of curvature $r_c(p_0)$. Then we can work out the probability $P_{sv}(p_0)$ up to the first-order approximation with respect to $r$, which writes as 
\begin{equation}
P_{sv}(p_0)=\frac{1}{2}+\frac{r}{2\pi r_c(p_0)}.
\end{equation}
Average out this quantity on interfaces, we get the final result 
\begin{equation}  \label{2dFsv}
F_{sv}(r)=s(\frac{1}{2}+\frac{r}{2\pi} \left\langle \frac{1}{r_c(p_0)}\right\rangle),
\end{equation}   
where $\left\langle 1/r_c(p_0) \right\rangle$ is the average of $1/r_c(p_0)$ on interfaces.\\
\indent Following the same procedure in three dimensions, we find
\begin{equation}
P_{sv}(p_0)=\frac{1}{2}+\frac{r}{4r_c(p_0)}.
\end{equation}
Thus we have
\begin{equation} \label{3dFsv}
F_{sv}(r)=s(\frac{1}{2}+\frac{r}{4} \left\langle \frac{1}{r_c(p_0)}\right\rangle).
\end{equation}
However, in three dimensions, the curvature varies when the normal plane rotates, and hence here $1/r_c(p_0)$ is to be interpreted to be the \textit {mean curvature} at the point. One should also notice that $1/r_c(p_0)$ is a signed quantity in general, although we only illustrate the positive situation in Fig. \ref{fig:derive}(b) for the sake of simplicity. Note our simple approach 
can be easily extended to derive the small-$r$ behavior in higher dimensions. In any $d$ dimension, we find 
\begin{equation} \label{dFsv}
F_{sv}(r)=s(\frac{1}{2}+\frac{r}{2B(\frac{d-1}{2},\frac{1}{2})} \left\langle \frac{1}{r_c(p_0)}\right\rangle), \\ 
\end{equation}
where $B(\frac{d-1}{2},\frac{1}{2})$ is the beta function. Using this approach, the connection between the small-$r$ behavior of $F_{sv}(r)$ and 
mean-curvature interfacial growth problems \cite{von1952metal} is intuitively clear.\\  
\indent {\it Remarks:}
Note that when the ``test" sphere of $p_0$ intersects with the nondifferentiable singularities, such as edges or corners, the derivation above breaks down. Thus, Eqs. (\ref{2dFsv}) and (\ref{3dFsv}) do not hold in general for interfaces that have singularities, even though the integrated mean curvature may still be defined and computed in these systems \cite{mecke1998integral}. Using the same approach, we obtain in Appendix A some results for certain systems in which the interfaces have singularities. A discussion of these issues can be found in Ref. \cite{ciccariello1982singularities}.  

\subsection{Phase-interchange relations for $F_{sv}$}
\indent Here we remark on phase-interchange relations involving the surface-void correlation function $F_{sv}$. For a two-phase medium, since the sum of indicator functions for phase 1 and phase 2 is unity everywhere, we have  
\begin{equation} \label{nouse1}
\left\langle \mathcal M(\mathbf x)[\mathcal I^{(1)}(\mathbf x+\mathbf r)+\mathcal I^{(2)}(\mathbf x+\mathbf r)]\right\rangle=\left\langle \mathcal M(\mathbf x)\right\rangle,
\end{equation}
implying that the sum of the two surface-void correlation functions for phases 1 and 2 equals the specific
surface, i.e.,
\begin{equation} \label{nouse2}
F_{sv}^{(1)}(\mathbf r)+F_{sv}^{(2)}(\mathbf r)=s.
\end{equation}
Furthermore, if the two phases are statistically the same, these surface-void correlation functions
are constants \cite{teubner1990scattering}, namely,
\begin{equation} \label{nouse3}
F_{sv}^{(1)}(\mathbf r)=F_{sv}^{(2)}(\mathbf r)=\frac{s}{2},
\end{equation}
which is a remarkable relation given that it applies to complex microstructures with such symmetries. This 
will be verified in Sec. VII.

\section{PRECISE ALGORITHMS TO COMPUTE BOTH $F_{ss}$ AND $F_{sv}$}

\begin{figure}[H] 
\centering
\includegraphics[width=7cm,height=6.7cm]{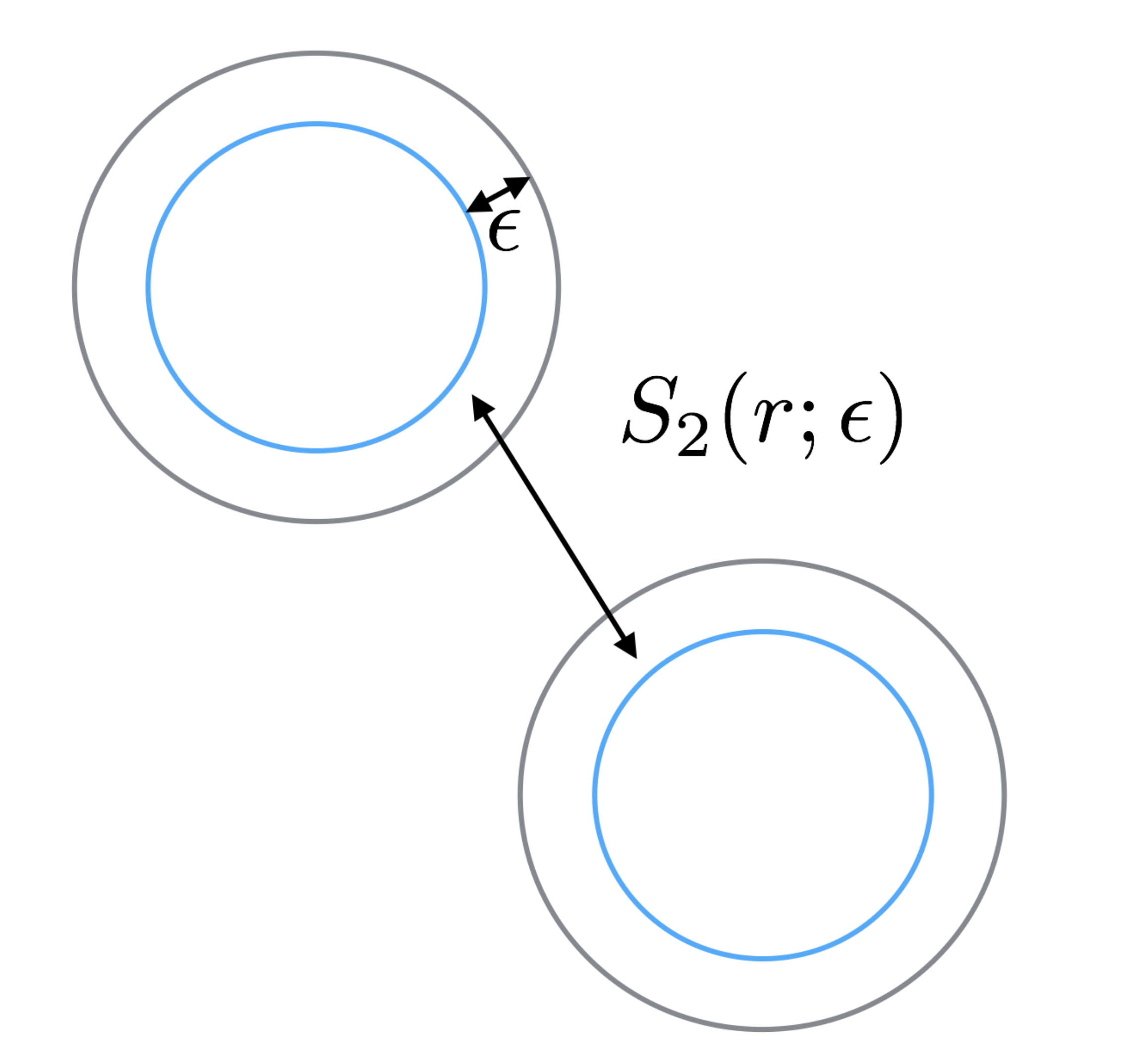}
\caption{An illustration of the previous algorithm that computes $F_{ss}(r)$, where $\epsilon$ is the dilation thickness. Then $F_{ss}(r)$ is computed by $S_2(r;\epsilon)/\epsilon^2$ as $\epsilon \rightarrow 0$.}
\label{fig:old}
\end{figure}

\indent Despite the fact that surface correlation functions contain crucial microstructural information, the technical difficulty involved in computing them has been a stumbling block in their widespread use. Methods have been devised to compute the surface correlation functions for dispersions of spheres that rely on dilating the interfaces \cite{seaton1986spatial, klatt2016characterization}. A schematic illustration of how the algorithm works for $F_{ss}$ is presented in Fig. \ref{fig:old}, where $\epsilon$ is the dilation thickness. The algorithm simply measures the two-point probability function $S_2(r;\epsilon)$ of the dilated phase and then one takes the appropriate limit of $\epsilon$. Surface-surface correlation functions have also been used as input information to reconstruct two-phase digitized materials by Jiao, Stillinger, and Torquato \cite{jiao2009superior}. Since reconstruction algorithms require numerous evaluations of evolving microstructures, the surface correlation functions were approximated to improve computational speed.\\

\subsection{Algorithmic Details}

\indent Here we describe efficient general algorithms that enable the precise determination of the surface-surface correlation function $F_{ss}(r)$ and the surface-void correlation function $F_{sv}(r)$ for most situations that one may encounter in simulations and experiments. We consider $d$-dimensional statistically homogeneous and isotropic two-phase systems within a cubic fundamental simulation cell of side length $L$ under periodic boundary conditions. We also assume that the interfaces are differentiable almost everywhere with exceptions for corners and edges only.\\
\indent The idea behind the algorithm is that one can reduce the complexity of the problem by extracting information from a cut of the $d$-dimensional statistically homogeneous and isotropic system with a $m$-dimensional subspace ($m=1,2,\ldots, d-1$) \cite{torquato2013random}. For example, the fully three-dimensional two-point correlation function $S_2(r)$ of such a two-phase system is the same as the one-dimensional two-phase system formed from the cut of the original system with an infinitely long line. Similar ideas can be exploited to compute surface correlation functions as well. A straight line intersects with the interface in $\mathbb{R}^d$ and leaves infinitely many intersection points, in principle. We can recover the fully three-dimensional surface correlation functions of the system by analyzing these intersections, but here we need to weight the points in accordance with the fact that the line cuts through the interface at different angles at each intersection point. In particular, because the interface projects to the line differently, each intersection point carries the weight $1/\cos \theta$, where $\theta$ is the acute angle between the straight line and the normal vector at the intersection point. From a ``dilation" point of view, the straight line will cut through the dilated phase and leave line segments with lengths $\epsilon/\cos \theta$, then $S_2(r;\epsilon)/\epsilon^2$ will reduce to the pair correlation function of intersection points with weights $1/\cos \theta$ in the limit of $\epsilon \rightarrow 0$.\\  
\indent Using this simple observation, the calculation of the surface-surface correlation function $F_{ss}(r)$ consists of the following steps: \\

{\it}

\noindent 1. Generate a straight line parallel to one of the edges of the box at a random position.   \\

\noindent 2. Find all the intersection points ($P_1$, $P_2$,...$P_n$) with interfaces of the system along this straight line. Store their positions $x_1$, $x_2$...$x_n$.\\

\noindent 3. Find the normal vectors at each intersection point and the angles (the acute one) between the straight line and these norm vectors $\theta_1$, $\theta_2$...$\theta_n$. Compute $1/\cos\theta_1$, $1/\cos\theta_2$...$1/\cos\theta_n$.\\

\noindent 4. Bin the distance between every pair of intersection points (suppose the size of each bin is $L_\text{bin}$). Add $1/(\cos\theta_i\cos\theta_j)$ to the corresponding bin.\\

\noindent 5. Normalize the value in each bin by dividing $2LL_\text{bin}$.\\

\noindent 6. Repeat the process from the beginning. \\

\noindent 7. Compute the average of the results.  \\

\begin{figure}[H] 
\centering
\includegraphics[width=8cm,height=4cm]{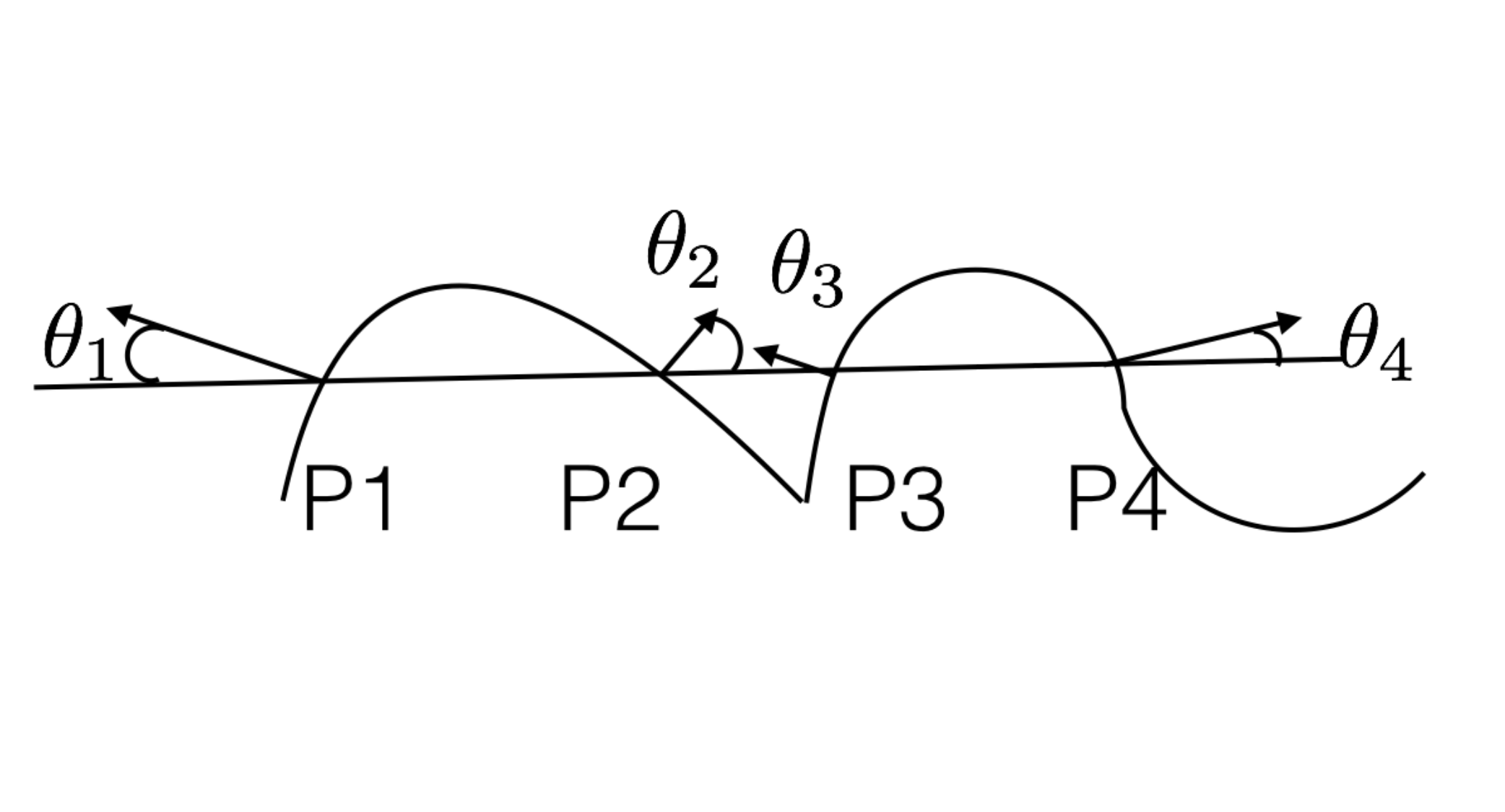}
\caption{A schematic plot that elucidates our algorithm that computes surface correlation functions. Here the sampling straight line intersects with the interface at the points $P_1$, $P_2$, $P_3$, and $P_4$.}
\label{fig:new}
\end{figure}

The calculation of the surface-void correlation function $F_{sv}(r)$ consists of the following steps:\\
{\it}

\noindent 1. Generate a straight line parallel to one of the edges of the box at a random position. \\

\noindent 2. Find all the intersection points ($P_1$, $P_2$,...$P_n$) with interfaces of the system alone this straight line. Store their positions $x_1$, $x_2$...$x_n$.\\

\noindent 3. Find the normal vectors at each intersection point and the angles (the acute one) between the straight line and these norm vectors $\theta_1$, $\theta_2$...$\theta_n$. Compute $1/\cos\theta_1$, $1/\cos\theta_2$...$1/\cos\theta_n$.\\

\noindent 4. Generate $t$ random points along the straight line. Determine whether each point is in the void phase or not. Suppose $Q_1$, $Q_2$,...$Q_m$ are the points in the void phase, store their positions $y_1$, $y_2$...$y_m$. \\

\noindent 5. Bin the distance between every pair of $P_i$ and $Q_j$ (suppose the size of each bin is $L_\text{bin}$). Add $1/\cos\theta_i$ to the corresponding bin.\\

\noindent 6. Normalize the value in each bin by dividing $2tL_\text{bin}$.\\

\noindent 7. Repeat the process from the beginning.\\

\noindent 8. Compute the average of the results.\\

\indent A schematic plot that elucidates our new algorithm is shown in Fig. \ref{fig:new}. For systems with hard-wall boundary conditions (the usual case for experimental images), the value in the $k$th bin should be multiplied by a factor $L/(L-kL_\text{bin})$ due to the fact that fewer pairs can be formed near both ends of the boundaries. One can also easily generalize the algorithm to anisotropic media and to higher-order correlation functions such as $F_{ssv}$ and $F_{svv}$ \cite{torquato2013random}.\\
\indent For a $d$-dimensional system consisting of $N$ particles (or voxels), the complexity for generating a single sampling line is $\mathcal O(N^{1/d})$. Computing each pair of intersection points on the line requires $\mathcal O(N^{2/d})$. The number of sampling lines is usually a preset number, and thus the overall complexity for computing surface-surface correlation function is $\mathcal O(N^{2/d})$. By a similar analysis, we know the complexity for computing surface-void correlation function is $\mathcal O(N^{1/d})$. Note that our algorithms are as efficient as the approximation method used in reconstructions \cite{jiao2009superior}, but with much better accuracy. Actually, as $N$ increases, fewer lines are needed, since each line contains more intersection points. If the total number of pairs we want to sample is fixed, both algorithms can give constant time complexity.

\subsection{Testing Against the Benchmark of Overlapping Spheres}
\begin{figure}[H]
\centering
\subfigure[\ $F_{ss}(r)$]{
\includegraphics[width=8cm, height=6cm]{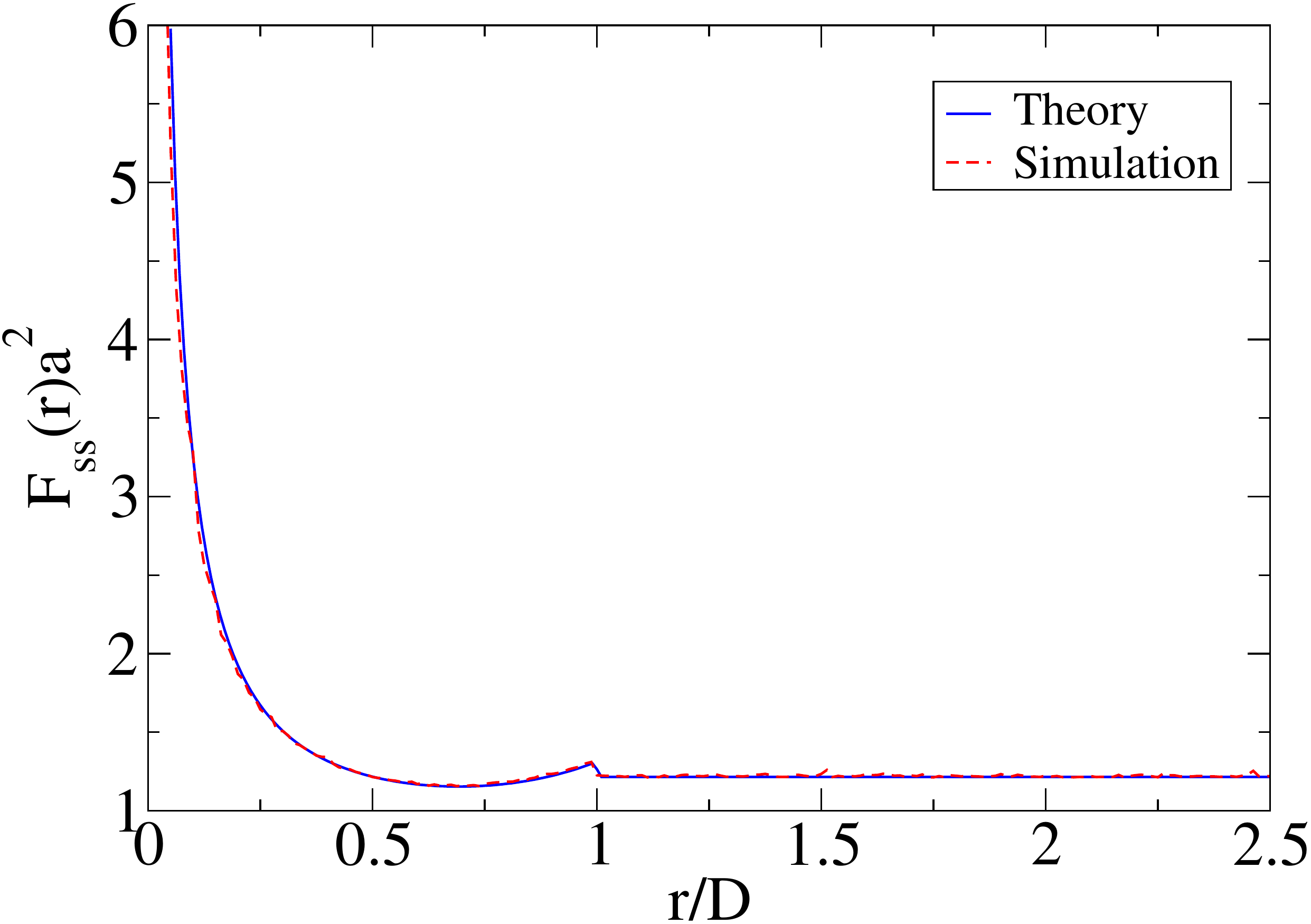}
}
\subfigure[\ $F_{sv}(r)$]{
\includegraphics[width=8cm, height=6cm]{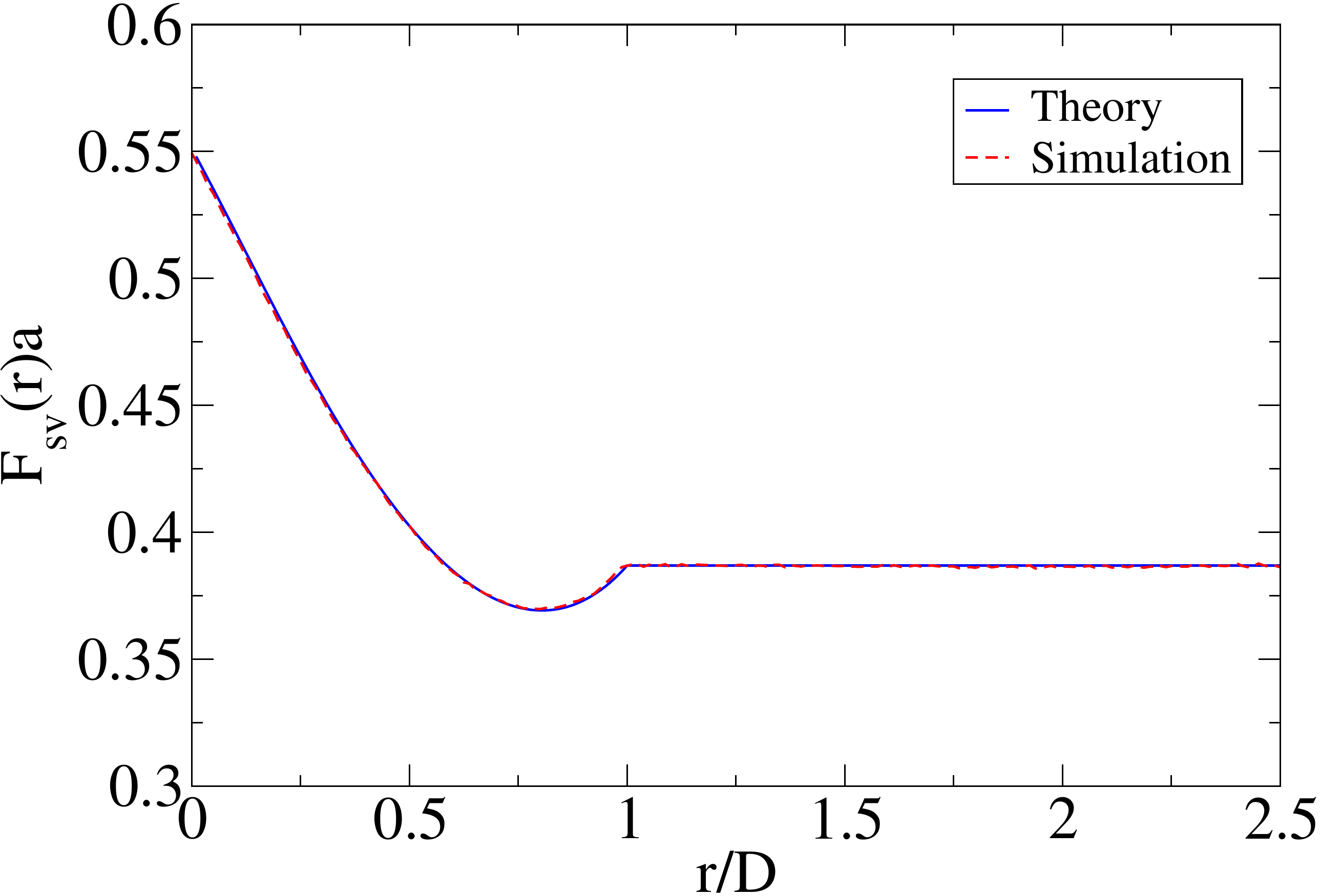}
}
\caption{Comparison of theoretical and simulation results of surface-surface correlation function $F_{ss}(r)$ and surface-void correlation function $F_{sv}(r)$ for overlapping spheres in three dimensions, where $D$ is the diameter of the sphere. The simulations are carried out using 250,000 overlapping spheres in a cubic box under periodic boundary conditions using 1000,000 line samples. The particle-phase volume fraction $\phi$ is 0.649. }
\label{fig:bench}
\end{figure}

\indent Three-dimensional overlapping sphere systems provide an excellent benchmark to test our algorithm, since the surface correlation functions are known exactly; see Eq. (\ref{overfss}) and Eq. (\ref{overfsv}). We generate a single but large configuration consisting of 250,000 overlapping spheres with a reduced density $\eta=1.047$ and particle-phase volume fraction $\phi=0.649$. We generate one million straight lines at random locations and on each line we generate 1000 random points ($t=1000$) in the case of computing $F_{sv}$. As we can see from Fig. \ref{fig:bench}, the theoretical and simulation results for the
surface correlation functions are in excellent agreement with one another, even at the nondifferentiable point $r=D$, indicating that the algorithm works remarkably well. As we discussed in Sec. III, the surface-surface correlation function $F_{ss}$ diverges at the origin. We also ran our algorithm at other particle-phase volume fractions and again find excellent agreement with the corresponding theoretical results.   

\section{COMPUTING SURFACE CORRELATION FUNCTIONS FOR DIGITIZED TWO-PHASE MEDIA}

Unlike continuous-space microstructures (e.g., overlapping spheres), where we know surfaces exactly, images of heterogeneous materials are necessarily digitized, which presents algorithmic challenges to identify surfaces and normal vectors. We devote this section to the discussion of how to apply the aforementioned algorithm to this practical setting. Considering that experimental images are generally gray scale, we first discuss the case in which the two-phase medium is obtained from a level cut of a digitized scalar field $F(\mathbf x)$ in $\mathbb{R}^d$ \cite{torquato2013random}. This common way to produce a two-phase medium enables us to identify interface normal vectors by the gradient of the scalar field. We then apply this idea to black and white images by first converting the given two-phase medium to a scalar field.  

\subsection{Two-phase Media Obtained From Level Cuts of Scalar Fields}

\begin{figure}[H]
\centering
\includegraphics[width=8cm,height=5.6cm]{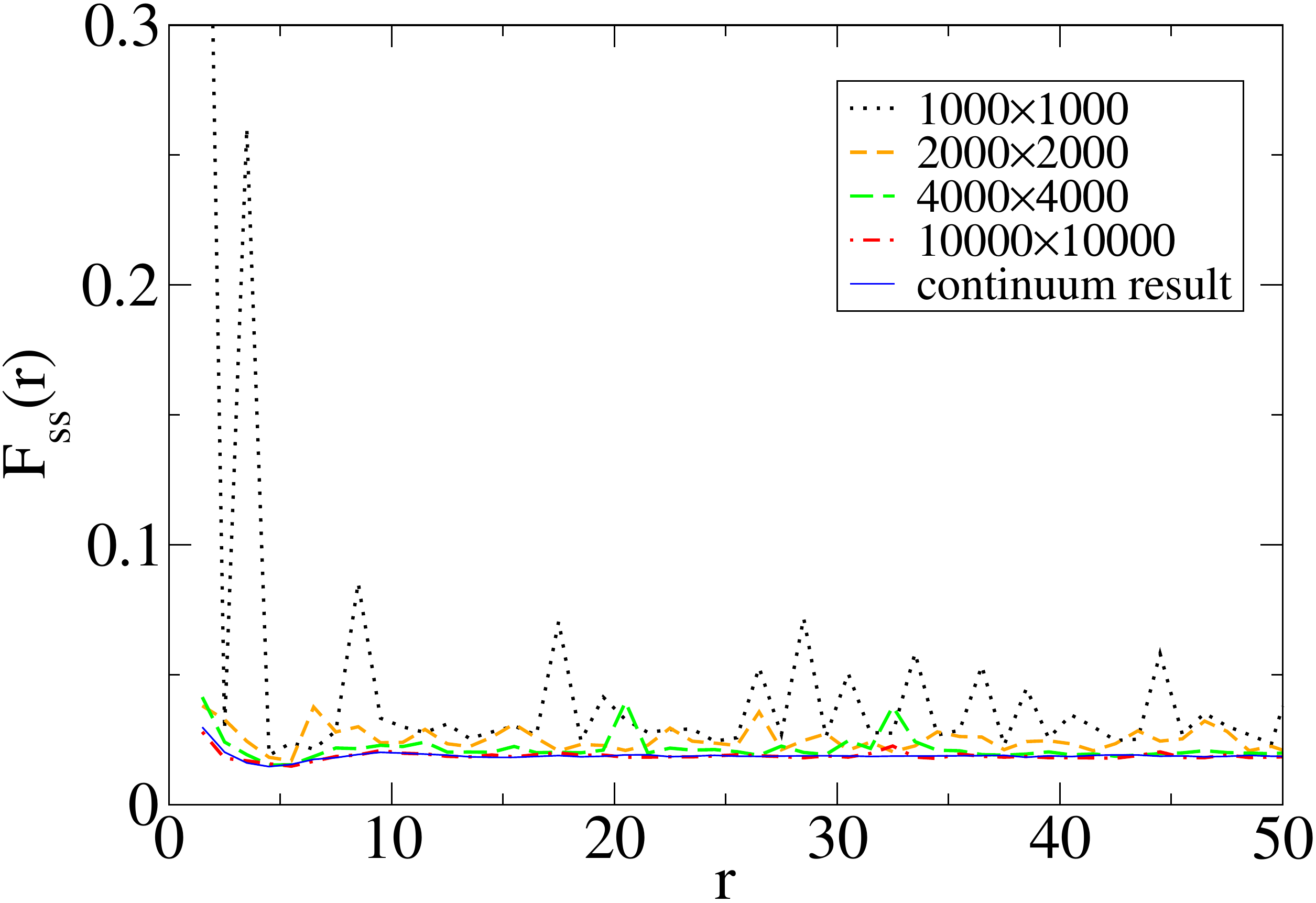}  
\caption{Simulation results of $F_{ss}(r)$ of a digitized Gaussian random field with level cut $F_0=0$ computed under different resolutions. The resolutions shown here are 1000$\times$1000, 2000$\times$2000, 4000$\times$4000 and 10,000$\times$10,000. The continuum result is computed directly using the analytic expression of the scalar field. One can see that as the resolution increases, the numerical result is closer to the continuum result.}
\label{fig:gaussian}
\end{figure}

\indent Suppose we set a threshold $F_0$ to convert a scalar field $F({\bf x})$ to a two-phase medium: regions that satisfy $F(\mathbf x)>F_0$ constitute phase 1, and regions that satisfy $F(\mathbf x)<F_0$ constitute phase 2. The phase indicator function $\mathcal I(\mathbf x)$ for phase 1 is given by
\begin{equation} \label{indicator2}
\mathcal I(\mathbf x)=\Theta[F(\mathbf x)-F_0], 
\end{equation}
where $\Theta(x)$ is the Heaviside step function. The interface between two phases is simply the contour defined by $F(\mathbf x)=F_0$. For any point on the contour, the normal vector is defined by the gradient of the scalar field, i.e., $\nabla F(\mathbf x)$.\\
\indent The algorithm can be implemented in essentially the same way as discussed in Sec. IV, but must be specialized to digitized two-phase media. In order to locate points of intersection
of the line with interfaces, we need to find where $F(\mathbf x)-F_0$ changes sign along a straight line, and then interpolate the position of the point. The gradient at the point can be computed approximately by the finite differences of its neighboring pixels. However, the most significant difference between dealing with continuous models and digitized media is that the number of sampling straight lines one can afford is bounded by the resolution in the later case. Indeed, for an $n\times n$ image, one can only sample at most $\mathcal O(n)$ times if the sampling straight lines are lined up with the grid. Thus the resolution of the image is crucial to obtain reliable results. \\
\indent Here we use Gaussian random field \cite{adler2009random} as an example to demonstrate the importance of resolution. The field is generated by a superposition of 10000 plane waves, as we employed elsewhere \cite{doi:10.1063/1.4989492}, to give a rather disordered structure. The results for the surface-surface correlation function are summarized in Fig. \ref{fig:gaussian}. Here we considered the field within a fixed square region but with different resolutions $1000\times 1000$, $2000\times 2000$, $4000\times 4000$ and $10000\times 10000$. We also include a continuum result which is calculated by directly solving the contour and computing the gradient analytically. It can be seen that as the resolution increases, the numerical results rapidly converge to the continuum result.        

\subsection{Significance of the $1/\cos{\theta}$ Threshold}
\begin{figure}[H]
\centering
\includegraphics[width=8cm,height=6cm]{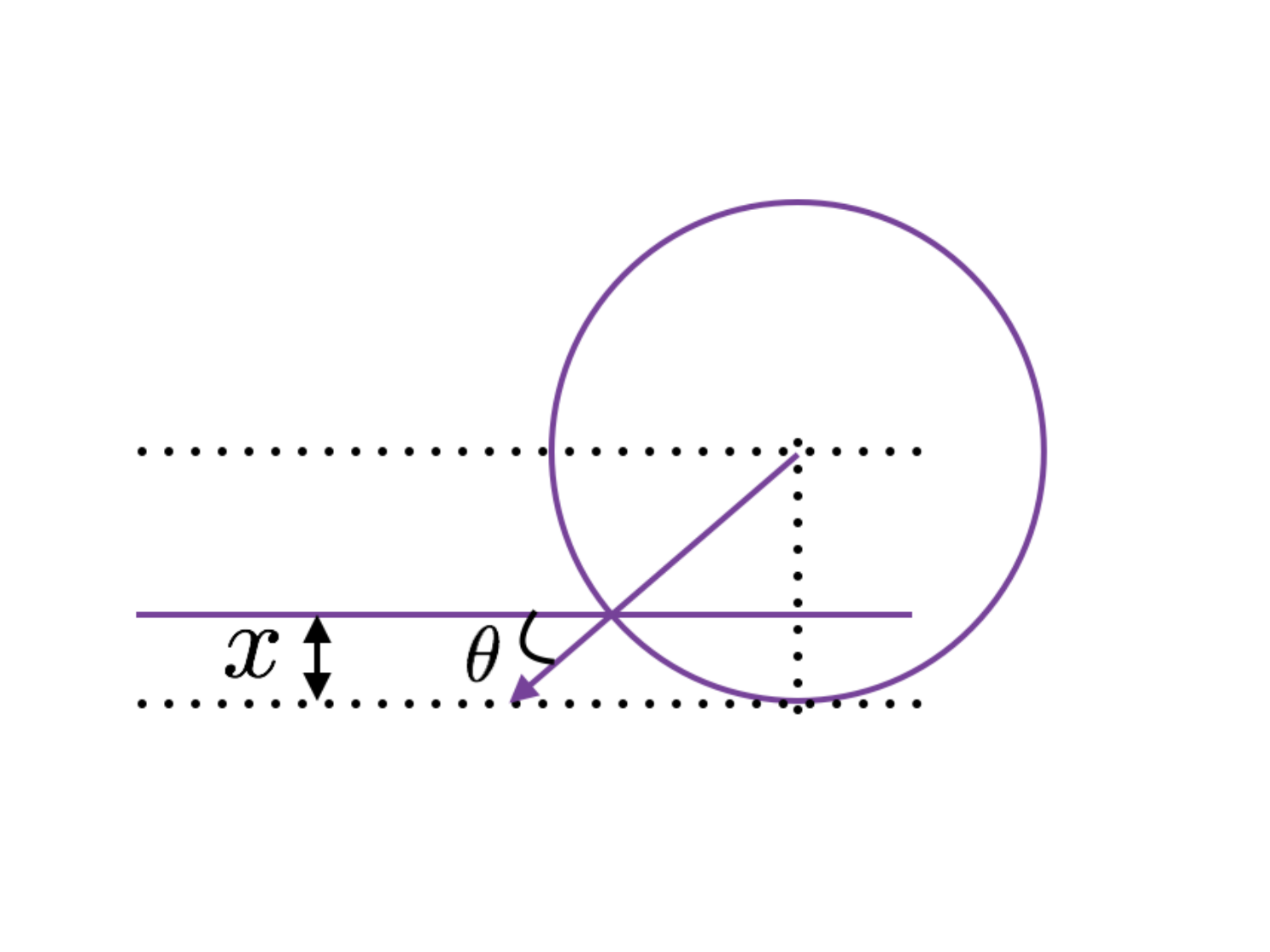}
\caption{An illustration of the sampling scenario. The unit circle is the interface and the solid straight line samples along the direction that is perpendicular to itself by varying $x$.}
\label{fig:cos}
\end{figure}

\indent Figure \ref{fig:gaussian} shows that the computed $F_{ss}$ fluctuates widely when the resolution is low. We discuss the origin of this behavior and how to deal with it in this subsection. \\ 
\indent To begin, consider the simple situation illustrated in Fig. \ref{fig:cos}, which involves a straight line sampling the boundary of a unit circle. The discussion of this case is instructive because the vicinity of the intersection point can be approximated by sphere surfaces in most cases, and when $r$ is large enough, $F_{ss}(r)$ is proportional to $\langle1/\cos{\theta}\rangle^2$. So for simplicity, the aim here is to estimate $\langle1/\cos{\theta}\rangle$ for the lower left quarter of the circle. Suppose the straight line samples from $x=0$ to $x=1$ uniformly along the direction that is perpendicular to itself; then we have

\begin{equation} \label{int} 
\int_{0}^{1}\frac{dx}{\cos{\theta}}=\int_{\frac{\pi}{2}}^{0}\frac {d(1-\sin{\theta})}{\cos{\theta}}=\int_{0}^{\frac{\pi}{2}}d\theta=\frac{\pi}{2}.
\end{equation}

\begin{figure}[H]
\centering
\includegraphics[width=8cm,height=5.6cm]{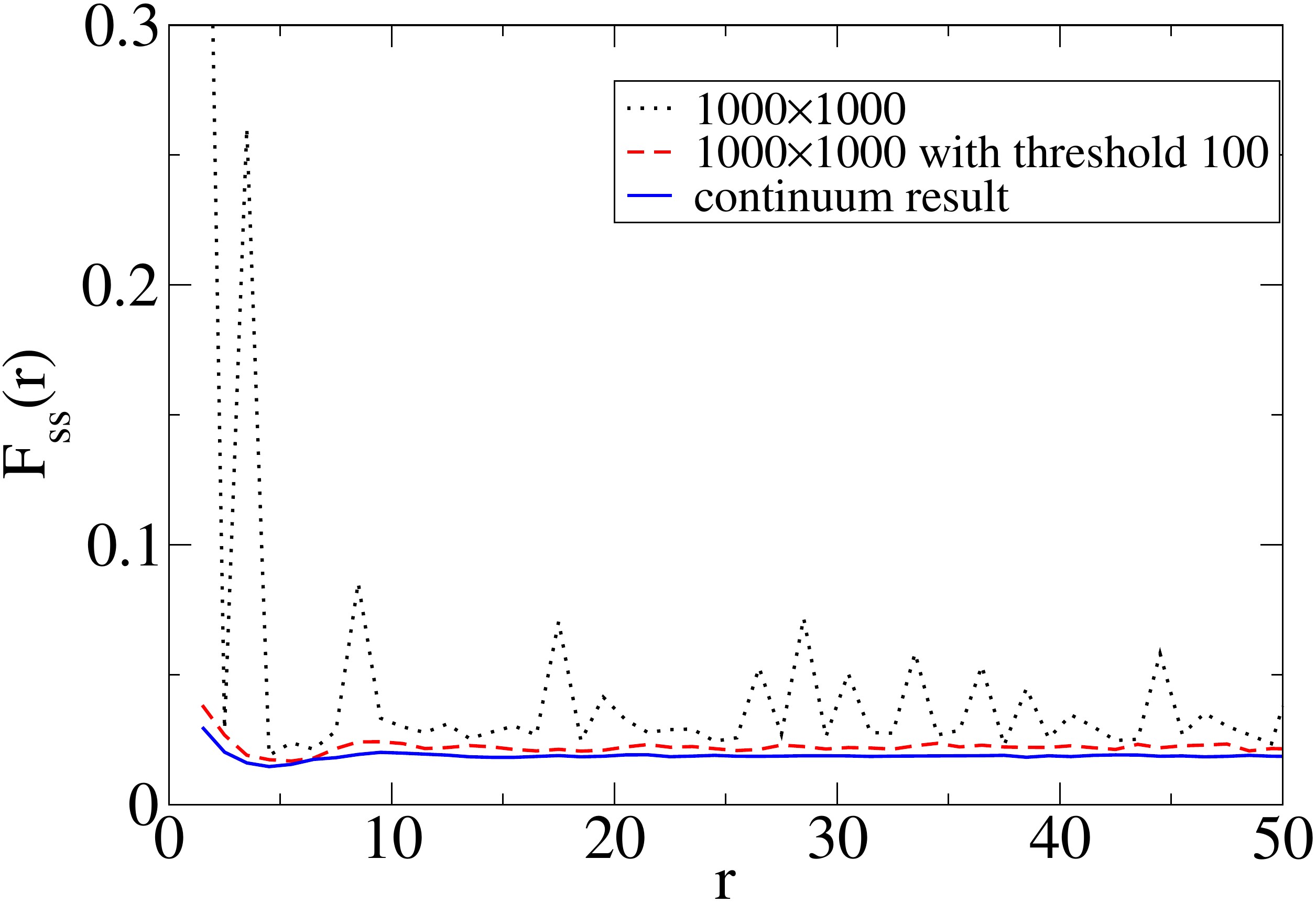}
\caption{A comparison of simulation results of $F_{ss}(r)$ of a digitized Gaussian random field with level cut $F_0=0$ computed with and without applying a threshold $\delta$. The resolution is 1000$\times$1000, and the threshold $\delta$ is 100. The continuum result is computed directly using the analytic expression of the scalar field. By applying a threshold, the fluctuations are largely suppressed and the result is much closer to the continuum result, even comparable to the ones with much higher resolutions.}
\label{fig:cutgauss} 
\end{figure}

\indent The integral correctly gives the surface area of the lower left quarter of the circle. Notice that although the integrand $1/\cos\theta$ is divergent at $x=0$, it is still integrable because the probability of hitting the vicinity of the singularity is proportionally infinitesimally small. However, it is easy to see that the variance of $1/\cos{\theta}$ is divergent since the integral of $1/\cos^2{\theta}$ diverges, which implies that large deviations can result when estimating the mean of $1/\cos{\theta}$. However, the simulation results suggest that increasing the number of sampling lines still reduces the fluctuations from the expected value, and a large sampling number yields good estimates, as one can see in Fig. \ref{fig:bench}. This suggests that the probability of getting a large deviation diminishes when the sampling number is increased. This is indeed the case, as we show in Appendix B.\\
\indent However, in the case of digitized media, one cannot increase the sampling number arbitrarily. On the other hand, the probability of hitting the vicinity where $\theta \approx \pi/2$ can be rounded to a relatively large fraction due to the finite resolution. For example, a curved interface can align parallel to the sampling line after the digitization. The consequence is that we are more likely to encounter large deviations, as one can see in Fig. \ref{fig:gaussian}, where the abnormal peaks [as well as the universal trend of overestimating $F_{ss}(r)$] are due to certain very large values of $1/\cos{\theta}$ encountered in the sampling. Although both problems can be alleviated by simply increasing the resolution, it is generally not known a prior that what resolution is required. Furthermore, obtaining high-resolution representations can also be computationally or economically costly, or simply beyond access due to the limitation of experimental techniques or available memory for a simulation. These restrictions force us to come up with a more efficient way to bypass the problems of digitized media. A straightforward way to remove this effect is to simply discard samples when they are larger than a certain threshold $\delta$, i.e., $1/\cos\theta >\delta$. The bias induced by this method is usually small and insignificant, but with this small compromise, one can significantly reduce fluctuations (see a detailed analysis in Appendix C). To demonstrate the effect of applying thresholds to digitized media, we take the lowest resolution representation ($1000\times 1000$) of the Gaussian random field in the last subsection and recompute $F_{ss}$ with a threshold $\delta=100$. The result is shown in Fig. \ref{fig:cutgauss}. Note that after applying a threshold, the fluctuations are dramatically suppressed and the result is much closer to the continuum result, even comparable to the ones with much higher resolutions in Fig. \ref{fig:gaussian}.\\  

\subsection{Converting Digitized Two-phase Media into Scalar Fields}

\indent We complete our discussion of two-phase media by discussing the case in which all of the information provided about the system is a binary digitized medium. Due to the jagged interface geometry, the transition from one phase to another is sharp and there is no easy way to estimate the direction of the norm vector as we did in the case of Gaussian random fields by computing the gradient of the scalar field.\\
\begin{figure}[H]
\centering
\subfigure[]{
\includegraphics[width=4.5cm, height=4.5cm]{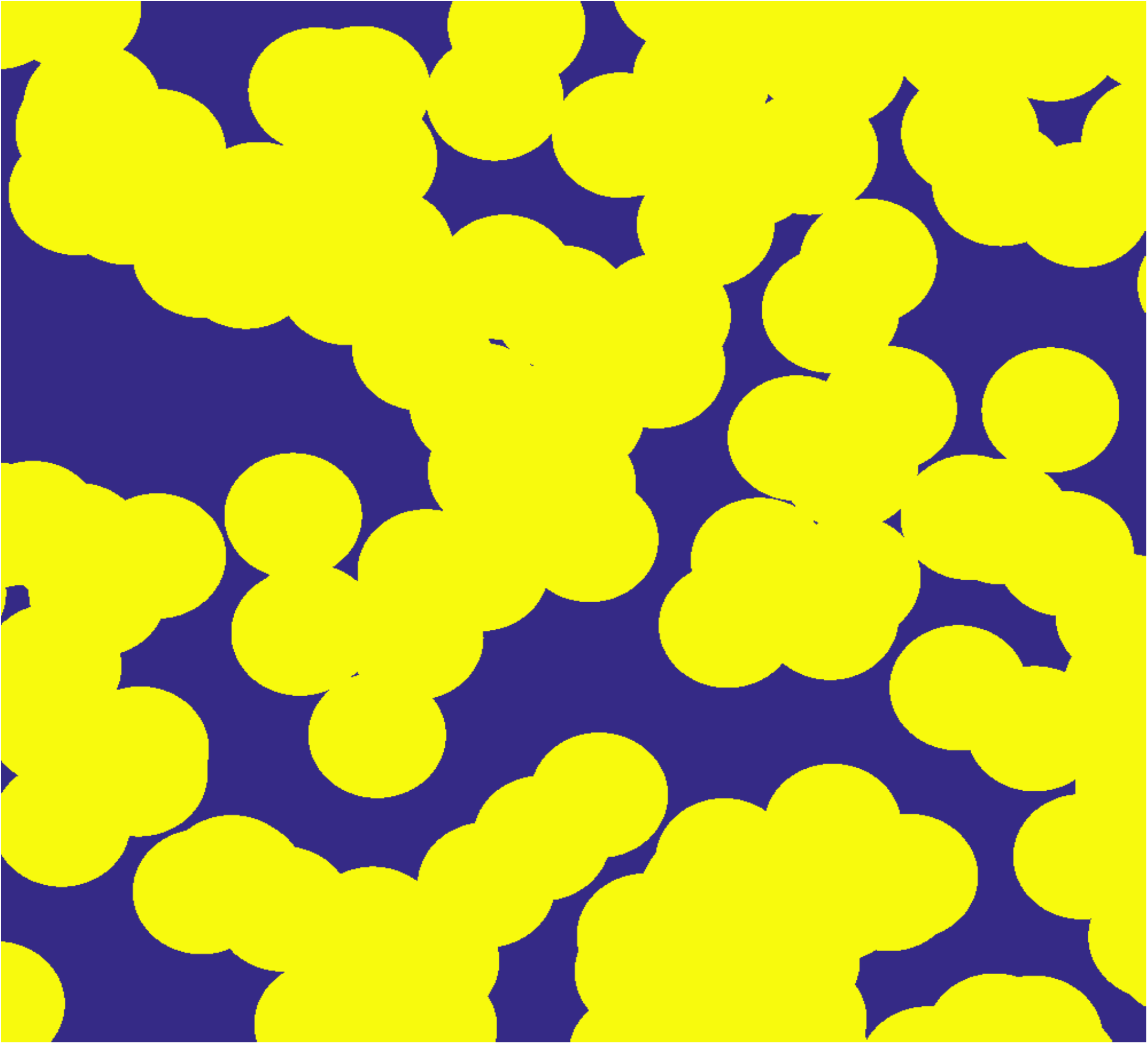}
}
\subfigure[]{
\includegraphics[width=4.5cm, height=4.5cm]{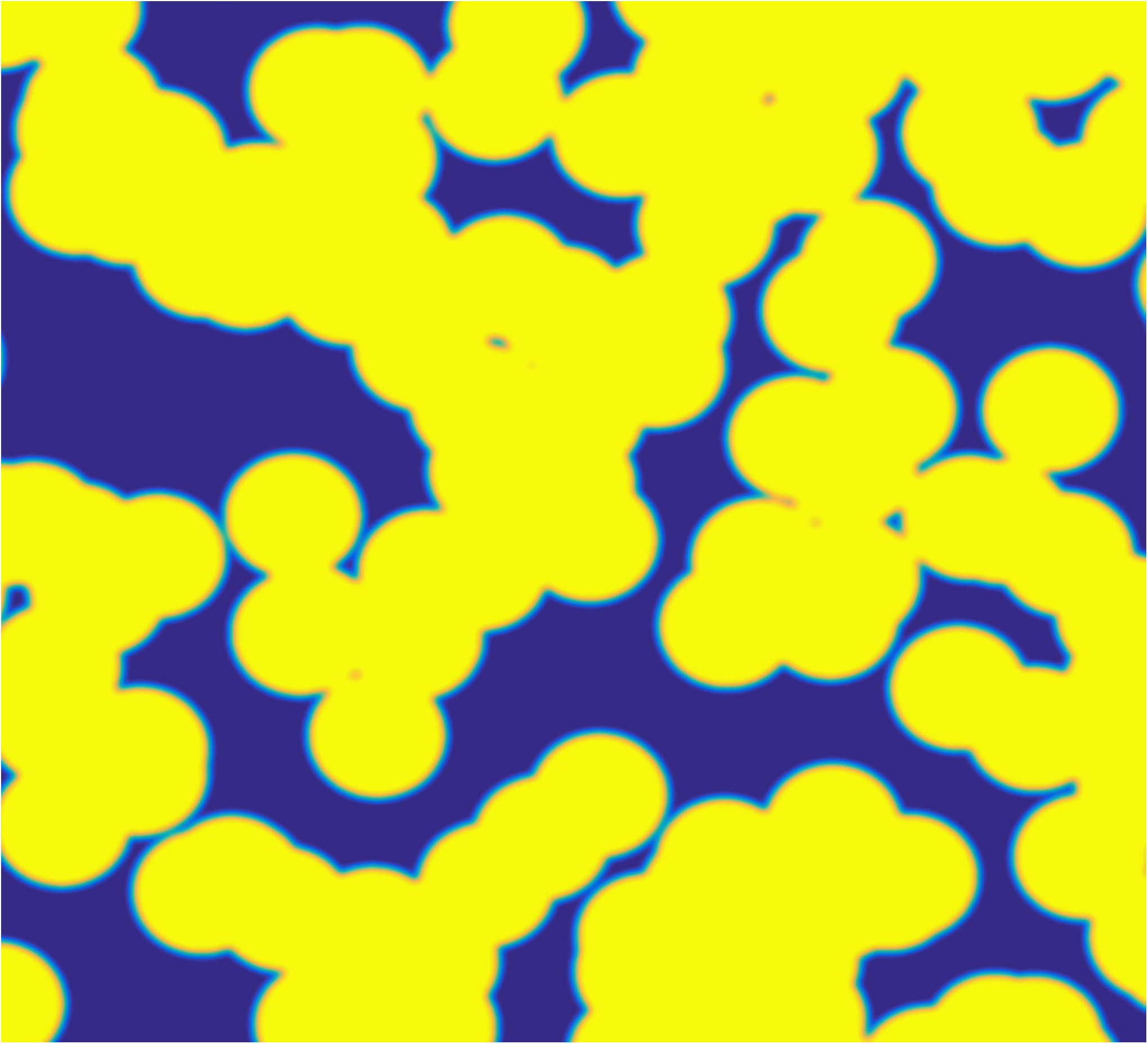}
}
\caption{(a) Digitized random overlapping disks with particle-phase volume fraction 0.677. (b) Corresponding scalar field of (a) by using a Gaussian kernel (\ref{eq:kernel}) with $b=0.042D$.}
\label{fig:digital}
\end{figure} 

\begin{figure}[H]
\centering
\includegraphics[width=8cm,height=5.6cm]{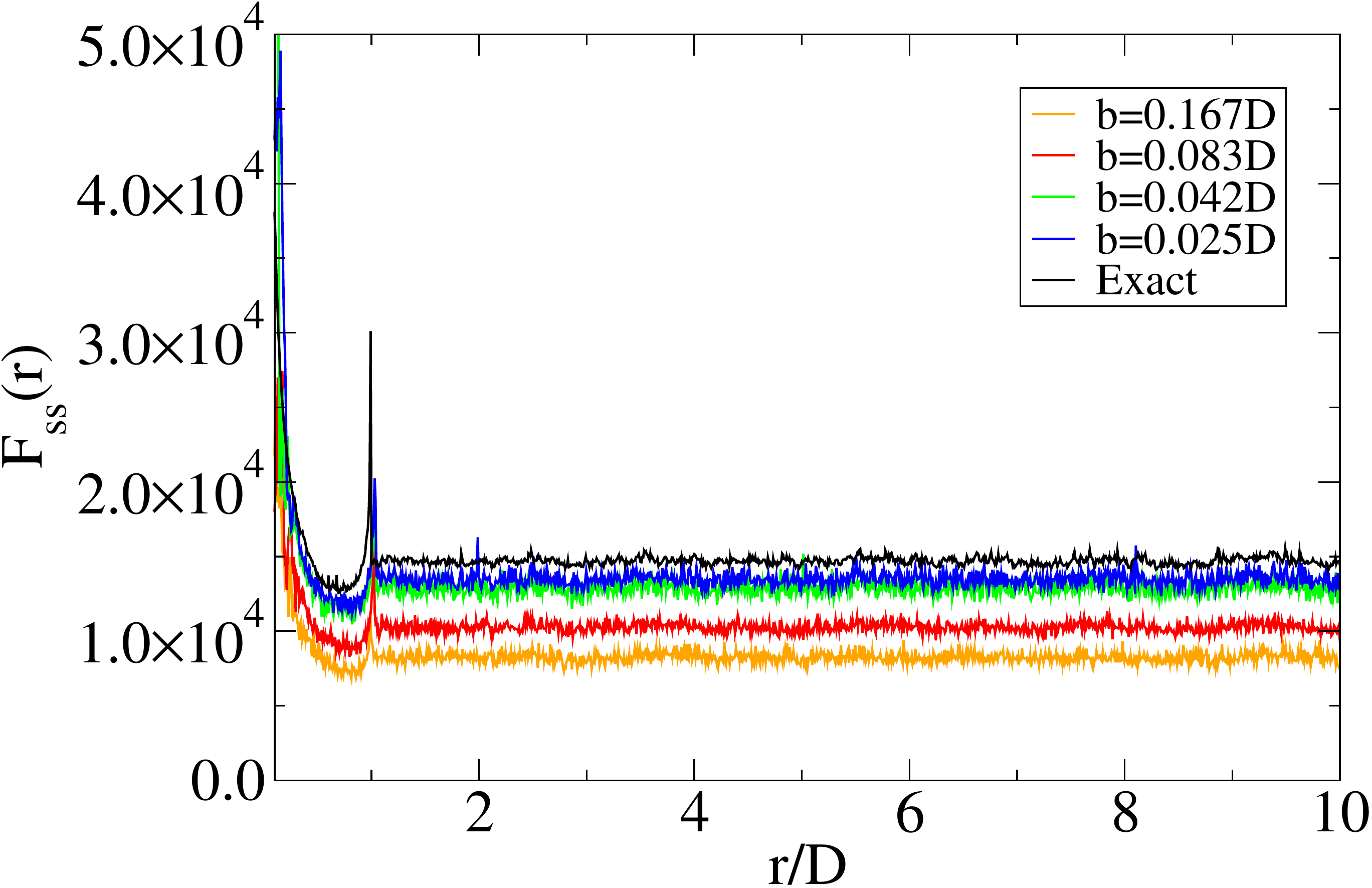}
\caption{Simulation results of $F_{ss}(r)$ of a digitized overlapping disk configuration by applying Gaussian kernels for different values of $b$.}
\label{fig:digitalfss}
\end{figure}

\indent Here we propose a straightforward method to deal with this situation. We first convert the two-phase medium to a coarse-grained scalar field, then convert it back to a two-phase medium by thresholding. In this way we can again use the algorithm introduced in Sec. V. A. We follow the procedure described in Refs. \cite{torquato2013random} and \cite{blumenfeld1993coarse}. By taking pixels in phase $i$ as source points, we can convert the two-phase medium into a scalar field $F(i,j)$ by convolving the indicator function $\mathcal I^{(i)}$ with a kernel or filter $K(\mathbf x)$. Then, the scalar field is
\begin{equation}
F(i,j,\{C\})=\sum_{k}\sum_{l}\mathcal I^{(i)}(i+k,j+l)K(k,l,\{C\}),
\end{equation}     
where $\{C\}$ represents the parameters of the kernel. One of the most common choices of kernels is the Gaussian filter,
\begin{equation} \label{eq:kernel}
K(\mathbf x; b)=\exp(-\frac{|\mathbf x|^2}{b^2}),
\end{equation} 
where $b$ is a length parameter that controls the size of the ``influence" region of the filter. By taking a level cut of the scalar field at a threshold $F_0$, we can then convert the scalar field back into a two-phase medium. The threshold $F_0$ is chosen to retain the original volume fraction of phase $i$.\\
\indent We include an example of overlapping spheres in two dimensions processed by this method. We prepare a digitized realization of 10,000 overlapping disks under periodic boundary conditions at a particle-phase volume fraction $\phi=0.677$. The resolution is chosen to be that the side length of a pixel is $1/120$ of the diameter $D$ of the disk. We apply the Gaussian filter mentioned above for different values of $b$ and compute the corresponding $F_{ss}$ of the converted scalar fields. A comparison of a portion of the system before and after applying the filter ($b=0.042D$) is shown in Fig. \ref{fig:digital}, one can see that the structure of the system is maintained while there is a transition region between two phases. The comparison of $F_{ss}$ computed with different filters is shown in Fig. \ref{fig:digitalfss} along with the exact result computed from the continuum model. It is noteworthy that although all the surface-surface correlation functions computed capture the shape of the exact one, they all tend to underestimate the actual function. The possible explanation is that the digitized version loses detailed interfacial information and hence the interface appears to be less curved, which leads to smaller surface areas. However, as $b$ decreases and the filter becomes more localized, the difference between the computed $F_{ss}$ and the exact one monotonically diminishes. The smallest value of $b$ shown in Fig. \ref{fig:digitalfss} is three times of the pixel width, one may expect that when the resolution is high enough, the curve of the digitized version will finally converge to the exact one of the underlying pattern. Although images obtained in experiments may not necessarily be of high resolution, generally they are gray-scale images, which means one can simply use the algorithm described for scalar fields directly.     
    
\section{Results for overlapping and nonoverlapping Sphere packings}

\begin{figure}[H]
\centering
\subfigure[]{
\includegraphics[width=6.5cm, height=4.5cm]{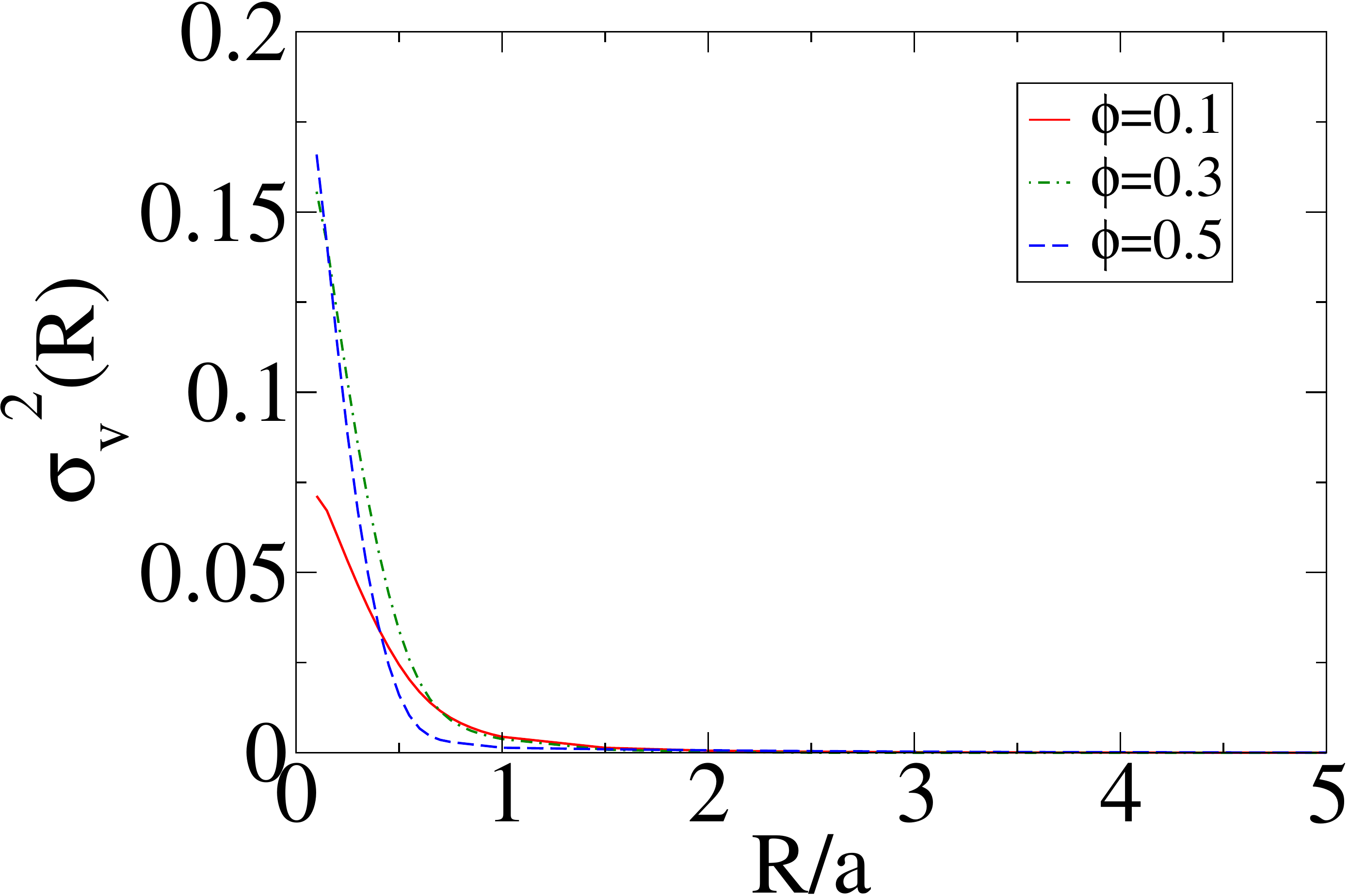}
}
\subfigure[]{
\includegraphics[width=6.5cm, height=4.5cm]{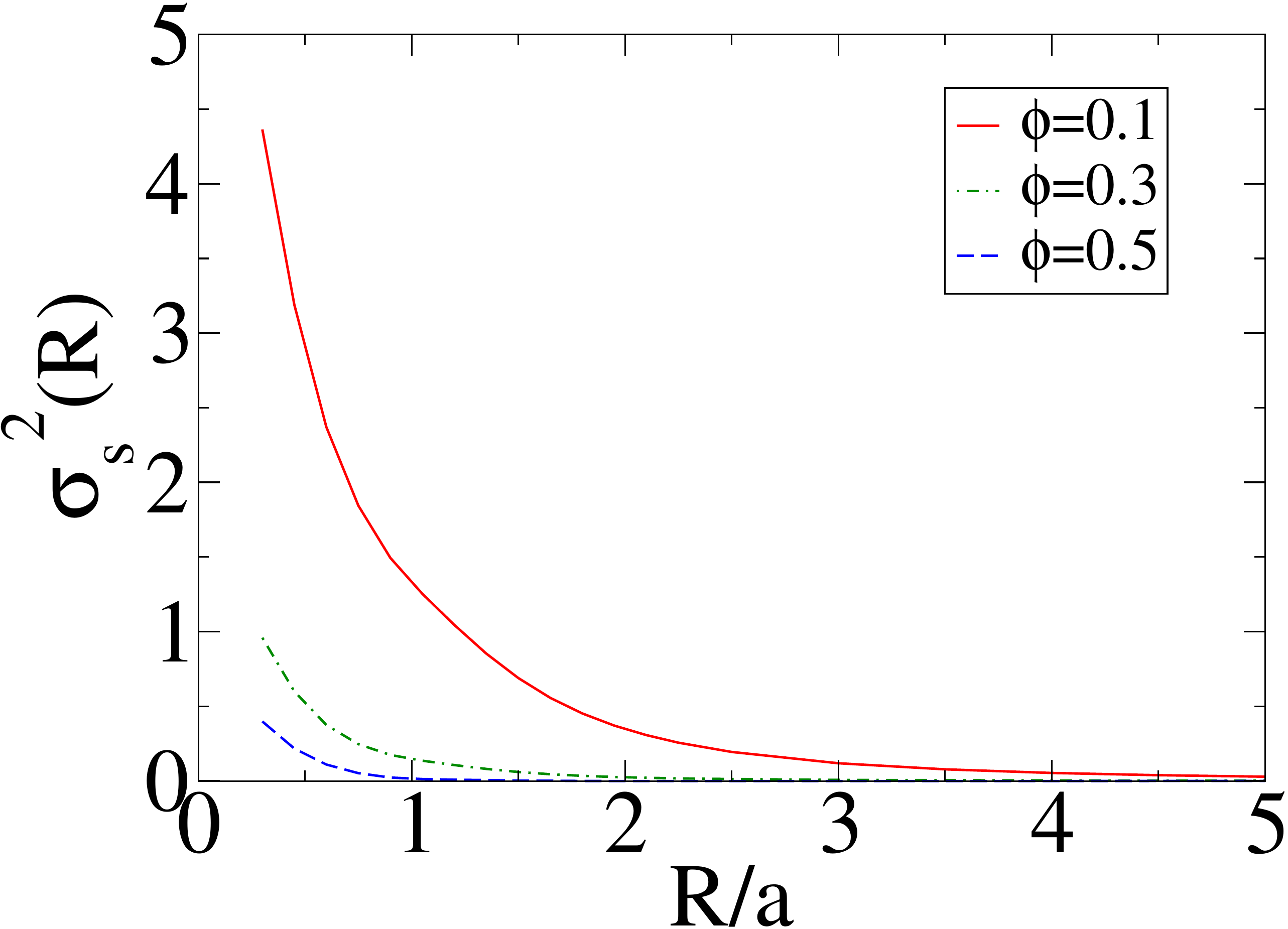}
}
\caption{(a) A comparison of volume-fraction variances of hard spheres at equilibrium at different packing fractions, it is noteworthy that there is a crossover at small $R$. (b) A comparison of surface-area variances of hard spheres at equilibrium at different packing fractions, which clearly reflect the increase of short-range order as packing fraction increases.}
\label{fig:hardfluc2}
\end{figure}

In this section, we compute the surface correlation functions of several particle systems, including overlapping spheres, hard-spheres in equilibrium and decorated ``stealthy" patterns. Given our abilities to compute the surface-surface correlation function, we can calculate 
local surface-area variances through Eq. (\ref{auto}), and compare them with local volume-fraction variances in these systems. \\
\begin{figure}[H]
\centering
\subfigure[]{
\includegraphics[width=6.5cm, height=4.5cm]{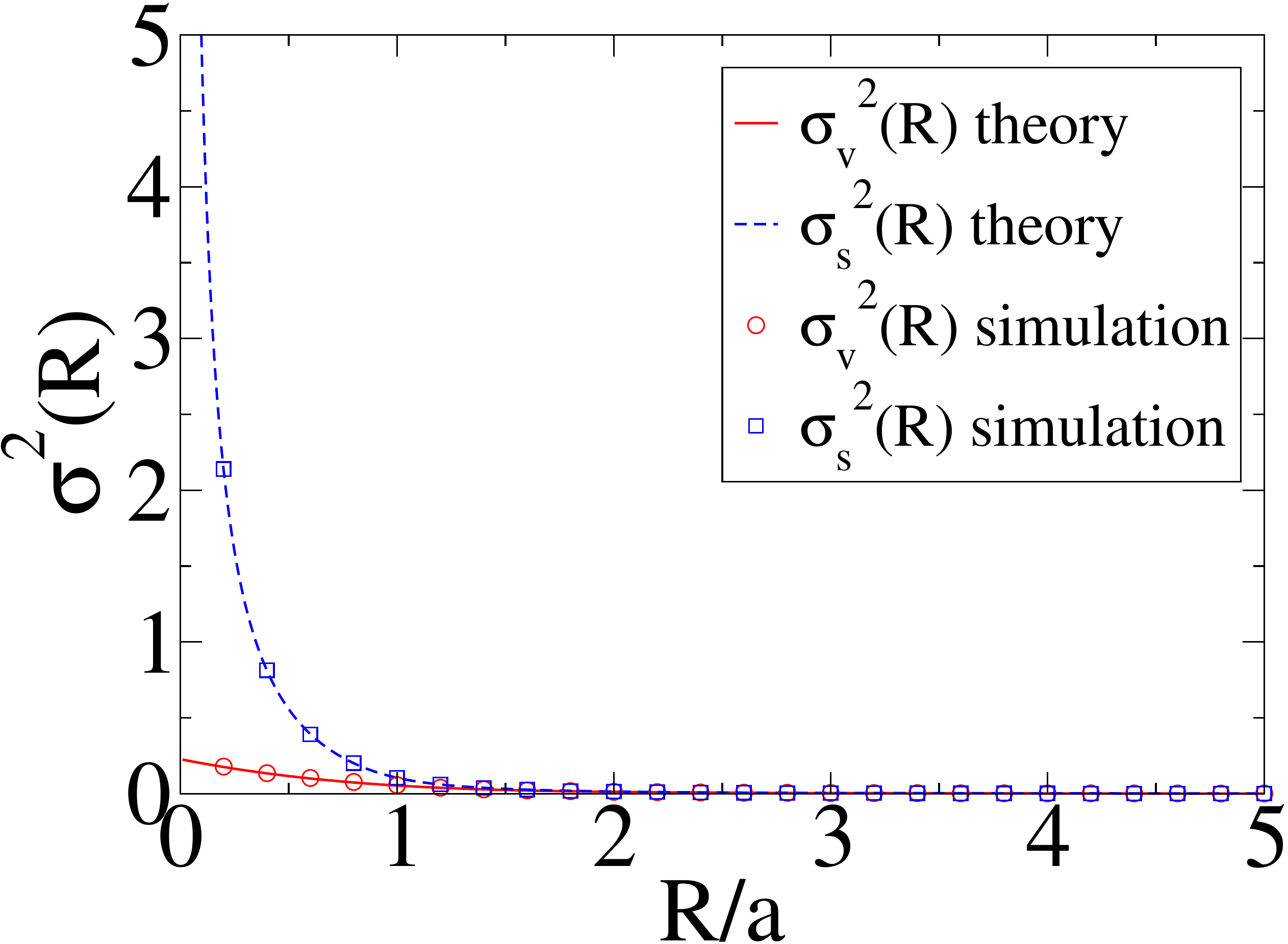}
}
\subfigure[]{
\includegraphics[width=6.5cm, height=4.5cm]{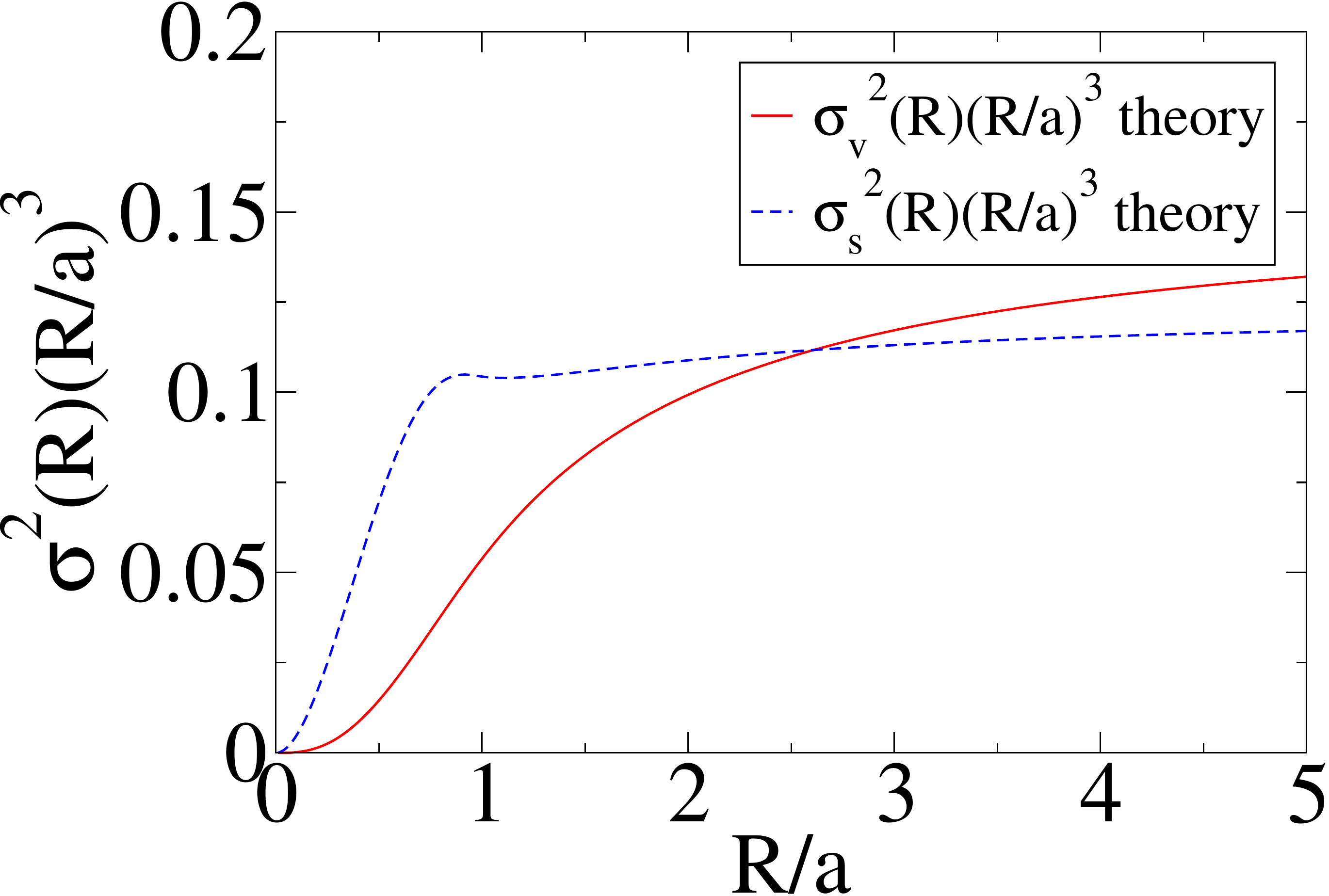}
}
\caption{(a) A comparison of the local surface-area variances $\sigma_{_S}^2(R)$ with the volume-fraction variances $\sigma_{_V}^2(R)$ for three-dimensional overlapping spheres of radius $a$ as functions of window radius $R$ at particle-phase volume fraction $\phi=0.649$. Note the surface-area fluctuation is much larger at small $R$, suggesting that it is a more sensitive microstructure descriptor. (b) A comparison of rescaled local surface-area variances with the volume-fraction variances from (a).}
\label{fig:overfluc}
\end{figure}
\indent We start by analyzing overlapping spheres in three dimensions. The volume-fraction variance $\sigma_{_V}^2(R)$ and surface-area variance $\sigma_{_S}^2(R)$ for the same system studied in Sec. IV are presented in Fig. \ref{fig:overfluc}(a), where $R$ is the radius of the spherical window and $a$ is the radius of particles. These quantities are computed by numerically computing the integrals in Eqs. (\ref{auto01}) and (\ref{auto}) as well as through Monte Carlo simulations. In the later method, we generate windows at random positions and calculate the volume-fraction and surface-area variances directly. Since in this model spheres can form very complex clusters, we evaluate the volume fraction and surface area inside each window by generating random points uniformly in the window or on the surface of spheres and counting their fractions inside or on the surface of the clusters correspondingly. The theoretical prediction and simulation results agree very well, as one can see in Fig. \ref{fig:overfluc}(a). It is also noteworthy that in Fig. \ref{fig:overfluc}(a) the surface-area variance is much larger compared to the volume-fraction variance. However, one can see that in Fig. \ref{fig:overfluc}(b), after multiplied by $R^3$ (in order to show the large-$R$ behavior of fluctuations), it is clear that there is a crossover of function values around $R=2.6a$. Further numerical experiments show that the crossover only happens when the particle-phase volume fraction is between 0.57 and 0.7, outside this interval the surface-area variance is always larger than the volume-fraction variance, suggesting that the surface-area variance is a more sensitive descriptor.      
\begin{figure*}[]
\centering
\subfigure[]{
\includegraphics[width=5cm, height=3.5cm]{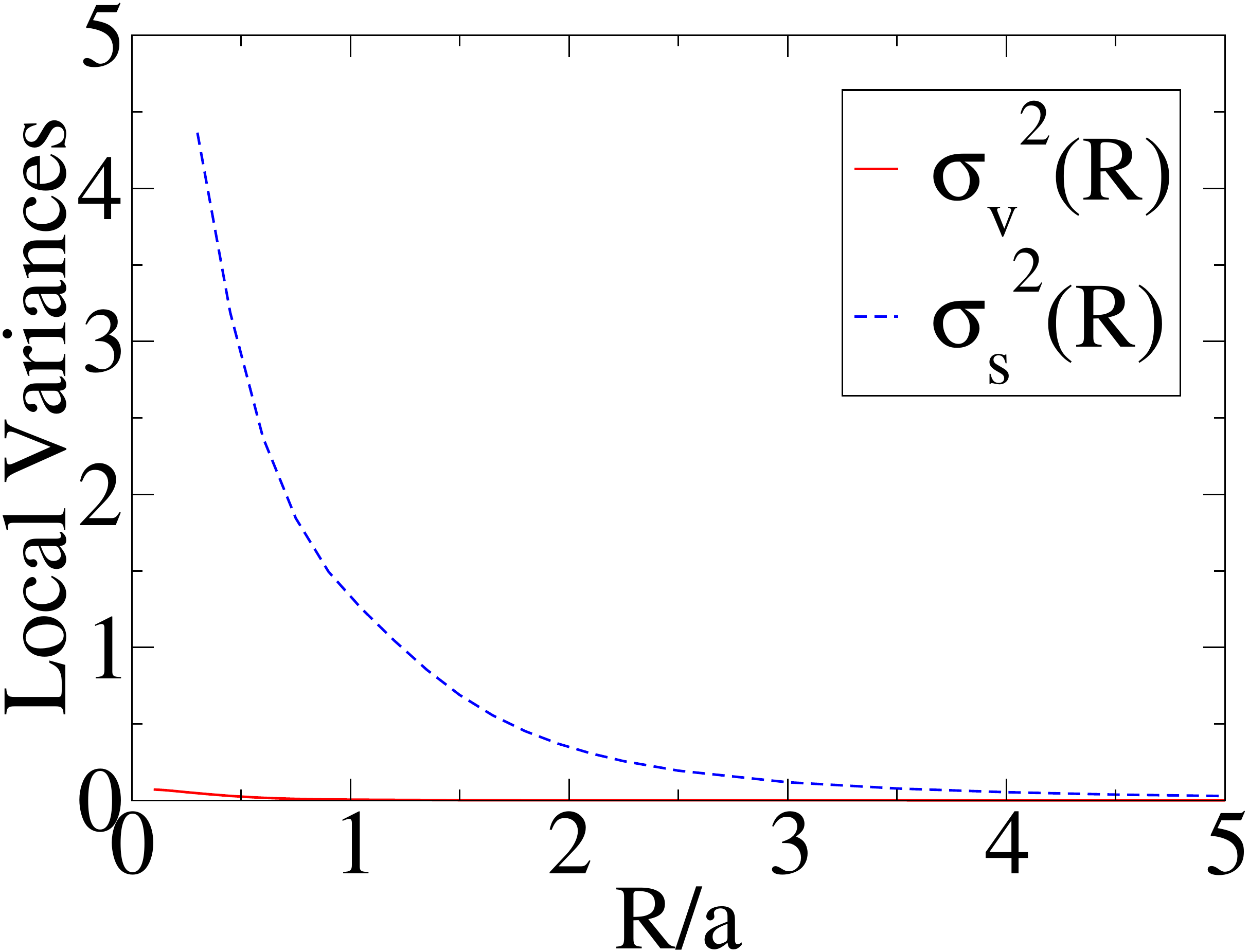}
}
\subfigure[]{
\includegraphics[width=5cm, height=3.5cm]{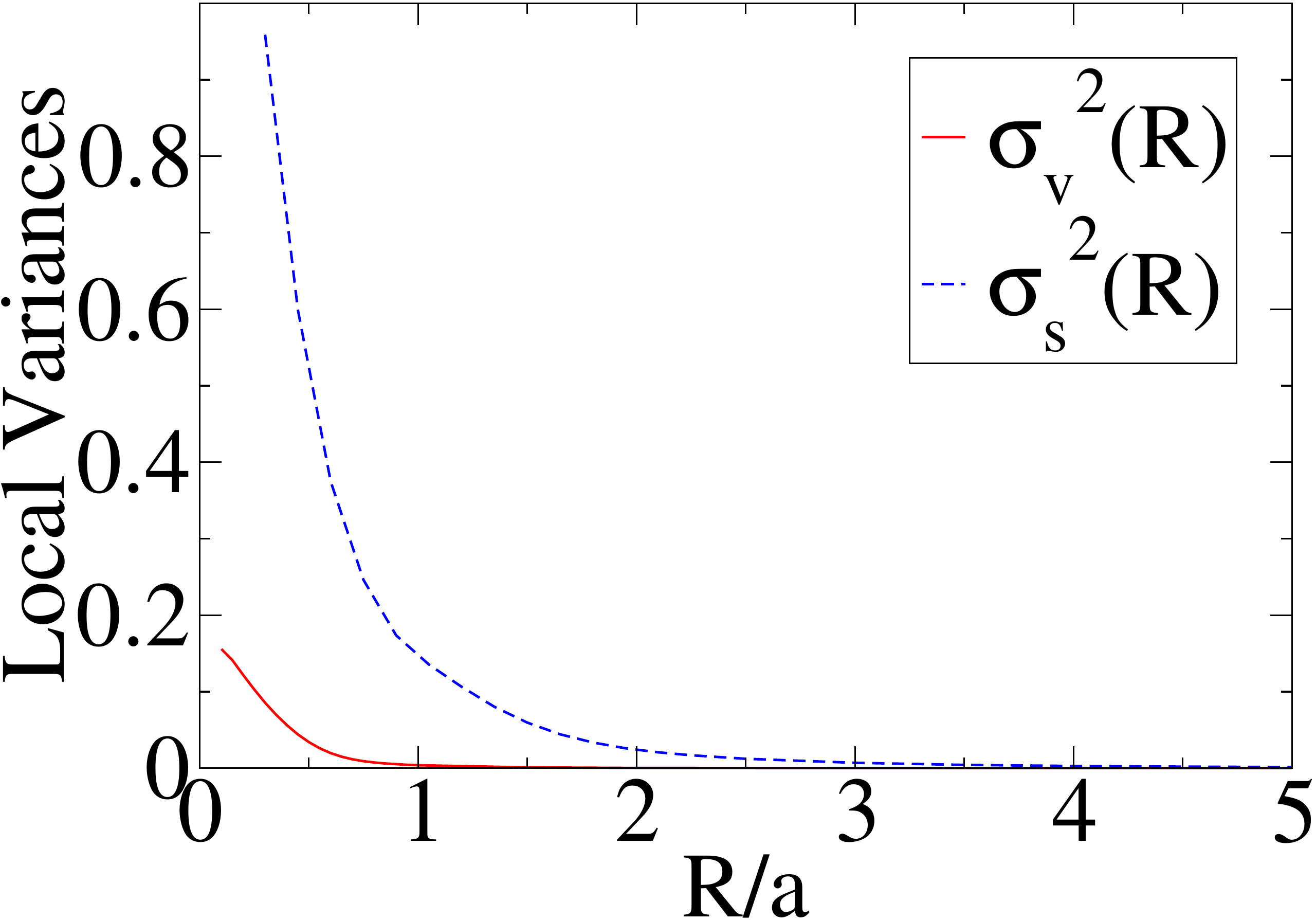}
}
\subfigure[]{
\includegraphics[width=5cm, height=3.5cm]{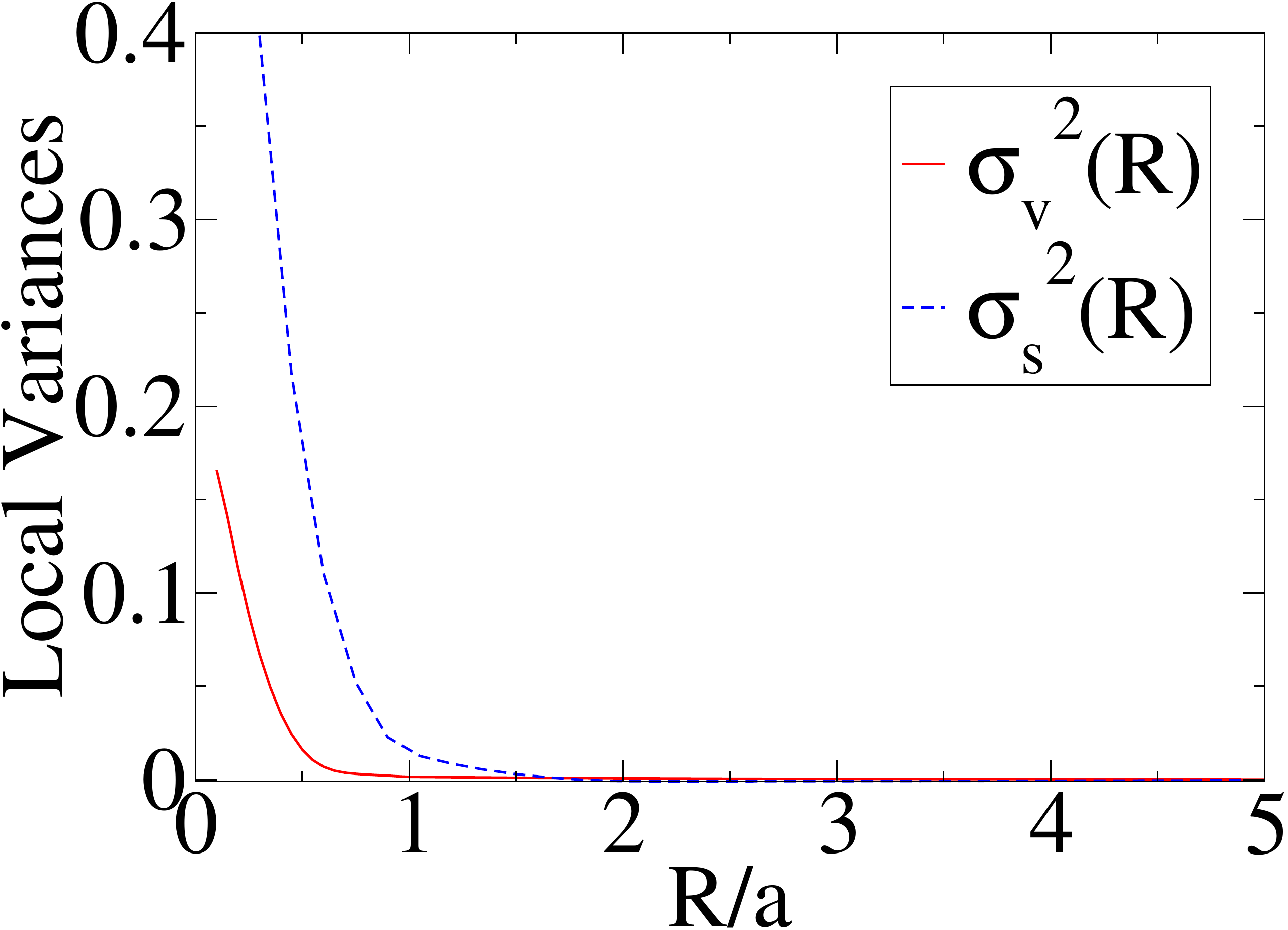}
}
\caption{A comparison of local surface-area variances $\sigma_{_S}^2(R)$ with volume-fraction variances $\sigma_{_V}^2(R)$ for three-dimensional hard spheres of radius $a$ as functions of window radius $R$ in equilibrium at different packing fractions $\phi$; (a) $\phi=0.1$. (b) $\phi=0.3$. (c) $\phi=0.5$.}
\label{fig:hardfluc1}
\end{figure*} 
\indent We further compare the surface-area variance $\sigma_{_S}^2(R)$ with the volume-fraction variance $\sigma_{_V}^2(R)$ of hard spheres in equilibrium in three dimensions at different packing fractions. The variances are again computed using Eqs. (\ref{auto01}) and (\ref{auto}); however, the autocovariance functions are not known analytically in this case. As mentioned previously, we use the results included in Ref. \cite{torquato1986interfacial}, which was computed using the Percus-Yevick approximation and the Verlet-Weis correction. We include our results in Fig. \ref{fig:hardfluc1}. Note that the surface-area variance is always larger than the volume-fraction variance across a large span of packing fractions. In Fig. \ref{fig:hardfluc2}, the surface-area variance and the volume-fraction variance are compared respectively at different packing fractions. As the packing fraction increases, the hard-sphere system becomes more short-range ordered \cite{torquato2000random}, thus the variances are expected to drop. The overall trend of $\sigma_{_S}^2(R)$ and $\sigma_{_V}^2(R)$ is consistent with this intuition. However, the volume-fraction variances experience another crossover at small $R$, while the surface-area variances drop monotonically and larger gaps can be seen between the curves for different packing fractions. These results strongly suggest that the surface-area variance is a more sensitive measure of microstructures of the system compared to the volume-fraction variance.     

\begin{figure}[H]
\centering
\subfigure[]{
\includegraphics[width=6.5cm, height=4.5cm]{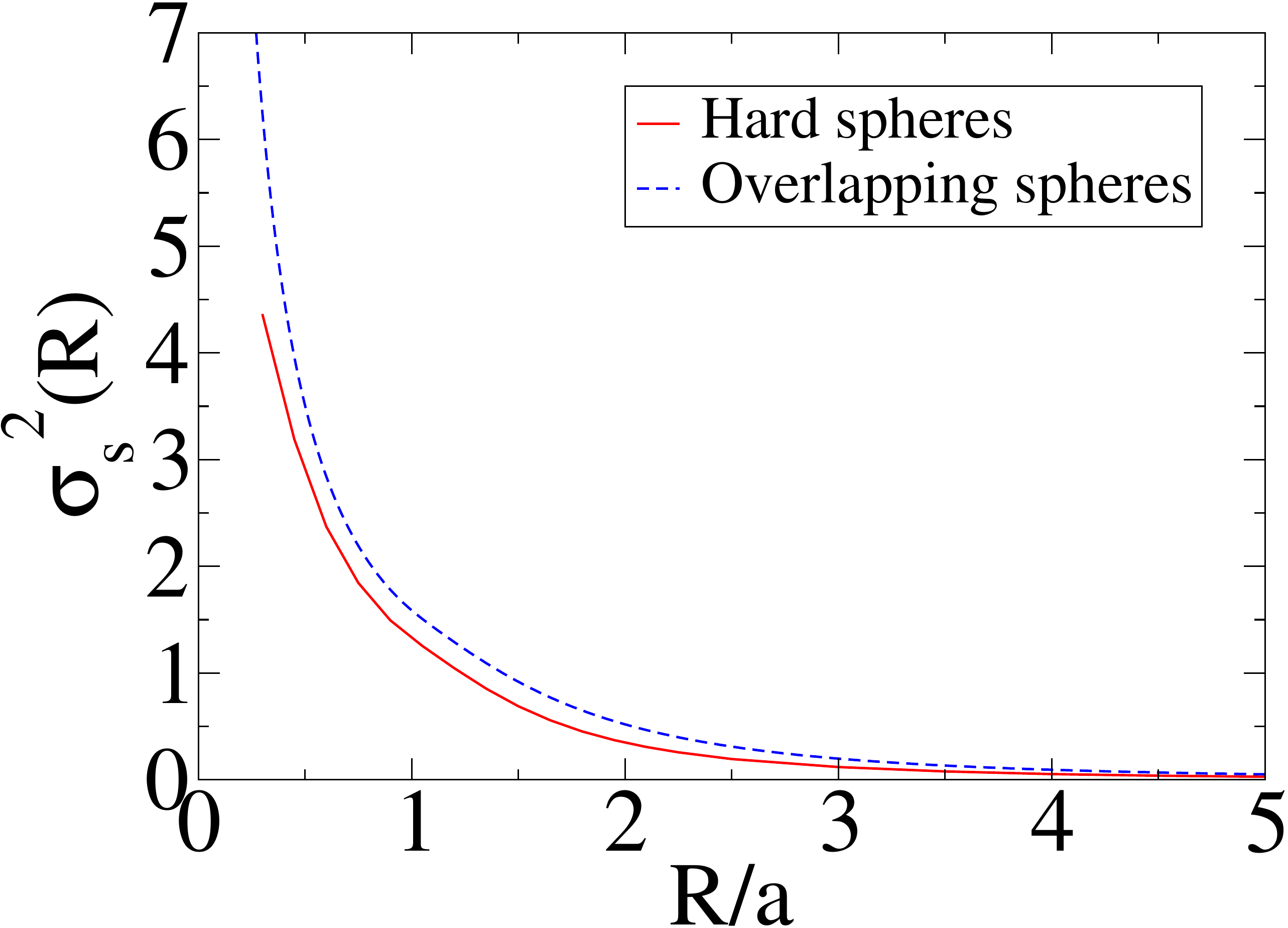}
}
\subfigure[]{
\includegraphics[width=6.5cm, height=4.5cm]{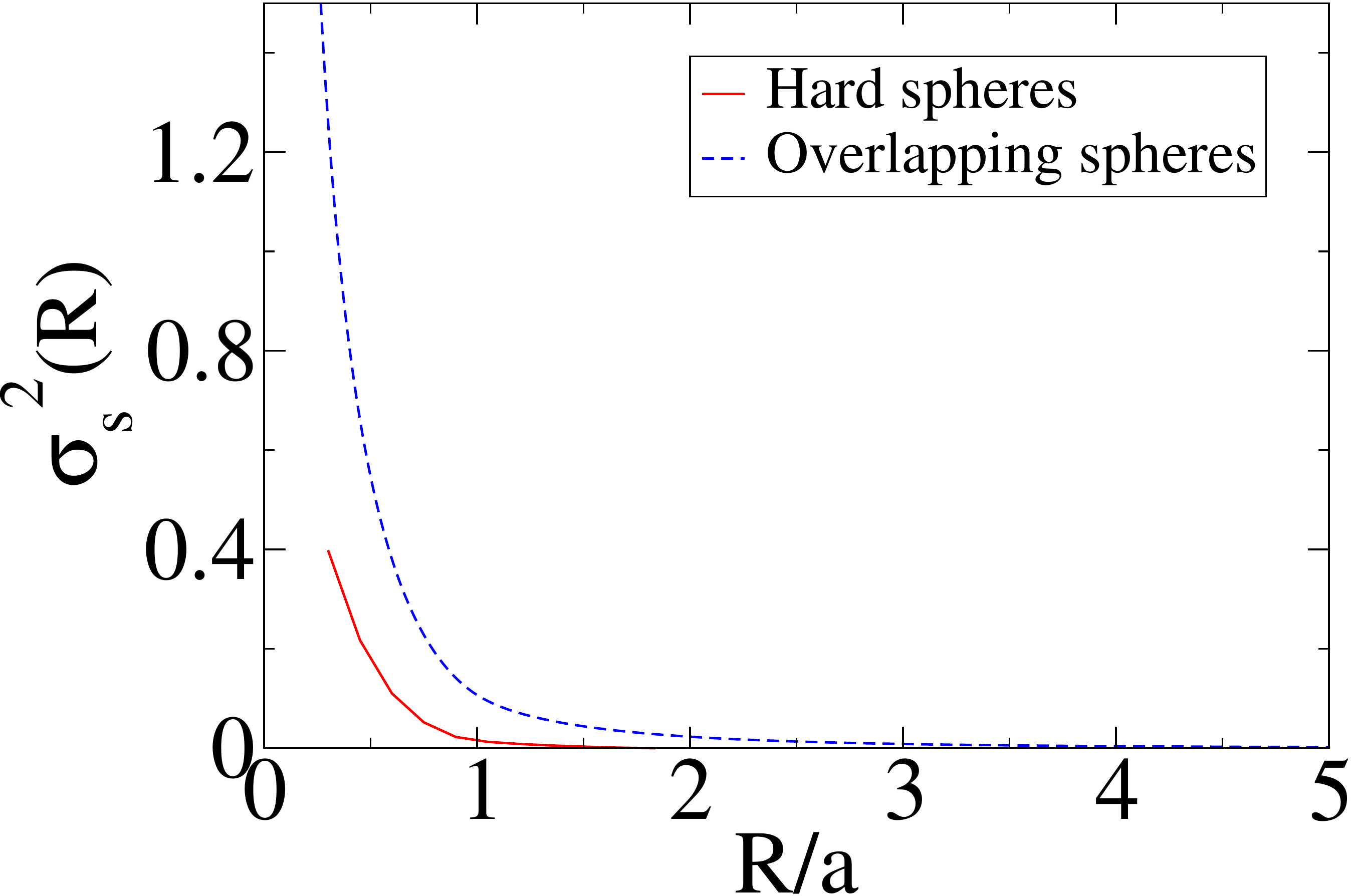}
}
\caption{A comparison of surface-area variances $\sigma_{_S}^2(R)$ of three-dimensional hard spheres in equilibrium and overlapping spheres of radius $a$ as functions of window radius $R$ at different volume fractions. (a) $\phi=0.1$. (b) $\phi=0.5$.}
\label{fig:overhard}
\end{figure}

Finally, we compare surface-area variances of hard spheres in equilibrium and overlapping spheres at different volume fractions in Fig. \ref{fig:overhard}. The fact that the hard-sphere systems always suppress surface-area fluctuations variances to a greater degree than those of overlapping spheres, which reflects the stronger pair correlations in the former system. 

\begin{figure*}[]
\centering
\subfigure[]{
\includegraphics[width=7.5cm, height=5cm]{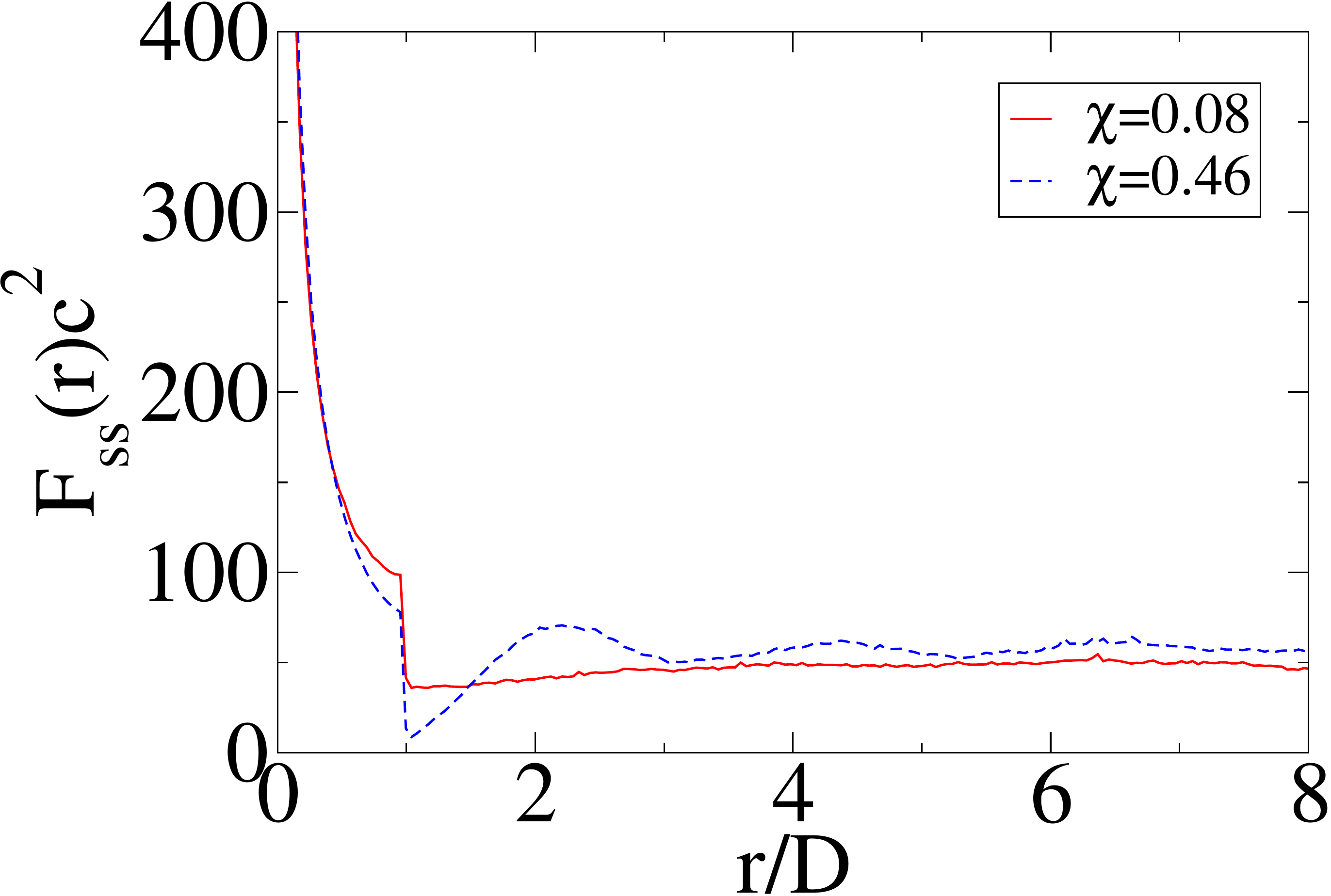}
}
\subfigure[]{
\includegraphics[width=7.5cm, height=5cm]{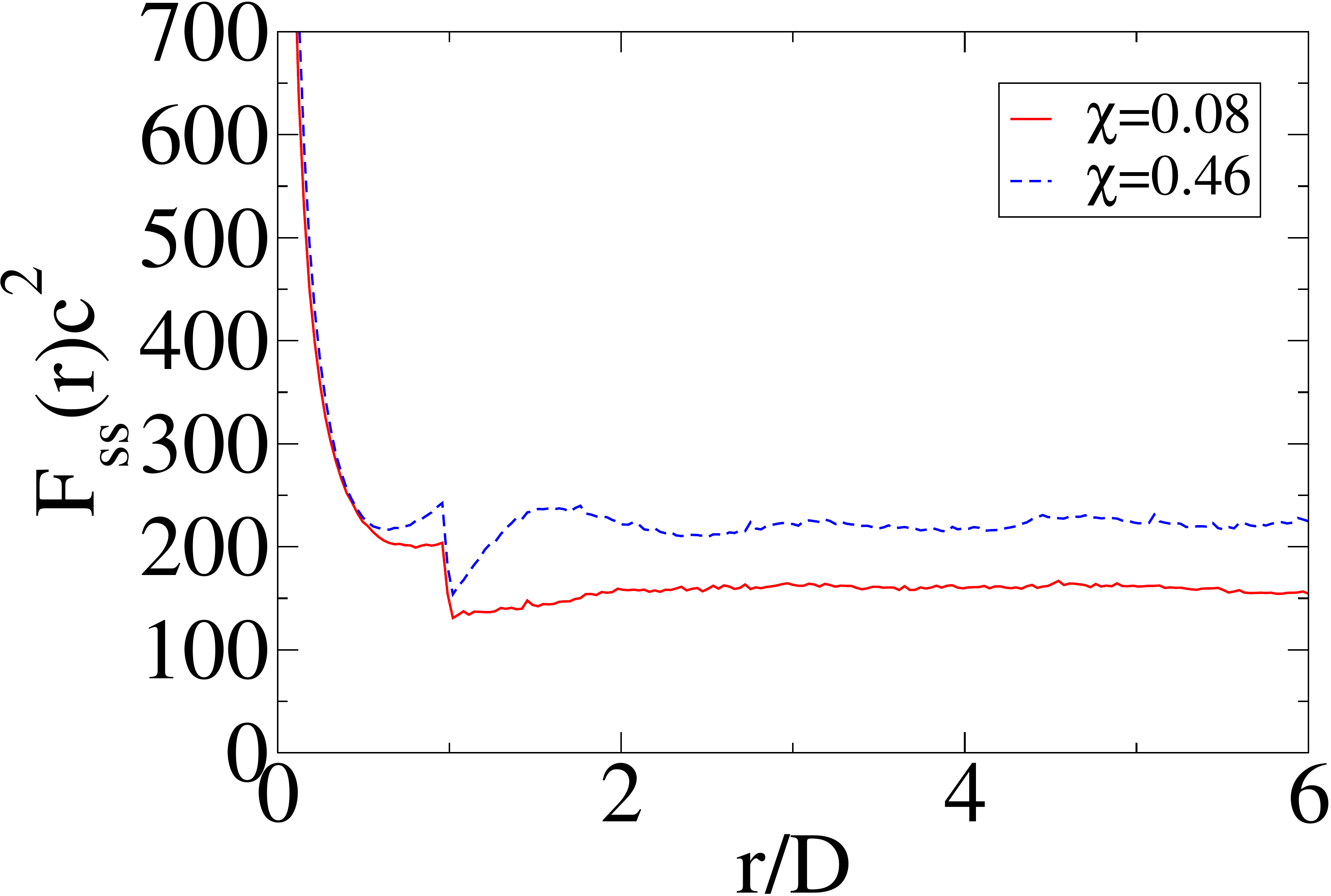}
}
\subfigure[]{
\includegraphics[width=7.5cm, height=5cm]{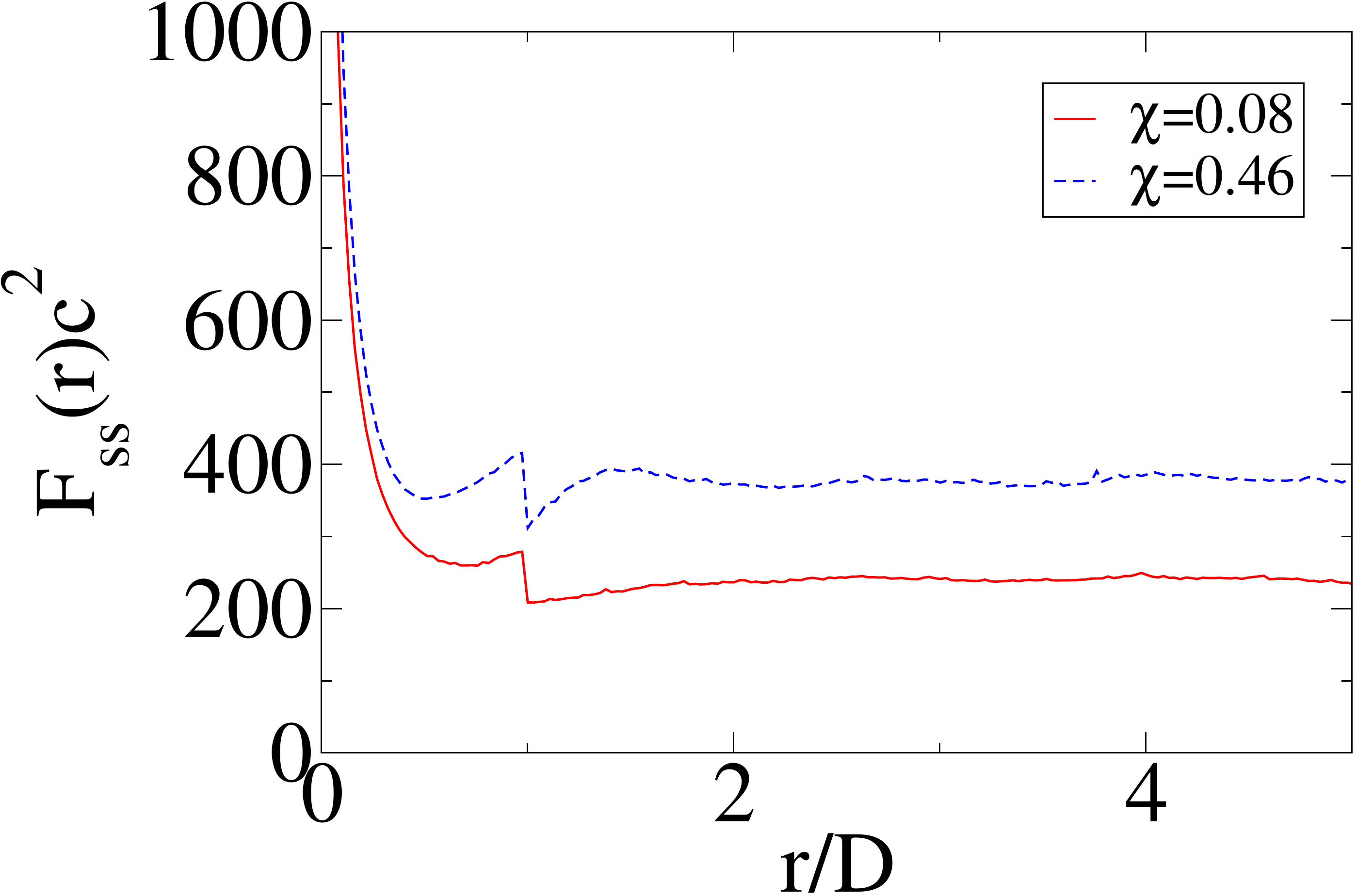}
}
\subfigure[]{
\includegraphics[width=7.5cm, height=5cm]{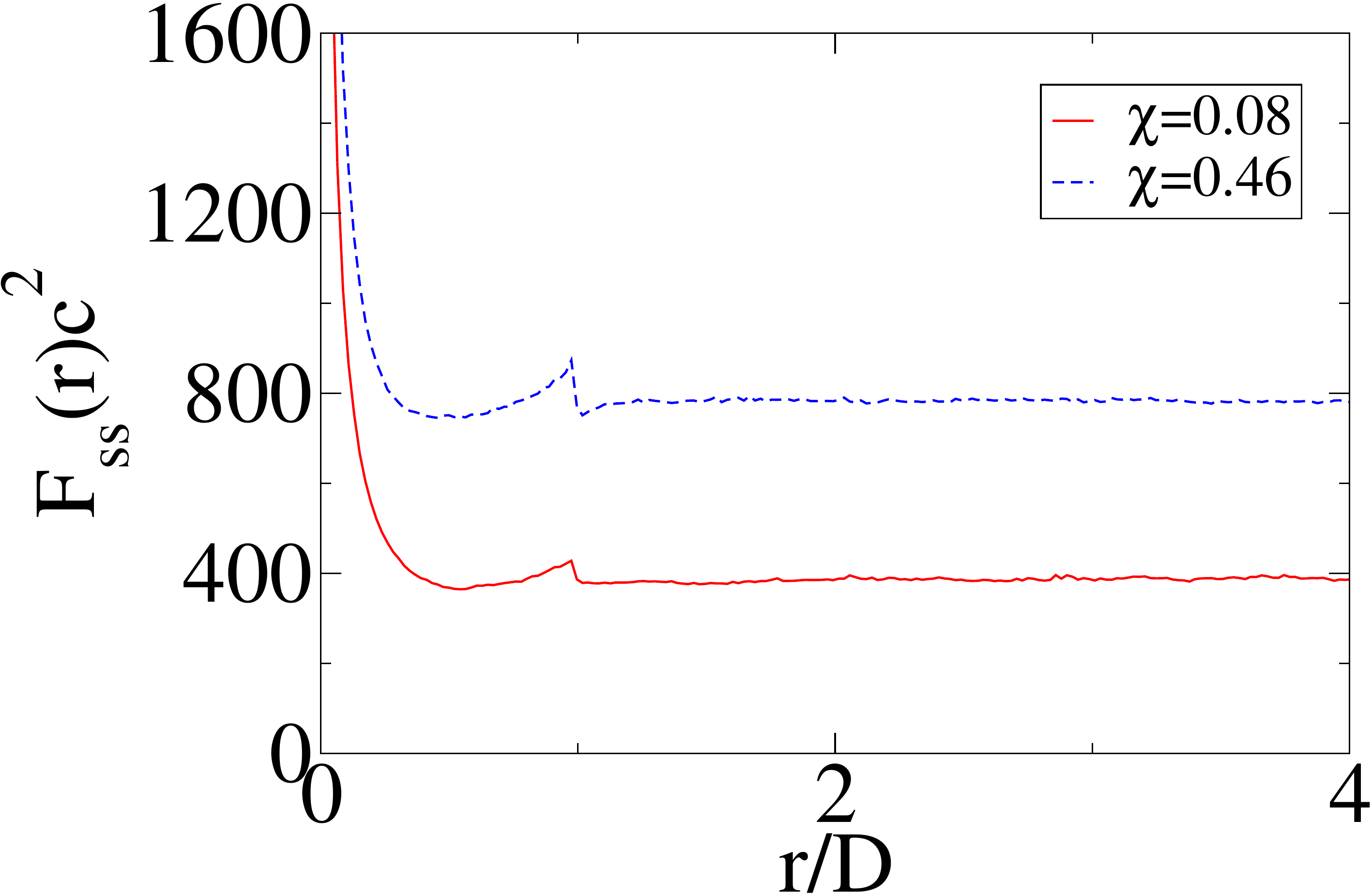}
}
\caption{Surface-surface correlation functions $F_{ss}$ of stealthy patterns decorated with spheres with different diameters: (a) $D=0.05c$. (b) $D=0.07c$. (c) $D=0.08c$. (d) $D=0.1c$.}
\label{fig:stealthy}
\end{figure*} 
Besides using $F_{ss}$ to compute surface-area variances, it can itself be used as a ``fingerprint" to detect important structural information, such as short-range order or the hyperuniformity of the system. Here we consider a special hyperuniform point patterns called ``stealthy" point patterns that were studied in a recent paper \cite{zhang2016transport}, and we follow the procedure of circumscribing each point with a sphere to make the system a two-phase medium. The ``stealthy" patterns are generated in a simulation box with basis vectors $(c,0,0)$, $(0,c,0)$ and $(c/2,c/2,c/2)$. We compare two sets of stealthy point patterns, with the parameter $\chi=0.08$ and 0.46 (In general, the system with larger $\chi$ will have more short-range order). Each point is decorated with a variable-sized sphere. We include our results for $F_{ss}$ in Fig. \ref{fig:stealthy}. When the decorated spheres are very small, such as the case in Fig. \ref{fig:stealthy}(a), they do not overlap with one another, and thus $F_{ss}$ should reveal structural features of the underlying point pattern. Clearly, the curve corresponding to $\chi=0.46$ in Fig. \ref{fig:stealthy}(a) exhibits stronger features, which is consistent with the fact that the pattern is more short-range ordered than that for $\chi=0.08$. As stated in Ref. \cite{zhang2016transport}, the system loses its hyperuniformity when spheres begin to touch each other. From Fig. \ref{fig:stealthy}, it is seen that as the diameter $D$ increases, the correlation function begins to lose its features, and ultimately these two ``stealthy" cases becomes indistinguishable from each other as well as the corresponding correlation function for overlapping spheres, shown in Fig. \ref{fig:bench}(a). The dramatic decrease around $r=D$ corresponds to the fact that the correlation between any two points on the same sphere cannot contribute to the function value beyond $r=D$, and thus reveals a characteristic length scale of the system. Moreover, although the two systems start almost at the same specific surface, the gap between two curves continues to increase, and in the end the system with $\chi=0.46$ has a much larger specific surface. This suggests that the system with $\chi=0.46$ has greater short-range order that keeps the decorated spheres from overlapping with one another, and thus leads to a larger specific surface.

\section{Results for Snapshots of Evolving Spatial Patterns}

In this section, we go beyond the analysis of well-known sphere models and extend the application of our algorithm to other important disordered patterns encountered in the physical and biological sciences. Specifically, we focus on time-dependent pattern formation processes that are governed by the Cahn-Hilliard equation and the Swift-Hohenberg equation. These patterns have recently been shown to be hyperuniform and could have important applications in material science \cite{doi:10.1063/1.4989492}. We determine correlation functions of snapshots of these patterns here.

\subsection{Spinodal decomposition patterns from the Cahn-Hilliard equation} 
\begin{figure*}[]
\centering
\subfigure[\ Critical quench]{
\includegraphics[width=6cm, height=6cm]{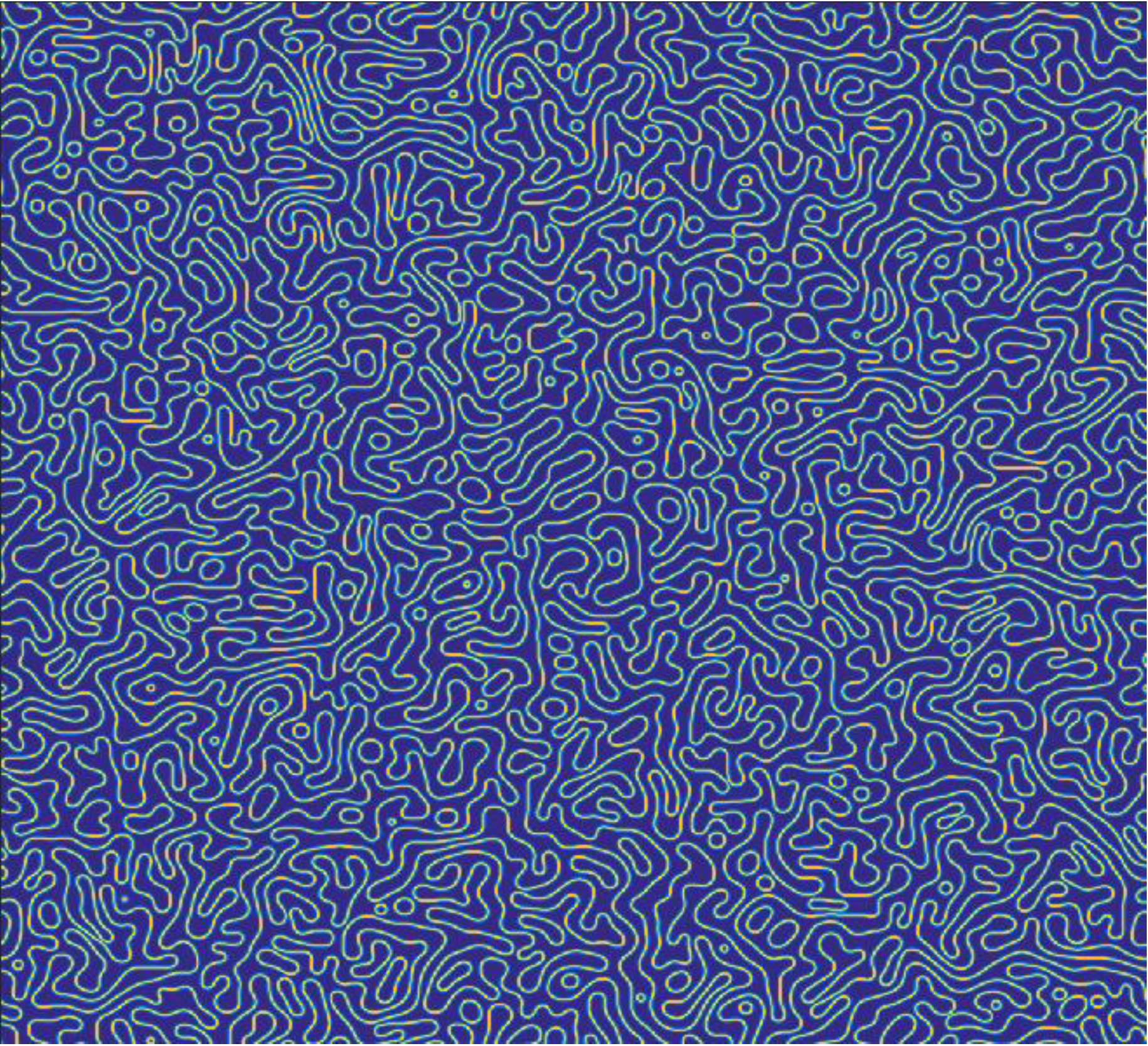}
}
\subfigure[\ Off critical quench]{
\includegraphics[width=6cm, height=6cm]{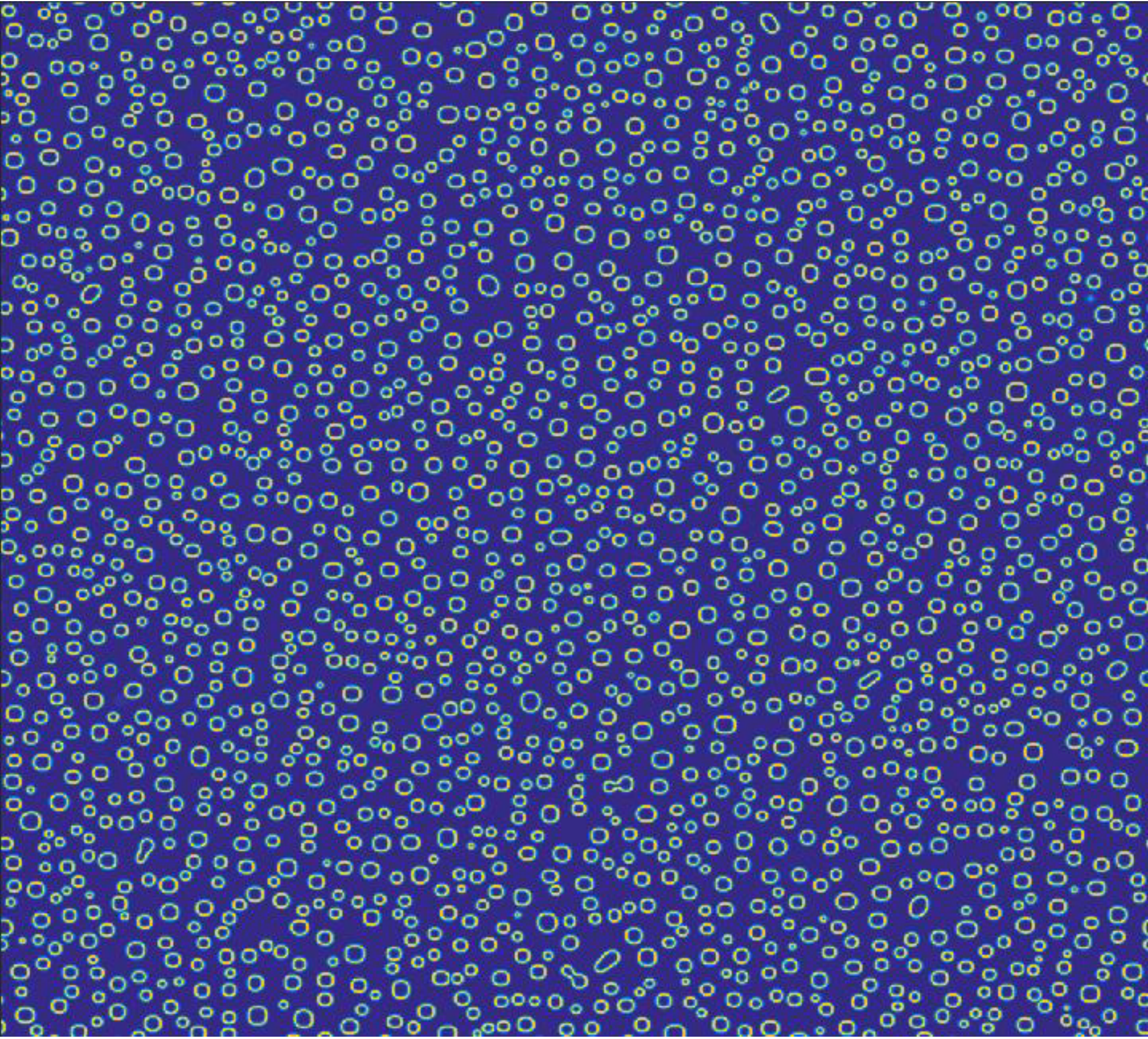}
}
\caption{The interfaces of two binary mixtures undergoing a phase separation at critical quench (volume fraction ratio of two phases is 1:1) and off critical quench (volume fraction ratio of two phases is 2:8), respectively. The system size is 1000$\times$1000.}
\label{fig:CH}
\end{figure*}

\begin{figure*}[]
\centering
\subfigure[\ Critical quench]{
\includegraphics[width=8cm, height=5cm]{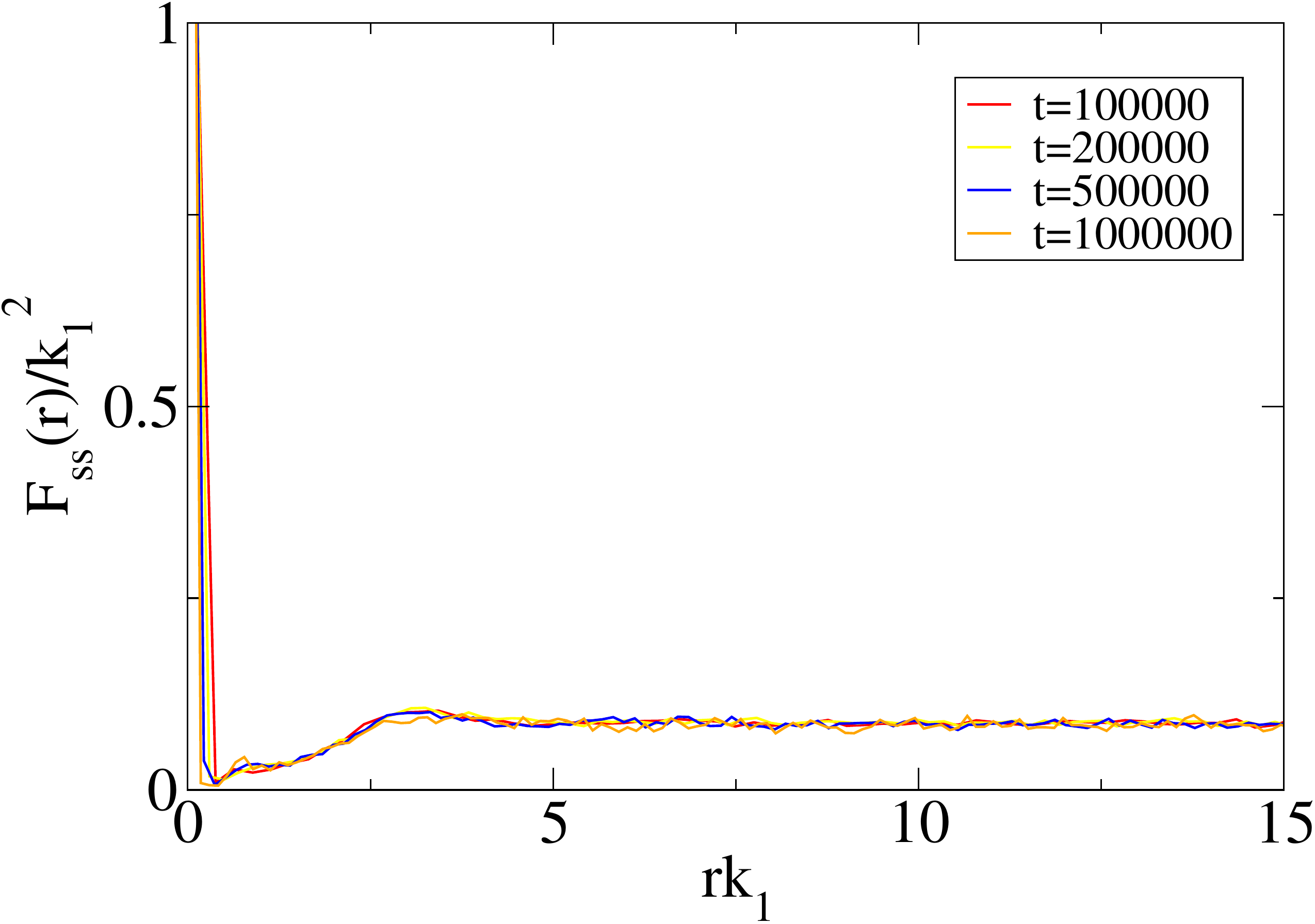}
}
\subfigure[\ Off critical quench]{
\includegraphics[width=8cm, height=5cm]{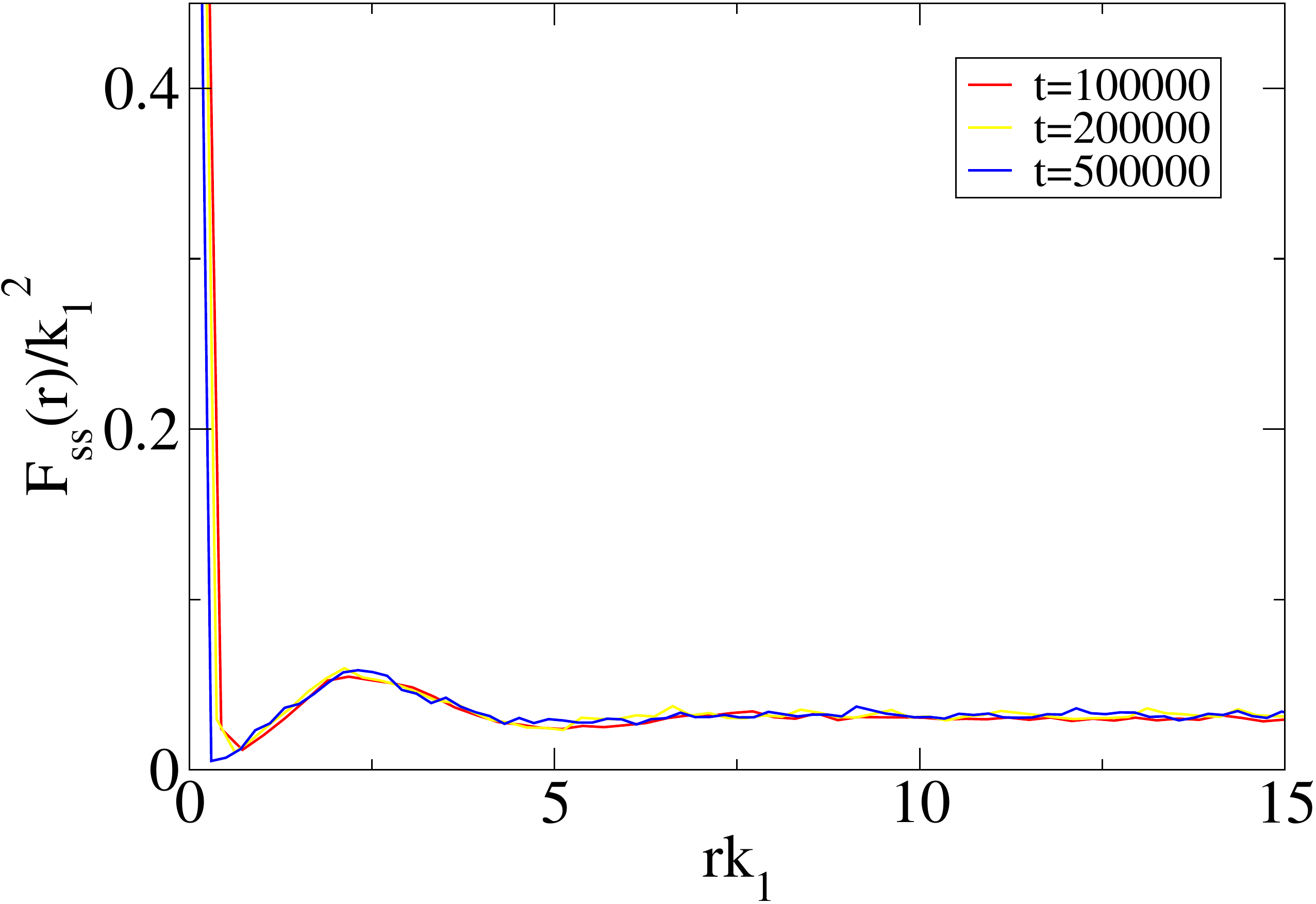}
}
\caption{(a) The scaled surface-surface correlation function $F_{ss}(r)/k_1(t)^2$ versus $rk_1(t)$ at different time stages associated with the spinodal decomposition pattern shown in the left panel of Fig. \ref{fig:CH}. (b) The scaled surface-surface correlation function $F_{ss}(r)/k_1(t)^2$ versus $rk_1(t)$ at different time stages associated with the spinodal decomposition pattern shown in the right panel of Fig. \ref{fig:CH}. One can see that they both collapse onto a single curve respectively after the rescaling. The shapes of two curves are significantly different.}
\label{fig:CHfss}
\end{figure*}

\begin{figure*}[]
\centering
\subfigure[\ Critical quench]{
\includegraphics[width=8cm, height=5cm]{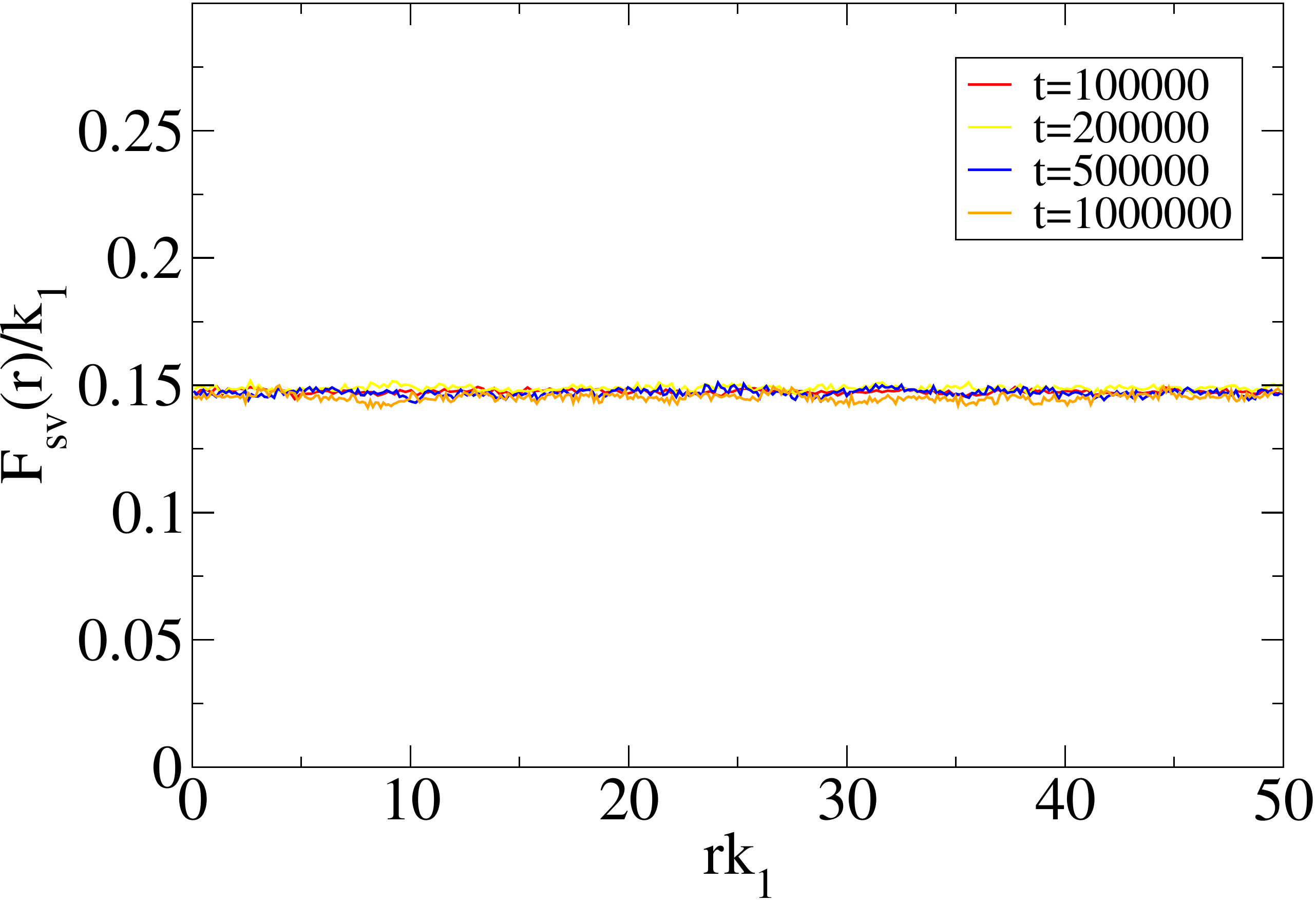}
}
\subfigure[\ Off critical quench]{
\includegraphics[width=8cm, height=5cm]{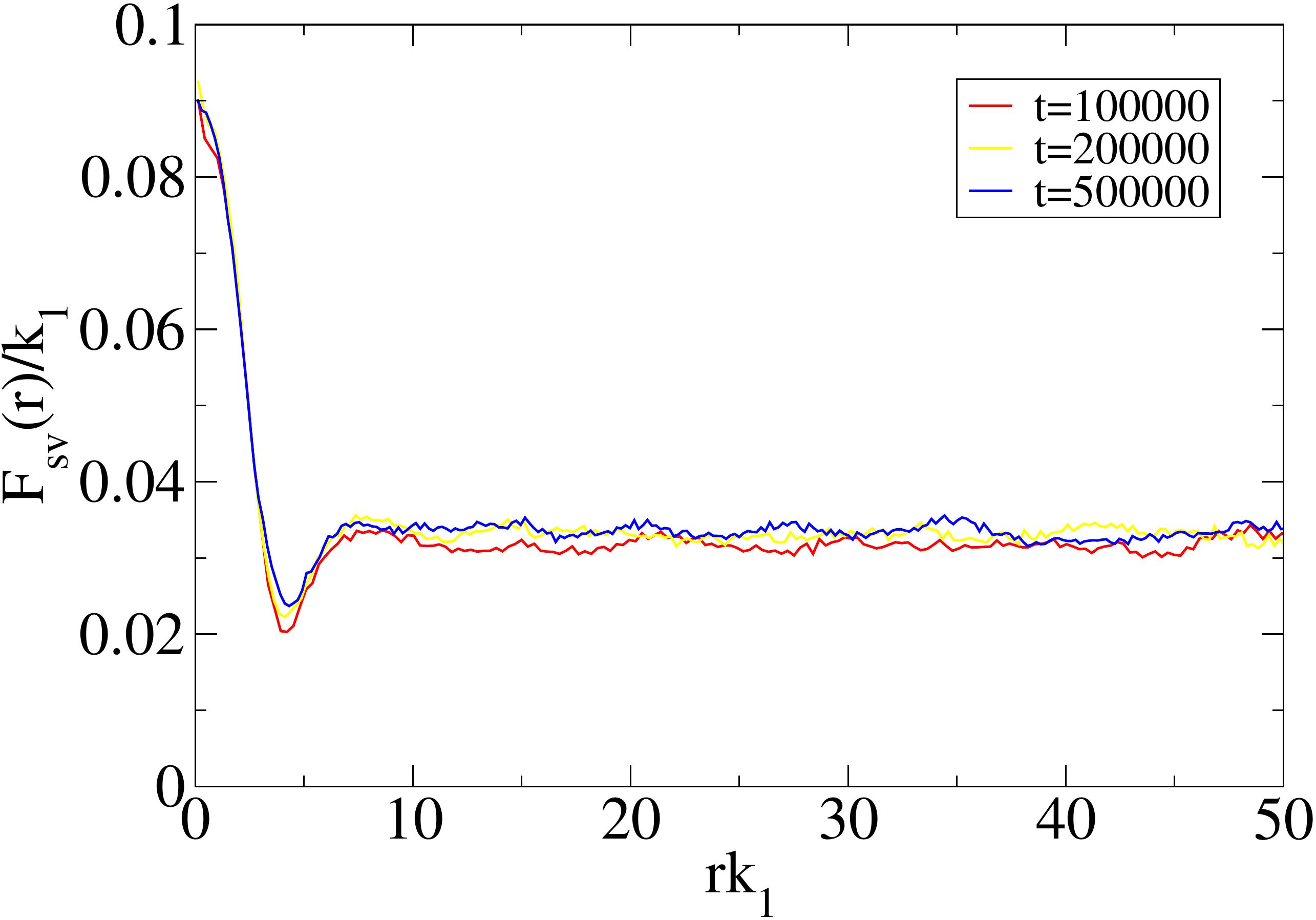}
}
\caption{(a) The scaled surface-void correlation function $F_{sv}(r)/k_1(t)$ versus $rk_1(t)$ at different time stages associated with the spinodal decomposition pattern shown in the left panel of Fig. \ref{fig:CH}. We see that $F_{sv}(r)$ is a constant (flat function), which is consistent with the exact expression (\ref{nouse3}). (b) The scaled surface-void correlation function $F_{sv}(r)/k_1(t)$ versus $rk_1(t)$ at different time stages associated with the spinodal decomposition pattern shown in the right panel of Fig. \ref{fig:CH}. One can see that they both collapse onto a single curve respectively after the rescaling. The shapes of two curves are significantly different.}
\label{fig:CHfsv}
\end{figure*}
\indent The Cahn-Hilliard equation was introduced to describe phase separation by spinodal decomposition  \cite{cahn1958free} and has been applied to model alloys \cite{rundman1967early}, polymer blends \cite{smolders1971liquid}, and even pattern formations in ecology \cite{liu2013phase}. In Fig. \ref{fig:CH}, we show two typical patterns generated by this equation. The left one is at critical quench, in which case the volume-fraction ratio for two phases is 1:1, while the right one is off critical quench and has volume-fraction ratio 2:8. The interface between the two phases is highlighted. \\   
\indent One important feature of the Cahn-Hilliard equation is that the system will enter a ``scaling regime'' after some time, and the system will remain statistically the same after scaled by a growing characteristic length. This provides an indirect way to check our algorithm on digitized media. We can compute the surface correlation functions at different times and then an appropriate scaling enables them to collapse onto a single curve.\\
\begin{figure*}[]
\centering
\subfigure[\ Critical quench]{
\includegraphics[width=8cm, height=5cm]{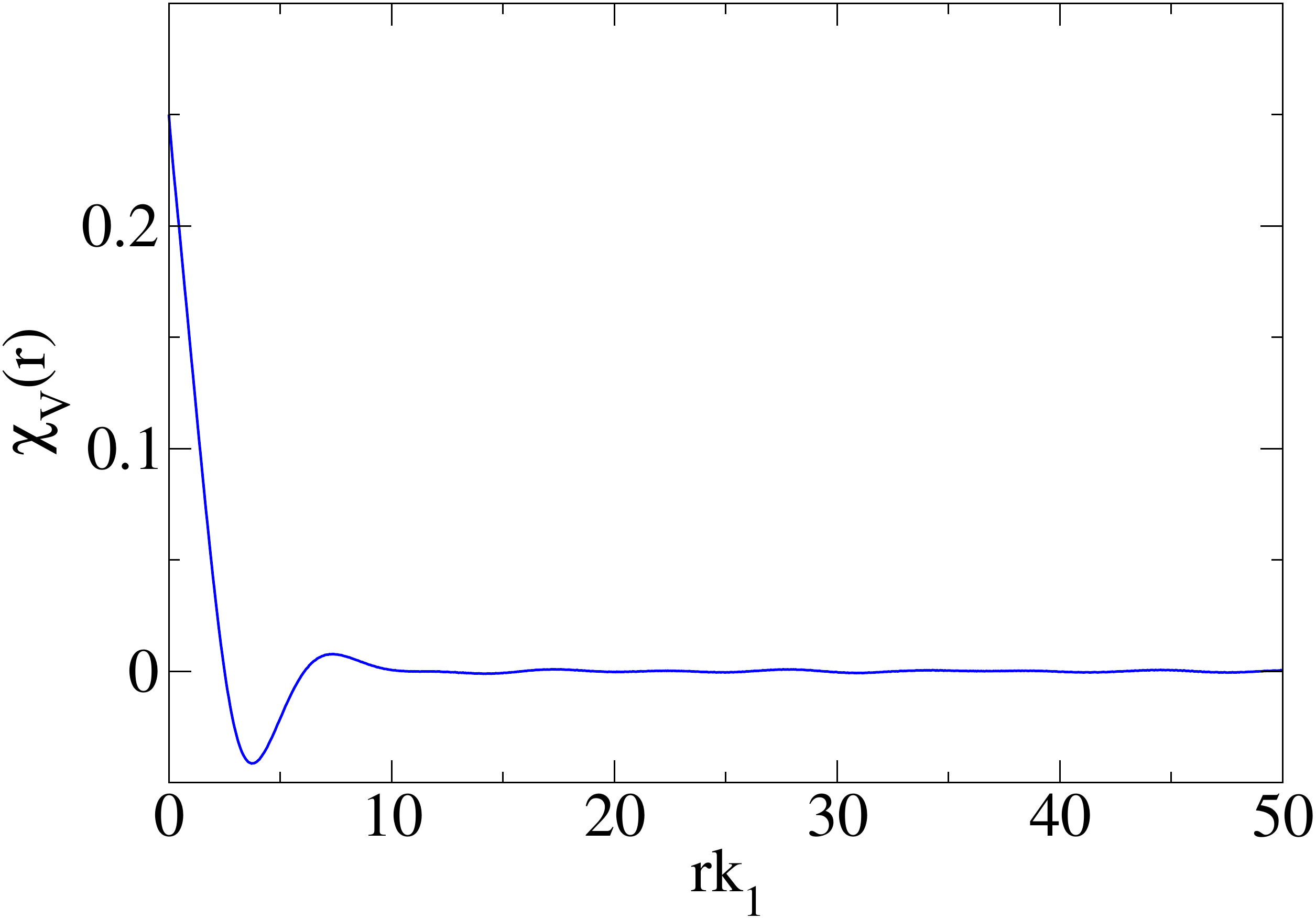}
}
\subfigure[\ Off critical quench]{
\includegraphics[width=8cm, height=5cm]{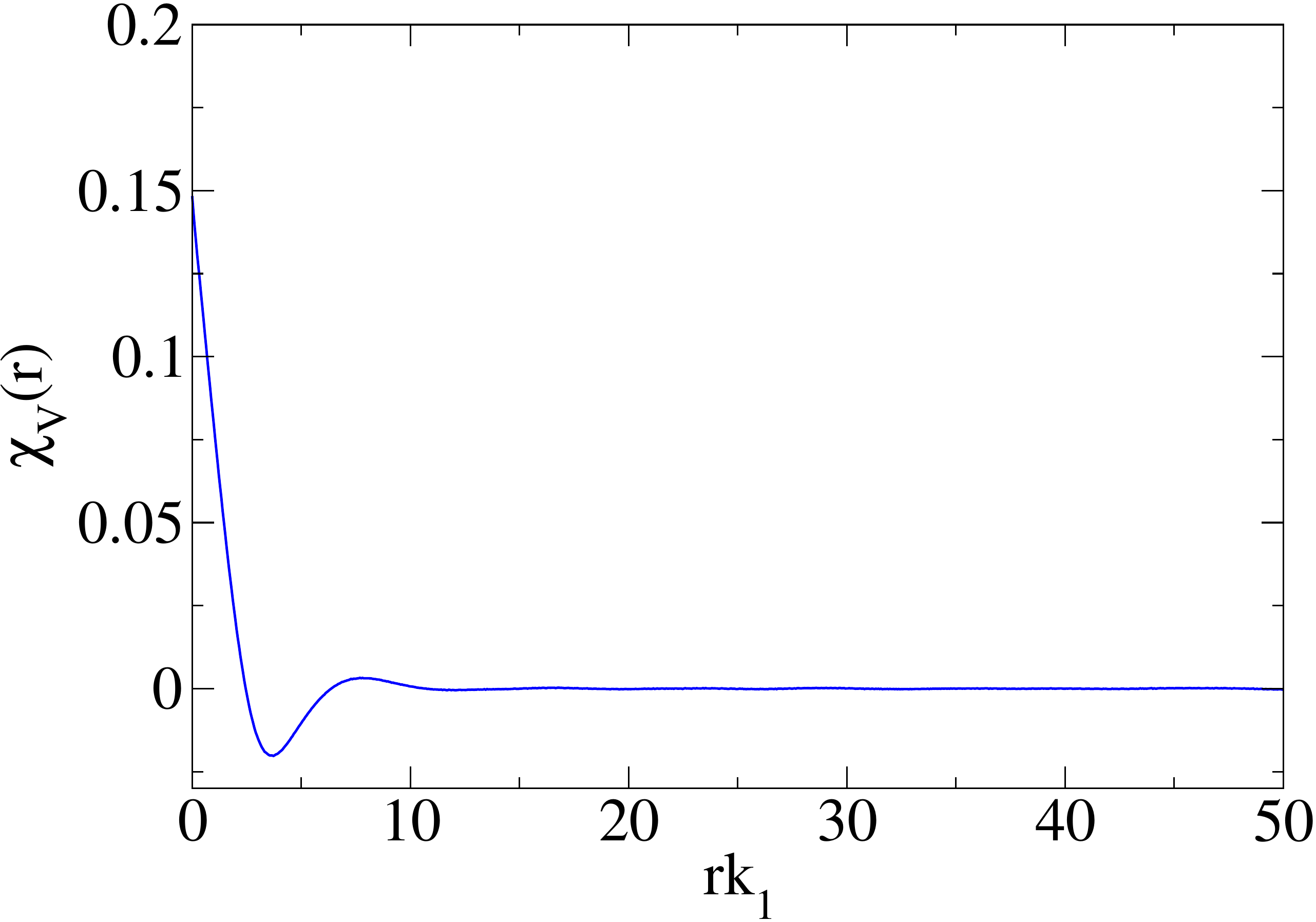}
}
\caption{(a) The autocovariance function $\chi_{V}(r)$ versus $rk_1$ at $t=100,000$ associated with the spinodal decomposition pattern shown in the left panel of Fig. \ref{fig:CH}, where $r$ is scaled by the characteristic wavenumber $k_1$. (b) The autocovariance function $\chi_{V}(r)$ versus $rk_1$ at $t=100,000$ associated with the spinodal decomposition pattern shown in the right panel of Fig. \ref{fig:CH}. There is no significant difference between these two curves in (a) and (b).}
\label{fig:CHchiv}
\end{figure*}
\indent The rescaled surface-surface and surface-void correlation functions at different times are shown in Figs. \ref{fig:CHfss} and \ref{fig:CHfsv} for critical and off-critical quenches. The curves for different times do collapse onto each other, as expected, further justifying the accuracy of our algorithm. Note that although the two systems shown
in Fig. \ref{fig:CH} appear to be structurally different, the corresponding standard autocovariance functions $\chi_V(r)$ in Fig. \ref{fig:CHchiv} are similar to one another. The inability to distinguish the structures of these two systems is easily overcome by complementing $\chi_V(r)$ with the information content of $F_{ss}(r)$ and $F_{sv}(r)$, as they differ greatly for these two systems. Specifically, one can see that in Fig. \ref{fig:CHfsv}, the surface-void correlation function for the critical quench has a flat slope at the origin [as predicted by Eq. (\ref{nouse3})], while the one for the off-critical quench has a downward slope at the origin. This can be well explained by the small-$r$ behavior of $F_{sv}$ that was derived in Sec. III. A.. From Eq. (\ref{2dFsv}), we know that the slope of $F_{sv}$ at origin is proportional to the \textit{mean curvature} of the system. In the case of critical quench, the surface consists of both concave and convex parts, whose contributions cancel each other out, and thus the \textit{mean curvature} is zero. In the case of off critical quench, where the matrix is the solid phase and the droplets are taken to be the void phase, the \textit{mean curvature} is apparently negative, and thus in Fig. \ref{fig:CHfsv}(b) we see the curve slopes down initially. This example again demonstrates the value of surface correlation functions in characterizing complex patterns.    

\subsection{Patterns from the Swift-Hohenberg equation}
\begin{figure}[H]
\centering
\includegraphics[width=8cm,height=6cm]{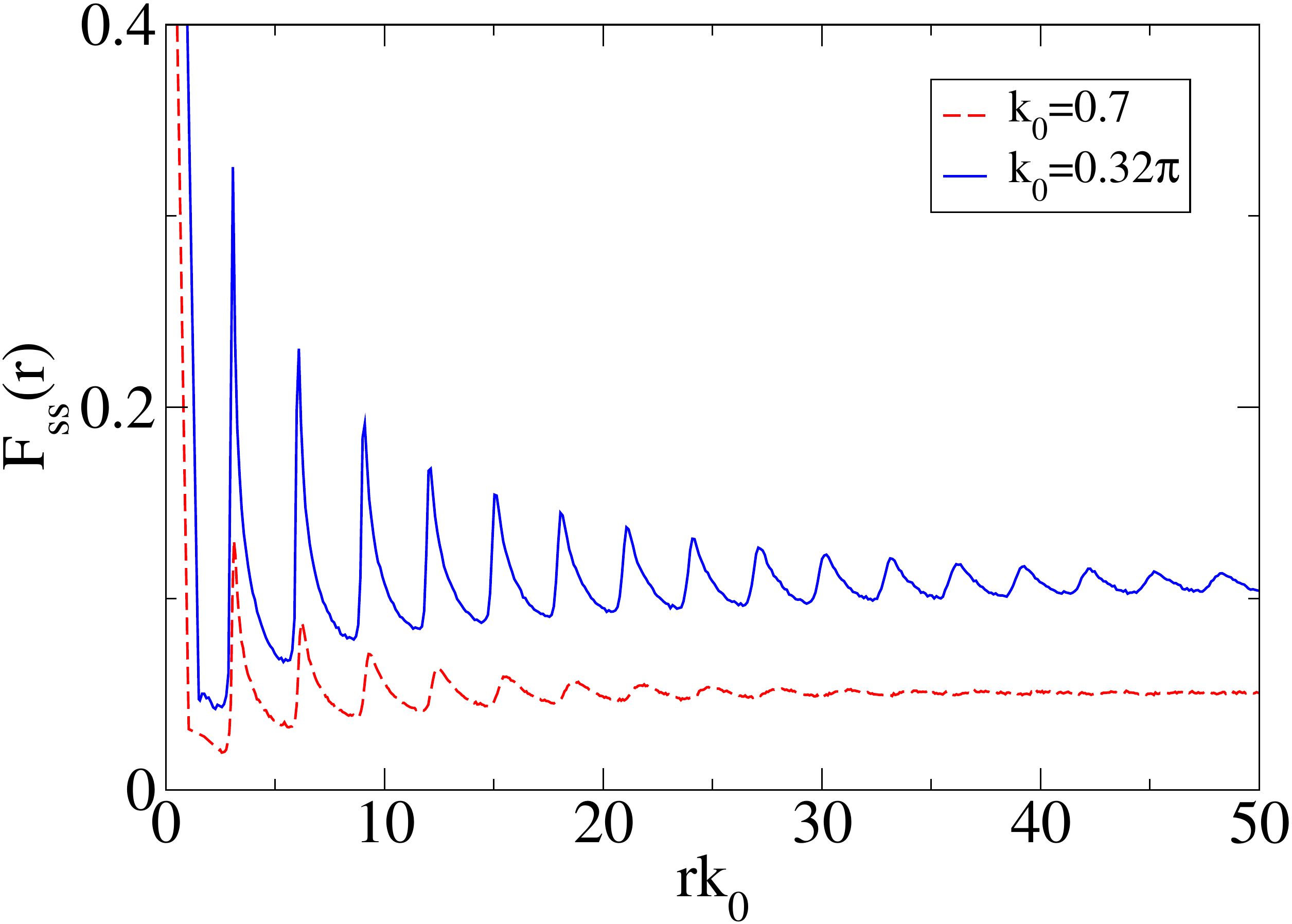}
\caption{Surface-surface correlation function $F_{ss}(r)$ of patterns generated from the Swift-Hohenberg equation with two different values of the parameter $k_0$.}
\label{fig:SHfss}
\end{figure}

\begin{figure}[H]
\centering
\includegraphics[width=8cm,height=5cm]{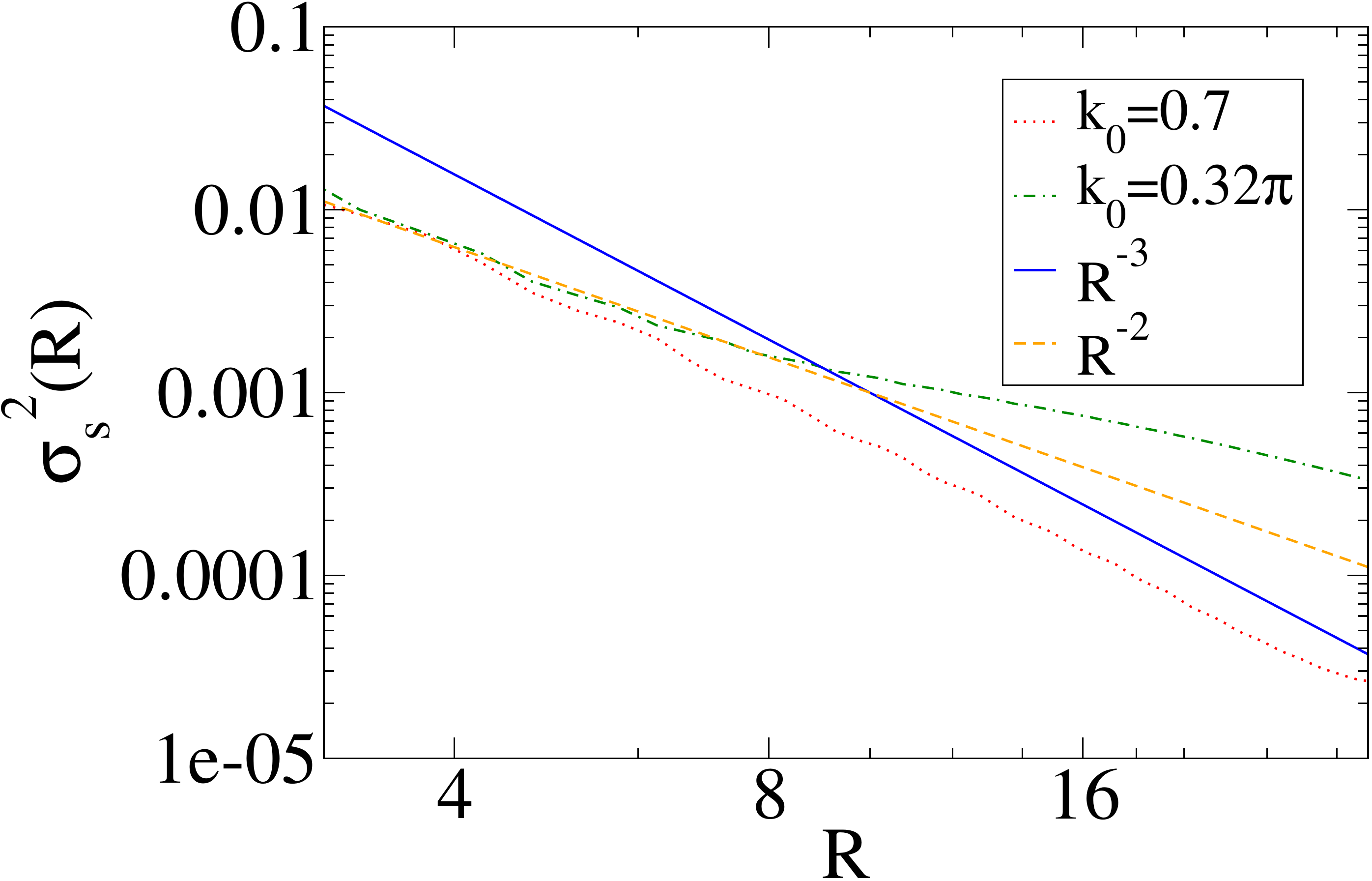}
\caption{Local surface-area variances $\sigma_{_S}^2(R)$ as functions of window radius $R$ computed from Eq. (\ref{auto}) for patterns generated from the Swift-Hohenberg equation with two different values of the parameter $k_0$, with comparison of different scalings.}
\label{fig:SHfluc}
\end{figure}

\indent The Swift-Hohenberg equation was developed to study Rayleigh-B{\' e}nard (RB) convection in hydrodynamics and later it became a subject of interest on its own in pattern formations \cite{cross2009pattern}. 
The pattern produced by this equation is usually labyrinth-like, and the width of the ``channel" is determined by a pre-selected wave number $k_0$. It has been shown that the patterns can have different degrees of hyperuniformity \cite{doi:10.1063/1.4989492} when some tuning parameters are changed, although they may appear to be structurally alike.\\
\indent Here we compute and compare two surface-surface correlation functions for two patterns generated under different $k_0$, namely $k_0=0.7$ and $k_0=0.32\pi$ in the same way in the authors' previous paper \cite{doi:10.1063/1.4989492}. It has been shown that the later one is more long-range ordered, which is also justified in our plot of $F_{ss}$ in Fig. \ref{fig:SHfss}. It is evident that the $F_{ss}$ for $k_0=0.32\pi$ is much more long-ranged than the one for $k_0=0.7$. Both curves have sharp spikes when $rk_0$ is integer times of $\pi$, which corresponds to the fact that the underlying patterns consist of stripes with width of $\pi/k_0$, leaving roughly parallel interfaces with the same spacing at short scales. Note that spikes in $F_{ss}(r)$ also occur in sphere systems but only at the single location $r=D$ (see Fig. \ref{fig:bench} and Fig. \ref{fig:stealthy} for examples), while the corresponding $S_2$ for these systems are smooth functions without sharp transitions (see Refs. \cite{torquato2013random} and \cite{yeong1998reconstructing} for plots). This again shows that surface-surface correlations can be superior in detecting short-scale microstructural features compared to that of the standard two-point correlation function $S_2(r)$.\\
\indent We also evaluate the local surface-area variances in these systems using $F_{ss}$ and Eq. (\ref{auto}). The results are shown in Fig. \ref{fig:SHfluc}. Note that the surface-area variance for $k_0=0.7$ scales like $R^{-3}$, implying hyperuniformity \cite{PhysRevE.94.022122}. However, the variance for $k_0=0.32\pi$ scales even slower than $R^{-2}$. The explanation is that in the case of $k_0=0.32\pi$, the corresponding wavelength is too small compared to the pixel size, which makes the numerical integration in Eq. (\ref{auto}) unreliable.

\section{CONCLUSIONS AND DISCUSSION}

\indent In this paper, we developed efficient general algorithms that enable the sampling
of the surface-surface correlation function $F_{ss}(r)$ and the surface-void correlation function $F_{sv}(r)$ with heretofore unattained precision. Our algorithms have advantages over the traditional ``dilation" method \cite{seaton1986spatial, klatt2016characterization}. First, the dilation method can only be easily implemented when the interfaces are relatively smooth and easy to be parameterized (e.g., packings of spheres and ellipsoids). However, our algorithms can be easily adapted to treat general complex interfaces. Second, the dilation method is difficult to implement for digitized media, which greatly limits its application to experimental data. By contrast, we have shown that our algorithms can be straightforwardly applied to digitized media. Third, as the dilation thickness $\epsilon$ approaches to zero, the probability of hitting the dilated phase will proportionally decrease, which requires a large number of samplings to ensure the accuracy, and hence greater computational
time. However, in the extreme situation that the information of a large but single system is available, our algorithms can yield accurate results from a single sample, since it is possible for the straight line to penetrate the interface a sufficiently large number of times. Moreover, our algorithms can be generalized to compute three-point surface correlation functions \cite{torquato2013random} straightforwardly. Application of our algorithms to a variety of model disordered microstructures reveals that surface-surface correlation function $F_{ss}(r)$ is a sensitive descriptor of small-scale structural features, especially compared to the information content of the standard two-point correlation function $S_2(r)$.\\
\indent We also showed that the extracted surface correlation functions can be used to compute accurately the surface-area variance, a quantity that can be a more sensitive measure of microstructural fluctuations compared to the volume-fraction variance. Through examples of spinodal decomposition patterns, we showed that surface correlation functions contain information that supplements that of $S_2$, and the small-$r$ behavior of $F_{sv}(r)$, which is determined by the \textit{mean curvature} of the system. In two dimensions, the \textit{total curvature} of a closed simple curve is a constant $2\pi$, implying that when the system approaches a percolation threshold, the absolute value of the \textit{mean curvature}  will drop dramatically due the formation of large clusters. This observation suggests that the surface-void correlation function $F_{sv}(r)$ may aid in detecting the onset of continuum percolation, which is an interesting topic for future exploration. We also showed how surface-surface correlation functions can be used to determine the hyperuniformity of two-phase media using patterns generated by the Swift-Hohenberg equation as examples.\\
\indent Lower-order correlation functions have been successfully used to infer the physical properties of random media as well as to reconstruct them. This bodes well for their use in machine learning in the area of material optimization \cite{stenzel2017big, roding2017functional}. We expect that the algorithms to compute precisely the surface-surface correlation function $F_{ss}(r)$ and the surface-void correlation function $F_{sv}(r)$ presented in this paper will equip the community with powerful computational tools to characterize the structure and physical properties of multiphase media, especially with respect to those physical processes that are intimately linked to the interfaces. In particular, our algorithms can be adapted in reconstruction algorithms \cite{yeong1998reconstructing, jiao2009superior} with heretofore unattained accuracy without sacrificing computational speed.\\
\indent A sample Matlab program that enables one to compute the correlation functions $F_{ss}$, $F_{sv}$ and $F_{vv}$ for three dimensional digitized media can be downloaded at Ref. \cite{link}.

\begin{acknowledgements}
\indent The authors are grateful to Michael Klatt, Ge Zhang, JaeUk Kim, Timothy Middlemas, and Duyu Chen for helpful discussions. This work was supported in part by the National Science Foundation under Award No. CBET-1701843.\\
\end{acknowledgements}

\begin{appendix}

\section{The small-$r$ behavior of surface correlation functions of systems with singularities}
\indent Since the surface-surface correlation function discussed in Sec. III is only approximated to the zeroth-order, we focus our attention here to the surface-void correlation function of systems with singularities. We include results for certain specific cases, namely, two-dimensional systems that all singularities are corners and three-dimensional overlapping spheres. \\
\indent For two-dimensional systems that all singularities are corners, suppose the angle formed by a corner to the solid phase is $\theta$, one can show that the $F_{sv}(r)$ for small $r$ can be written as 
\begin{equation}
F_{sv}(r)=s(\frac{1}{2}+\frac{r}{2\pi} \left\langle \frac{1}{r_c}\right\rangle_{\Omega})+2\rho_c \left\langle \cot \frac{\theta}{2}\right\rangle r, 
\end{equation}
where $\Omega$ is the set of all the points on the interfaces that are differentiable, and $\rho_c$ is the density of corners in the system. One interesting implication of this formula is that the expression of the surface-void correlation function for packings of equilateral polygons is the same as the expression for packings of their inscribed circles.\\
\indent Singularities in three dimensions are much more complex to analyze. Here we only take overlapping spheres as an example. Surprisingly, this is a nontrivial model of a heterogeneous material, since the lack of spatial correlation implies that the particles may overlap to form complex clusters, and leave many nondifferentiable edges in the system. Using the same geometric approach in Sec. III, by naively plugging in $a$ as $r_c$ in Eq. (\ref{3dFsv}), we should have 
\begin{equation} 
F_{sv}(r)=s(\frac{1}{2}+\frac{r}{4a}),
\end{equation}  
where $a$ is the radius of spheres and $s=3\eta e^{-\eta}/a$ is the specific surface of overlapping spheres \cite{torquato2013random}. However, expanding Eq. (\ref{overfsv}) to the first order directly gives us
\begin{equation} 
F_{sv}(r)=s(\frac{1}{2}+\frac{r}{4a}-\frac{3\eta r}{8a}).  
\label{expand}
\end{equation}
\begin{figure}[H]
\centering
\includegraphics[height=6cm]{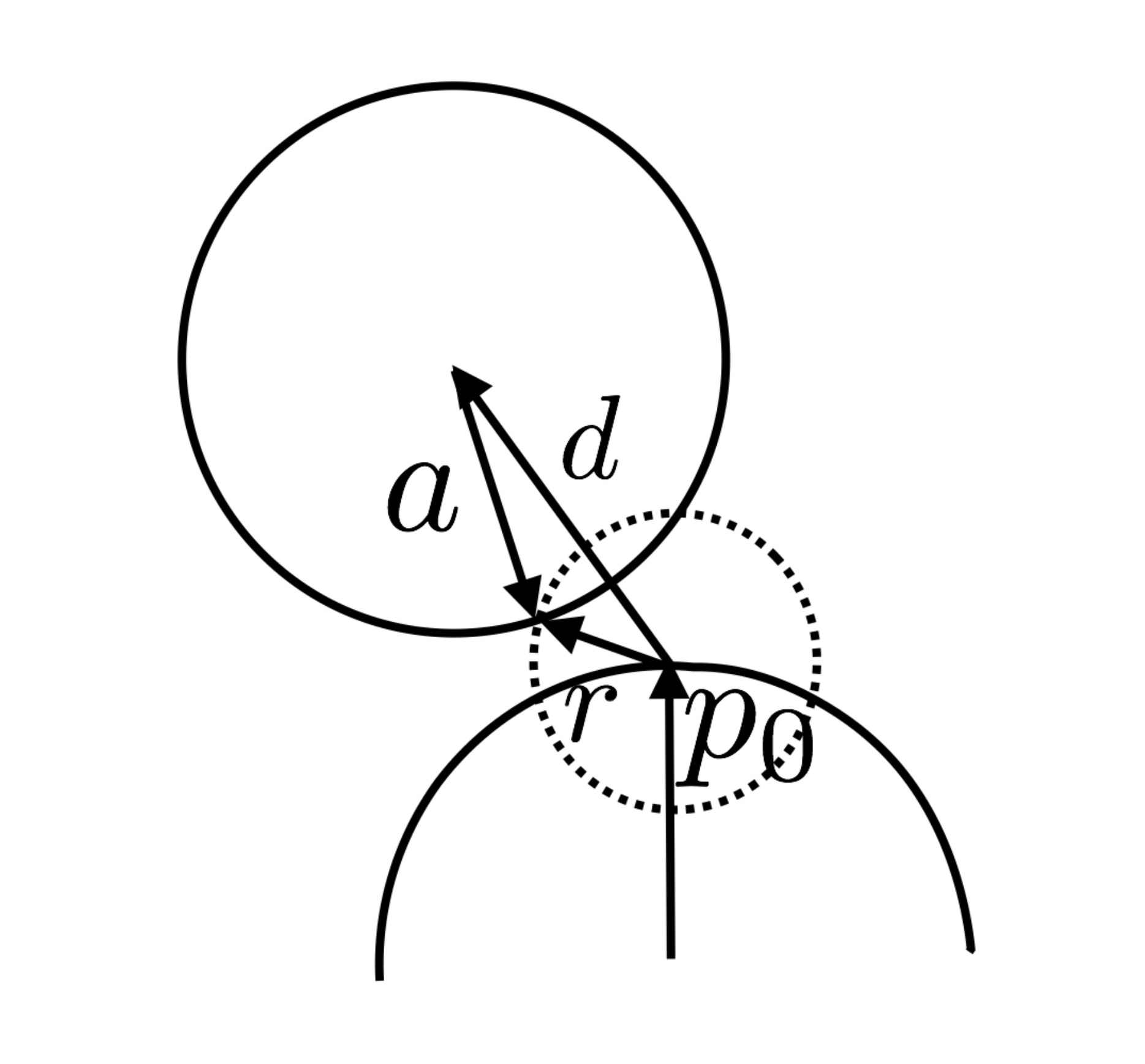}
\caption{ An illustration of evaluating the two-body correction to the small $r$ behavior of $F_{sv}(r)$, where the upper left sphere is the ``invading" sphere and the sphere in dotted line is the ``test" sphere.}
\label{fig:smallr}
\end{figure} 
\indent The extra negative term $-3s\eta r/8a$ implies that we have overestimated the probability of falling into the void phase $P_{sv}$ by neglecting the fact that another sphere (see the upper left sphere in Fig. \ref{fig:smallr}, which we  henceforth call the ``invading" sphere) can approach to and intersect with our ``test" sphere (see the ``dotted" sphere in Fig. \ref{fig:smallr}) and reduce its fraction of surface area covered in the void phase. Indeed, any sphere whose centroid lies in the concentric shell with radius $a$ to $a+r$ around the reference point $p_0$ will intersect with the ``test" sphere (we do not consider spheres that are closer than $a$ since then the reference point 
would no longer be on the interface). Here we evaluate the reduced fraction of surface area in the void phase of the ``test" sphere.\\
\indent When the ``invading" sphere overlaps with both the ``test" sphere and the interface it can be difficult to evaluate the extra surface area of the ``test" sphere covered by the ``invading" sphere. Luckily, since the volume of the shell is $4\pi a^2r$, which is already first order, we can approximate the interface around the reference point as a flat plane that divides the ``test" sphere into two hemispheres. By symmetry we know the extra surface area covered on average is just $1/2$ of the total surface area covered by the ``invading" sphere on average. Suppose the distance between the center of the ``invading" sphere and the reference point is $d$, then the surface area of the spherical crown that is covered is
\begin{equation} 
S=2\pi r^2(1- \frac{d^2+r^2-a^2}{2dr}).  
\end{equation} 
Letting $d=a+x$, the fraction of surface area of the ``test" sphere that is covered by the ``invading" sphere is
\begin{equation} 
\frac{S}{4\pi r^2}=\frac{1}{2}(1-\frac{(a+x)^2+r^2-a^2}{2(a+x)r})\approx \frac{1}{2} (1-\frac{x}{r}).  
\end{equation} 
Then on average the total fraction that is covered by the ``invading" sphere is
\begin{equation} 
\int_{0}^{r} \frac{1}{2} (1-\frac{x}{r})\times 4\pi (a+x)^2\rho dx = \pi a^2r\rho+\mathcal O(r^2). 
\end{equation} 
Finally, by symmetry, we know the correction term to $F_{sv}(r)$ is $-s\pi a^2r\rho/2$ or
 $-3s\eta r/8a$, as in Eq. (\ref{expand}). This correction term will disappear in the dilute limit, since spheres will not overlap with one another.\\   
\indent One can carry out the same analysis for impenetrable spheres, but the calculation will be much more involved. In this case, the other sphere can only approach from the void phase, and the nonoverlapping condition will restrict its direction to a small range, which will in the end make the correction term of the order $\mathcal O(r^2)$ as long as $g_2(D^{+})$ is not a delta function, where $D$ is the diameter of a sphere. Thus our general formula will apply to hard spheres in equilibrium or random sequential addition (RSA) packings \cite{zhang2013precise}. 

\section{The probability of getting an abnormal peak}
It is instructive to estimate the probability $P\{\frac{1}{N}\sum\limits_{i=1}^{N}\frac{1}{\cos{\theta_i}}>\frac{\pi}{2}+e\}$, where $N$ is the number of sampling and $e$ stands for a given error.
To do so, first we can sort $1/\cos{\theta}$ such that $1/\cos{\theta_i}\geq1/\cos{\theta_{i+1}}$ for $i=1, 2, ..., N-1$. Define
\begin{equation}    \label{defines}  
T=\operatorname*{min} \{j |\sum\limits_{i=1}^{j}\frac{1}{\cos{\theta_i}}>N(\frac{\pi}{2}+e) \},
\end{equation}
 \begin{figure*}[]
\centering
\subfigure[]{
\includegraphics[width=7.5cm,height=5cm]{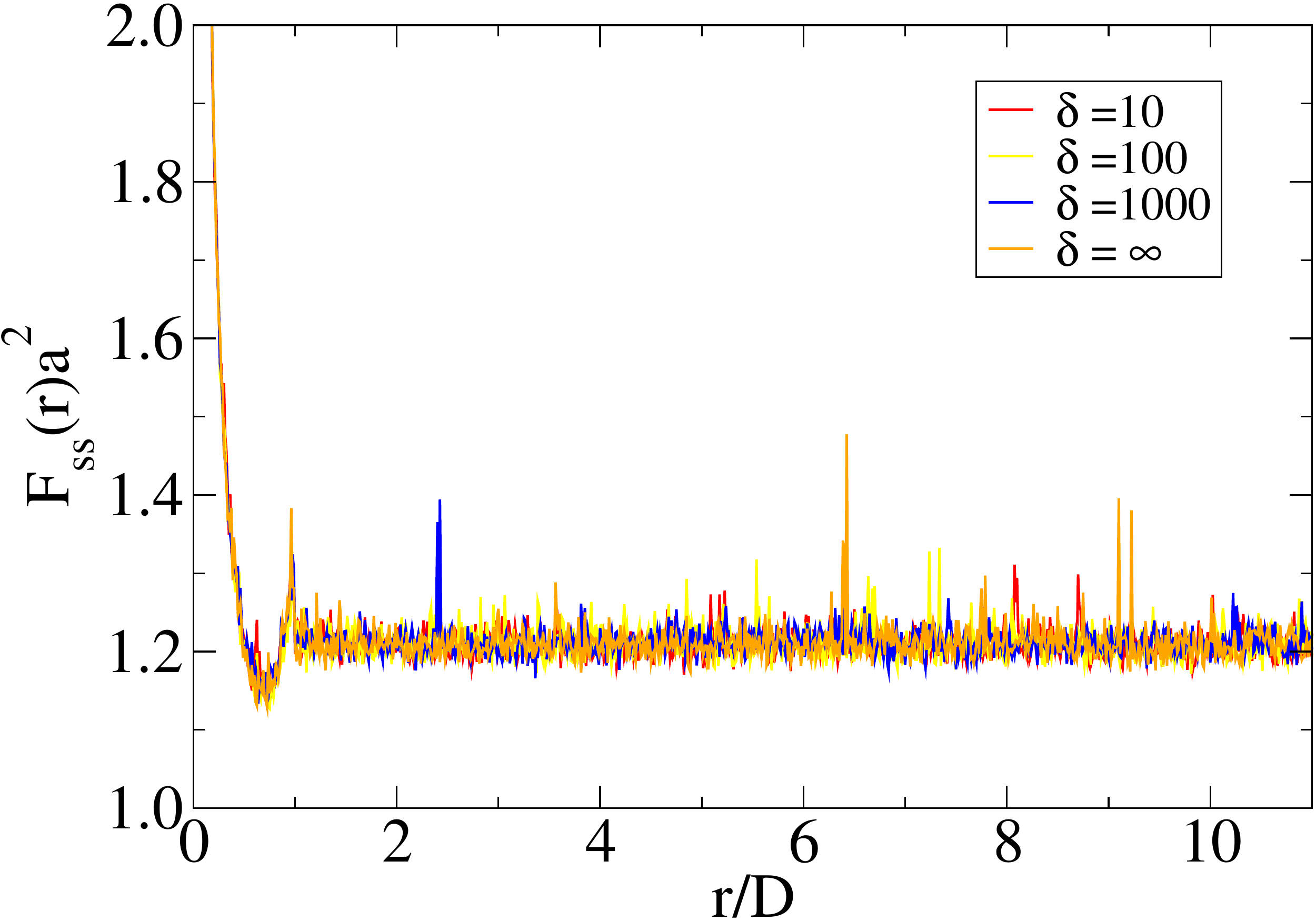}
}
\subfigure[]{
\includegraphics[width=8.5cm,height=5cm]{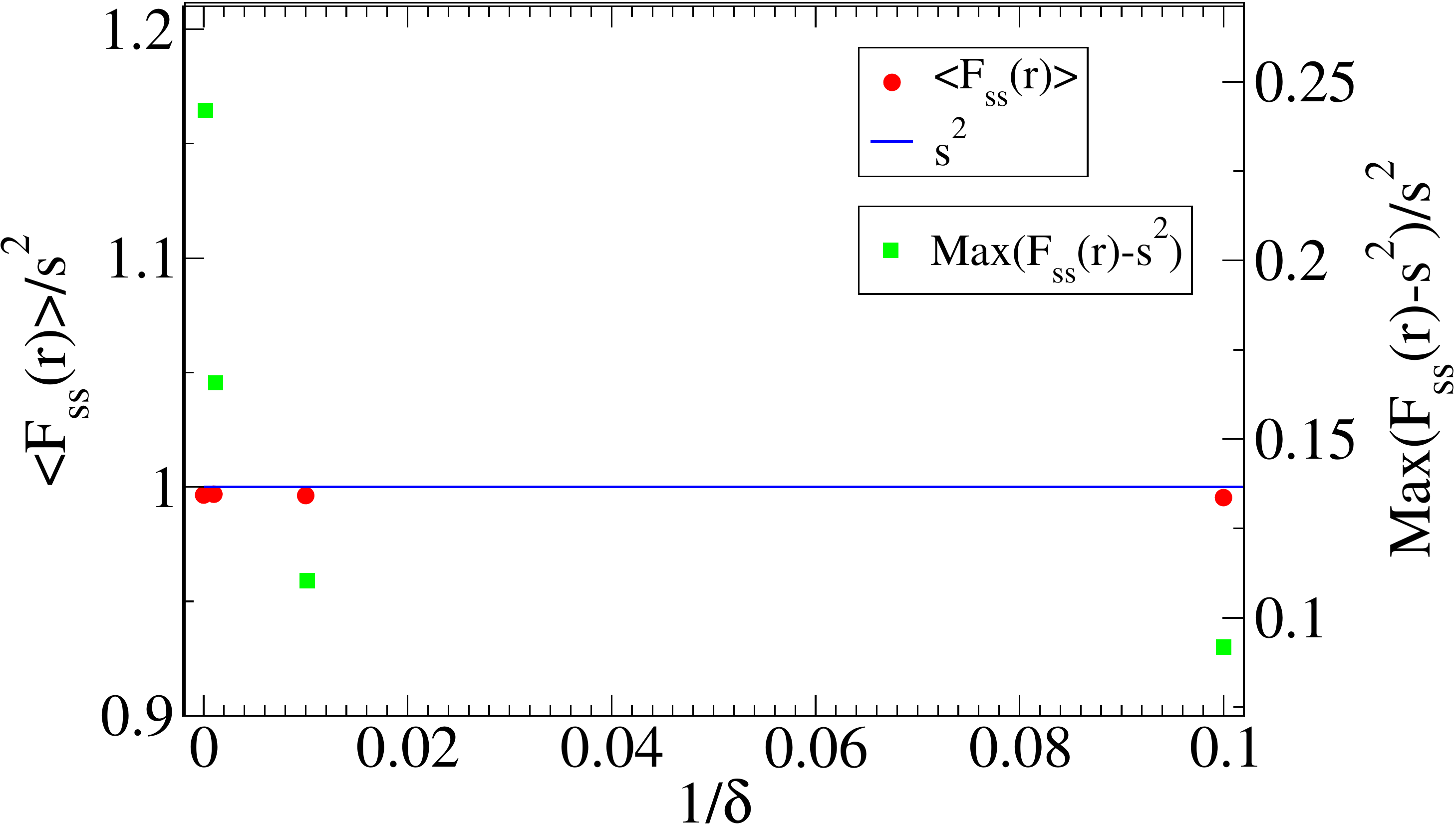}
}
\caption{(a) Simulation results of $F_{ss}(r)$ for systems of three-dimensional overlapping spheres computed with different thresholds 10, 100, 1000 and without threshold. We use the same system in Sec. IV, except that only 10,000 sampling lines are used. Clearly, applying thresholds does not change the overall shape of the curve, but significantly reduce the fluctuations. (b) Red circles: Average of $F_{ss}(r)$ in the interval $[D,11D]$ for different thresholds. The blue line is $s^2$, which is the theoretical value of $F_{ss}(r)$ when $r>D$. Green squares: the largest deviation of $F_{ss}(r)$ from $s^2$ in the interval $[D, 11D]$ for different thresholds. The average of $F_{ss}(r)$ is always very close to the expected value (within $1\%$ for all cases). However, by applying a more stringent threshold, there is a trend to reduce the abnormal peaks significantly. } 
\label{fig:cutover} 
\end{figure*}
which is the smallest number of elements needed to make the inequality hold. Then we can decompose the probability by conditioning on $T$, i.e., 
\begin{equation} \label{decompose}
P\{\frac{1}{N}\sum\limits_{i=1}^{N}\frac{1}{\cos{\theta_i}}>\frac{\pi}{2}+e\}=\sum\limits_{i=1}^{N}P\{T=i\}.
\end{equation}
We can write down each term explicitly. When $N$ is large enough, we have
\begin{gather}
\begin{align}
&P\{T=1\}=\int_{0}^{\arcsin{\frac{1}{N(\frac{\pi}{2}+e)}}}\sin{\theta}{\cos^{N-1}{\theta}}d\theta \propto \frac{1}{N^2},  \\
&P\{T=2\}=\int_{\arcsin{\frac{1}{N(\frac{\pi}{2}+e)}}}^{\arcsin{\frac{2}{N(\frac{\pi}{2}+e)}}}\sin{\theta}d\theta \notag \\
&\int_{\theta}^{\arcsin{\min\{1,\frac{1}{N(\frac{\pi}{2}+e)-\frac{1}{\sin{\theta}}}}\}}\sin{\phi}{\cos^{N-2}{\phi}} d\phi \propto \frac{1}{N^3}  \label{p2}
\end{align}
\\ \vdots \notag
\end{gather}
\indent The reason why Eq. (\ref{p2}) has this scaling behavior is because ${\cos^{N-2}{\phi}}$ is effectively zero when $\phi$ is larger than $1/\sqrt N$. One can continue this process and it is easy to see that the major contribution to the summation in Eq. (\ref{decompose}) comes from the first a few terms. Thus one can see that by increasing $N$ one can significantly reduce the chance of getting an abnormal peak in the simulation, which is why the results in Fig. \ref{fig:bench} are very smooth, especially compared to the ``$\delta=\infty$" case in Fig. \ref{fig:cutover}(a), which has a much smaller sampling number.

\section{The effect of setting a threshold}
\indent We still take Fig. \ref{fig:cos} as an example. By setting a threshold $\delta$, the new expected value of $1/\cos{\theta}$ is  
 \begin{equation}  \label{cutmean}
 \langle \frac{1}{\cos{\theta}}\rangle=\int_{\arccos{\frac{1}{\delta}}}^{0}\frac {d(1-\sin{\theta})}{\cos{\theta}}=\frac{\pi}{2}-\arcsin{\frac{1}{\delta}}.
 \end{equation}
 The relative error is then $2/{\pi}\arcsin{{1}/{\delta}}$. When $\delta \gg 1$, it is approximately $2(\pi \delta)^{-1}$. We can easily control this error by using a moderate threshold. For example, by setting $\delta=100$, the error is already below $0.7\%$. However, with this little compromise we can have a finite second moment, i.e.,    
 \begin{equation} \label{cutvar}
 \langle \frac{1}{\cos^2{\theta}}\rangle=\int_{\arccos{\frac{1}{\delta}}}^{0}\frac {d(1-\sin{\theta})}{\cos^2{\theta}}=\ln{\delta(1+\sqrt{1-\frac{1}{{\delta}^2}})}.
  \end{equation} 
Thus, one can reduce the fluctuation to any level simply by adding more samples. When $\delta \gg 1$, the variance is approximated by $\ln\delta$. In the case of $\delta=100$, one can reduce the relative standard deviation to $1\%$ by using around 12000 samples. It is noteworthy that the mean and variance have different asymptotic behaviors, the error diminishes as $1/\delta$, while the variance only grows as $\ln \delta$, which gives us a great flexibility to choose $\delta$.\\
 \indent The suppression of abnormal peaks by using a threshold $\delta$ can also be deduced from Eq. (\ref{decompose}). Since now $1/\cos \theta$ is bounded, it requires at least $N(\pi/2+e)/\delta$ terms to have the inequality in Eq. (\ref{defines}), which means the leading $N(\pi/2+e)/\delta$ terms in Eq. (\ref{decompose}) vanish, leaving the probability significantly smaller than the case without a threshold (actually by Hoeffding's inequality the probability will decrease exponentially).\\
\indent We test this idea on two models: overlapping spheres and Gaussian random fields. In the former case, we consider the same system in Sec. IV, while restricting ourselves to a relatively small amount of sampling lines (10,000) and compare the computed $F_{ss}$ with different thresholds $\delta=10$, 100, 1000, and $\infty$ (no threshold). The result is shown in Fig. \ref{fig:cutover}(a). Clearly, in all these cases, $F_{ss}$ 
fluctuates around a common curve, while as the threshold is tightened, fluctuations are suppressed. 
To show this point quantitatively, we focus on the behavior of $F_{ss}$ in the interval $[D, 11D]$. As noted in Sec. IV, when $r>D$ the surface-surface correlation function $F_{ss}$ is a constant $s^2$ in theory and should be a flat line in the plot; while in the simulations $F_{ss}$, fluctuates around a baseline. To compare the simulation and theoretical results, we compute the average of surface-surface correlation function $\langle F_{ss}(r) \rangle$ and the largest deviation from $s^2$ in this interval. As shown in Fig. \ref{fig:cutover}(b), the average of $F_{ss}(r)$ is always very close to the expected value (within $1\%$ error for all cases). However, by applying a more restrictive threshold, there is a trend to reduce the abnormal peaks significantly. We indeed get much smoother curves by making a very minor sacrifice of accuracy. \\

\end{appendix}

\end{document}